\documentclass[aps,prc,preprint,nofootinbib]{revtex4-2}

\usepackage{sidecap}

\usepackage{epsfig}
\usepackage{amsmath,amssymb,amsthm,mathtools}
\usepackage{graphicx}

\usepackage{xcolor}
\definecolor{ao}{rgb}{0.0, 0.5, 0.0}

\usepackage[hidelinks]{hyperref} 

\DeclareMathAlphabet{\mathpzc}{OT1}{pzc}{m}{it}

\definecolor{ao}{rgb}{0.0, 0.5, 0.0}

\begin{document}

\renewcommand{\textfraction}{0.00}


\newcommand{\vAi}{{\cal A}_{i_1\cdots i_n}}
\newcommand{\vAim}{{\cal A}_{i_1\cdots i_{n-1}}}
\newcommand{\vAbi}{\bar{\cal A}^{i_1\cdots i_n}}
\newcommand{\vAbim}{\bar{\cal A}^{i_1\cdots i_{n-1}}}
\newcommand{\htS}{\hat{S}}
\newcommand{\htR}{\hat{R}}
\newcommand{\htB}{\hat{B}}
\newcommand{\htD}{\hat{D}}
\newcommand{\htV}{\hat{V}}
\newcommand{\cT}{{\cal T}}
\newcommand{\cM}{{\cal M}}
\newcommand{\cMs}{{\cal M}^*}
\newcommand{\vk}{\vec{\mathbf{k}}}
\newcommand{\bk}{\bm{k}}
\newcommand{\kt}{\bm{k}_\perp}
\newcommand{\kp}{k_\perp}
\newcommand{\km}{k_\mathrm{max}}
\newcommand{\vl}{\vec{\mathbf{l}}}
\newcommand{\bl}{\bm{l}}
\newcommand{\bK}{\bm{K}}
\newcommand{\bb}{\bm{b}}
\newcommand{\qm}{q_\mathrm{max}}
\newcommand{\vp}{\vec{\mathbf{p}}}
\newcommand{\bp}{\bm{p}}
\newcommand{\vq}{\vec{\mathbf{q}}}
\newcommand{\bq}{\bm{q}}
\newcommand{\qt}{\bm{q}_\perp}
\newcommand{\qp}{q_\perp}
\newcommand{\bQ}{\bm{Q}}
\newcommand{\vx}{\vec{\mathbf{x}}}
\newcommand{\bx}{\bm{x}}
\newcommand{\tr}{{{\rm Tr\,}}}
\newcommand{\bc}{\textcolor{blue}}
\newcommand{\rc}{\textcolor{red}}

\newcommand{\beq}{\begin{equation}}
\newcommand{\eeq}[1]{\label{#1} \end{equation}}
\newcommand{\ee}{\end{equation}}
\newcommand{\bea}{\begin{eqnarray}}
\newcommand{\eea}{\end{eqnarray}}
\newcommand{\beqar}{\begin{eqnarray}}
\newcommand{\eeqar}[1]{\label{#1}\end{eqnarray}}

\newcommand{\nn}{\nonumber}
\newcommand{\tcr}{\textcolor{red}}
\newcommand{\dif}{\mathrm{d}}
\newcommand{\RAA}{$R_{AA}$}

\newcommand{\half}{{\textstyle\frac{1}{2}}}
\newcommand{\ben}{\begin{enumerate}}
\newcommand{\een}{\end{enumerate}}
\newcommand{\bit}{\begin{itemize}}
\newcommand{\eit}{\end{itemize}}
\newcommand{\ec}{\end{center}}
\newcommand{\bra}[1]{\langle {#1}|}
\newcommand{\ket}[1]{|{#1}\rangle}
\newcommand{\norm}[2]{\langle{#1}|{#2}\rangle}
\newcommand{\brac}[3]{\langle{#1}|{#2}|{#3}\rangle}
\newcommand{\hilb}{{\cal H}}
\newcommand{\pleft}{\stackrel{\leftarrow}{\partial}}
\newcommand{\pright}{\stackrel{\rightarrow}{\partial}}

\title{Radiative energy loss in a temperature‑evolving quark-gluon plasma with dynamical constituents}

\author{Bithika Karmakar}
\affiliation{Faculty of Nuclear Sciences and Physical Engineering,Czech Technical University in Prague, B\v{r}ehov\'a 7, 11519 Prague 1, Czech Republic }
\author{Magdalena Djordjevic}
\email{magda@ipb.ac.rs}
\affiliation{Institute of Physics Belgrade, University of Belgrade, 11080 Belgrade, Serbia}
\affiliation{Serbian Academy of Sciences and Arts, 11000 Belgrade, Serbia}
\begin{abstract}
We present a theoretical formalism for calculating first‑order‑in‑opacity radiative energy loss that incorporates the spatial and temporal temperature evolution of the quark–gluon plasma (QGP) in a finite‑size quantum chromodynamics medium with dynamical (i.e., moving) constituents. The derived expressions allow for general temperature profiles within a local‑equilibrium (slowly varying) approximation, enabling detailed evaluations of radiative energy loss across different medium‑evolution scenarios. Importantly, the resulting kernel applies to single partons ($R=0$) and, after imposing the
finite-cone selection, to the radiative out-of-cone contribution from a jet-initiating parton,
providing a unified starting point for precision QGP tomography.
\end{abstract}


\pacs{25.75.-q, 25.75.Nq, 12.38.Mh, 12.38.Qk}

\maketitle
\vspace*{.6cm}
\section{Introduction}
\label{sec1}
Quantum chromodynamics (QCD) predicts that, at extremely high energy densities, a new form of matter can be created. This form of matter, called quark-gluon plasma (QGP)~\cite{QGP1,QGP2,QGP3,QGP4}, consists of interacting quarks, antiquarks and gluons, which are no longer confined. According to the current cosmology, it is believed that QGP existed during the few microseconds after the Big Bang~\cite{Stock}. Today, this form of matter can be produced in the experiments involving ultra-relativistic heavy-ion collisions (so called Little Bangs), where heavy ions with extremely high kinetic energy are collided. Therefore, creating QGP in a laboratory  [Large Hadron Collider (LHC) at CERN, and Relativistic Heavy Ion Collider (RHIC) at Brookhaven National Laboratory] can enable us to both study the properties of this new form of matter, and also learn more about origin of matter at its most basic level. It is now widely accepted that QGP is discovered in RHIC and LHC experiments involving heavy-ion collisions, and the current challenge is to map its properties.

In this regard, we note that bulk QGP properties are traditionally explored through low-$p_\perp $ ($p_\perp  \lesssim 3$GeV) theory and data, which describe the collective motion of 99.9\% of QCD matter. Although crucial for the discovery of QGP~\cite{RHICRAA,LHCRAA}, rare high-$p_\perp $ ($p_\perp  \gtrsim 8$GeV) hadrons (and jets) have rarely been used to explore bulk QGP properties, mainly because of the challenges associated with accurately describing the interactions between high-$ p_\perp $ partons and the QGP—an essential step to using such probes for this purpose. However, some important bulk QGP properties are known to be difficult to constrain using low-$ p_\perp $ observables and corresponding theory/simulations. We thus advocate for \textit{high-$ p_\perp $ QGP tomography}, where low- and high-$ p_\perp $ physics jointly constrain the bulk QGP parameters.

The key idea behind QGP tomography is that different bulk medium parameters lead to different temperature profiles during QGP expansion and cooling. High-$p_\perp$ partons traversing the medium in different directions experience varying temperature evolutions and path lengths, resulting in different energy losses for both light and heavy flavor partons. For example, hadron quenching~\cite{QGP2}, observed through suppression patterns in high-$p_\perp$ $R_{AA}$~\cite{ALICE_CH_RAA,ALICE_D_RAA,ATLAS_CH_RAA,CMS_CH_RAA}, $v_2$, and higher harmonics~\cite{ALICE_CH_vn,ALICE_D_vn,ATLAS_CH_vn,CMS_CH_vn,CMS_D_vn}, is directly linked to the energy loss of fast partons traversing the QGP~\cite{Achenbach:2023pba,Apolinario:2022vzg,Zigic:2022xks,Zigic:2021rku,JETSCAPE:2020mzn,JETSCAPE:2020shq,JETSCAPE:2022hcb,Stojku:2020wkh,Stojku:2021yow,Soloveva:2021quj,Song:2015sfa,Xing:2023ciw,He:2022evt,Karmakar:2024jak,Karmakar:2023ity,Ke:2022gkq,Chien:2015vja}. Hadron observables in the $p_\perp$ range (8--35~GeV) are particularly powerful for tomography, as they exhibit strong flavor dependence and enhanced sensitivity to the QGP temperature. Comparing theoretical predictions with experimental data across both light and heavy flavors enables inference of temperature profiles---and thus bulk QGP parameters---consistent with both low- and high-$p_\perp$ data. Importantly, energy loss increases with temperature, making high-$p_\perp$ observables sensitive to hotter regions of the plasma, thereby providing a crucial complement to low-$p_\perp$ analyses for precision characterization of the QGP.

As mentioned above, theoretical predictions of hadron and jet quenching rely heavily on precise calculations of medium-induced energy loss. Traditionally, however, such calculations have been performed under one or more simplifying assumptions—such as a constant-temperature QCD medium, high-$ p_\perp $ partons produced at infinite past (instead of at the time of the heavy-ion collision), static scattering centers, or vacuum-like propagators (see~\cite{BDMPS,Z,ASW,GLV,DG,HT,HTM,AMY,MD_PRC,DH_PRL}, as well as various extensions, e.g.,~\cite{Andres,Andres2,Mehtar-Tani1,Mehtar-Tani2,Sievert1,Sievert2,Barata:2023qds,Sadofyev:2022hhw}). While these approaches have provided valuable insights, they inherently limit applicability for QGP tomography, where the plasma is produced at a finite time, spans a finite size, and exhibits strong spatial and temporal temperature variations, with dynamical (i.e., moving) constituents.

To address these limitations, we developed a dynamical energy loss formalism~\cite{MD_PRC,DH_PRL} for a nonevolving QGP, modeling the medium as a quasiparticle system at a constant effective temperature. Notably, an interacting quasiparticle system has been shown to provide a reasonable microscopic description of the QGP created at LHC and RHIC energies~\cite{Quasiparticle1,Quasiparticle2,Quasiparticle3}. In our formalism, the QGP constituents are treated as moving (dynamical) particles, rather than static scattering centers; this microscopic use of the term ``dynamical'' is distinct from the macroscopic space-time evolution introduced below through the local temperature profile. The calculations are based on the finite-temperature Hard Thermal Loop (HTL) approach, where infrared divergences are naturally regulated with no artificial cutoffs. This formalism laid the groundwork for understanding radiative and collisional energy loss in a finite-size, finite-temperature QCD medium composed of dynamical QGP constituents~\cite{MD_PRC,DH_PRL}. Importantly, radiative and collisional energy losses are calculated within the same theoretical framework and are applicable to both light and heavy flavor probes.

Consequently, unlike the widely used static approximation~\cite{StaticApprox}, where only the chromo-electric component contributes to energy loss, both chromo-electric and chromo-magnetic components naturally appear in our approach and are shown to be comparable in magnitude~\cite{MagneticComparison}. The significant contribution of the magnetic sector indicates that the finite magnetic mass~\cite{Maezawa:2010vj,Hart:2000ha,Borsanyi:2015yka} may play an important role in the energy loss mechanism. Within our framework~\cite{Djordjevic:2011dd}, the magnetic mass is consistently included through sum rules~\cite{SumRules}, enabling incorporation of non-perturbative effects without relying on explicit Hard Thermal Loop (HTL) propagators. This analytical generalization led to what we refer to as the \textit{generalized HTL approach}. 

Additionally, we phenomenologically implemented running coupling effects to further enhance the predictive power of the model~\cite{Djordjevic:2013xoa}. As angular-averaged $R_{AA}$ is known to be weakly sensitive to medium evolution~\cite{RAASensitivity}, it serves as an excellent test of the underlying parton energy loss framework. We extensively tested our formalism against a wide range of angular-averaged $R_{AA}$ data, showing good agreement across different hadron species (both light and heavy flavors), collision energies, and centralities~\cite{Djordjevic:2013xoa,DDB,MD_HF1,MD_HF2}, using a fixed parameter set and no additional tuning.

Our dynamical energy loss model also successfully addresses the long-standing heavy-flavor puzzles at both RHIC and the LHC~\cite{MD_HF1,MD_HF2}, while maintaining clear predictive power~\cite{Djordjevic:2015hra}. Furthermore, systematic validation~\cite{BD_JPG} confirmed that each incorporated effect is essential for an accurate description of parton–medium interactions, indicating that all ingredients must be retained for reliable modeling of high-$p_\perp$ phenomena. The strong agreement with experimental $R_{AA}$ data further supports the conclusion that our framework accurately captures the physics of high-$p_\perp$ parton energy loss in the QGP.

However, since high-$p_\perp$ anisotropic flow observables ($v_2$, $v_3$, etc.) are particularly sensitive to the medium's evolution, it is essential to extend our dynamical energy-loss formalism~\cite{MD_PRC,DH_PRL} to an evolving QGP with time- and space-dependent temperature. Specifically, we derive radiative energy-loss expressions by evaluating all relevant Feynman diagrams in a finite-size QGP using finite-temperature field theory and the Hard Thermal Loop (HTL) approach, under the assumption of local thermal equilibrium. The resulting framework enables calculations for general temperature profiles within a local-equilibrium (slowly varying) approximation, where the temperature profile is the only required input, while all other parameters are fixed by theory or standard literature values. This allows direct use of temperature profiles from bulk medium simulations without additional tuning, making high-$p_\perp$ observables precision tools for constraining QGP properties.

Notably, the energy loss expression derived in this manuscript forms the theoretical foundation 
of our DREENA framework~\cite{Zigic:2022xks,Zigic:2021rku,Stojku:2020wkh,Stojku:2021yow,
Karmakar:2024jak,Karmakar:2023ity}, 
which has been systematically applied to high-$p_\perp$ hadron data to extract bulk QGP 
properties by combining low- and high-$p_\perp$ theory with experimental results. The present 
manuscript provides the full derivation underlying these applications, thereby addressing an 
important methodological advancement and ensuring transparency and rigor in the theoretical 
foundations of DREENA. 

Moreover, the same radiative energy-loss formalism provides a natural starting point for jet 
observables by selecting gluons emitted outside a cone of opening angle $R$ around a 
jet-initiating parton. The derived expression therefore connects the single-parton limit 
($R=0$) with the finite-$R$ radiative out-of-cone contribution in a temperature-evolving 
medium. Here the cone parameter $R$ should be understood as a parton-level cone-selection 
parameter, not as a reconstructed-jet radius obtained after full in-medium shower evolution. 
In this work, our primary focus remains on hadron observables; we use the cone parameter 
mainly to illustrate qualitative trends, while deferring detailed quantitative predictions for 
reconstructed-jet observables to future studies. In this context, the present derivation 
corresponds to the single-emitter (jet-core) limit and does not model early vacuum-like hard 
splittings or full jet-substructure evolution. We note that complementary approaches in the 
literature address additional aspects such as multi-emitter coherence and full shower evolution 
(e.g., multi-stage frameworks such as JETSCAPE~\cite{JETSCAPE:2022jer}, 
coherence/antenna-based approaches~\cite{Casalderrey-Solana:2012evi}, 
Jet-Med~\cite{Caucal:2018dla}, and hybrid models~\cite{Casalderrey-Solana:2018wrw}); 
these are beyond the scope of the present work.

The structure of this paper is as follows: In Section~II (Methods and Results), we derive the first-order-in-opacity, medium-induced radiative energy--loss expression under the assumption of a temperature-evolving QCD medium. In Secs.~II.A--II.C, we present the derivation under the pure HTL assumption, while in Sec.~II.D we generalize the result to include finite magnetic mass and running coupling effects, consistent with the discussion above. Appendix~A outlines the assumptions underlying our calculations, and Appendices~B--L provide the detailed derivations of the relevant Feynman diagrams corresponding to Secs.~II.A--II.C. To facilitate the presentation, we follow the procedure and notation from~\cite{MD_PRC,DH_PRL}, which allows direct comparison to previous work and helps the reader track how medium evolution is incorporated step by step into the dynamical energy--loss formalism. While there are overlaps in methodology and figures with earlier studies, this approach ensures that extending the formalism to a more complex, temperature-evolving medium is clear and accessible.

In Section~III, we complement the analytic derivation with a compact set of illustrative 
observables, demonstrating how the formalism captures key qualitative features of partonic
radiative energy loss and finite-$R$ radiative out-of-cone contributions---including their
characteristic dependence on the traversed path length---in a temperature-evolving medium,
with a comparison to a constant-temperature case. Finally, in the Conclusion, we discuss the implications of our results for QGP tomography, summarize our findings, and provide an outlook for future research.

\section{Methods and Results}
\label{sec2}
\subsection{Medium-induced gluon radiation spectrum}

This section outlines the calculation of the medium-induced gluon radiation spectrum
for a high-$p_\perp$ parton of finite mass $M$. In the finite-$R$ application, this parton is
interpreted as the jet-initiating parton; $M$ denotes the mass of the emitting parton, not a
reconstructed-jet mass. We consider a high-$p_\perp$ parton produced at $\tau_0 = 0$ at an arbitrary position $(x_0, y_0)$, propagating in an arbitrary direction characterized by the angle $\phi$. While the specific initial temperature $T(x_0, y_0)$ and high $p_\perp$ parton direction are not the primary focus in this study, they serve as starting parameters for describing the temperature evolution along the path of the high-$p_\perp$ parton. As the high $p_\perp$ parton propagates through the medium, it encounters a temperature profile that evolves as $T(x_0 + \tau \cos\phi, y_0 + \tau \sin\phi)$, where $\tau$ represents the proper time~\cite{Zigic:2021rku}.

When the temperature experienced by the high-$p_\perp$ parton drops below the transition temperature $T_c=155$~MeV~\cite{Tcritical}, we assume that the high $p_\perp$ parton has exited the QGP~\cite{Zigic:2021rku}. At this point, the calculation of the energy loss is terminated. The total path-length $L$ of the high-$p_\perp$ parton in this study corresponds to the distance traveled by the parton from its creation point $(x_0, y_0)$ at $\tau_0 = 0$ until it leaves the medium, i.e., until the condition $T(x_0 + \tau \cos\phi, y_0 + \tau \sin\phi) < T_c$ is met.

This derivation focuses solely on the dependence of energy loss on the temperature profile along the path of the high-$p_\perp$ parton, without delving into the specifics of initial conditions or high $p_\perp$ parton orientation. Such a general treatment provides a robust analytical baseline, essential for developing frameworks like DREENA~\cite{Zigic:2022xks,Zigic:2021rku} to calculate high-$p_\perp$ observables. For the above expressions, the position along the trajectory is directly proportional to the proper time $\tau$. Consequently, we introduce the integration variable $d\tau$, where $\tau$ represents the proper time along the path of the high-$p_\perp$ parton.

Medium-induced radiative energy loss arises from gluon radiation caused by collisional interactions between a high-$p_\perp$ parton and the medium. In this study, we derive the first-order opacity radiative energy loss, considering a single collisional interaction between the high-$p_\perp$ parton and the medium, accompanied by the emission of one gluon~\cite{MD_PRC,DH_PRL}. In a separate study~\cite{Stojku:2023ell}, we numerically evaluated higher-order opacity corrections within our framework and found that their impact on hadron $R_{AA}$ is small. Therefore, the first-order opacity approximation provides a good description within the current formalism.

As in~\cite{DH_Inf,MD_PRC,DH_PRL}, we describe the medium as a quark-gluon plasma at temperature $T$ and zero baryon density, consisting of $n_f$ effective massless quark flavors in equilibrium with gluons. The key distinction between~\cite{MD_PRC,DH_PRL} and this study lies in the treatment of the medium's thermal equilibrium. In~\cite{MD_PRC,DH_PRL}, a global thermal equilibrium was assumed, where the temperature $T$ is constant throughout the QGP evolution. In contrast, this study accounts for local thermal equilibrium, incorporating temperature variations as the QGP evolves in space and time.

The formalism for energy loss in a finite-size dynamical QCD medium is detailed in
Appendices~\ref{app_M101C}-\ref{app_M12RL}, where the diagrams are evaluated using
finite-temperature field theory~\cite{Kapusta,Le_Bellac} and HTL-resummed
propagators~\cite{Le_Bellac}. In this study, the temperature is treated as a slowly varying
background field within the leading local-equilibrium approximation. Accordingly, the HTL
self-energies retain their standard equilibrium isotropic tensor structure and are evaluated at
the local temperature $T(\tau)$; at this order, we do not introduce additional projection
operators or tensor structures associated with the temperature-gradient vector $\partial_\mu T$.
Such gradient-induced corrections correspond to higher-order terms in the Wigner/gradient
expansion and are beyond the scope of the present work.

In Appendices~\ref{app_M101C}-\ref{app_M12RL}, we provide analytical derivations for Feynman diagrams~\cite{MD_PRC} that contribute to the first-order opacity energy loss. Each diagram represents a jet source $J$ producing a high-$p_\perp$ parton, which subsequently radiates a gluon with momentum $k=(\omega, k_z, k \bk)$ while exchanging a virtual gluon of momentum $q=(q_0, q_z, \bq)$ with the medium. The high $p_\perp$ parton emerges with momentum $p=(E, p_z, \bp)$. Following~\cite{GLV}, we assume $J$ varies slowly with momentum, so $J(p + k + q) \approx J(p + k) \approx J(p)$. Since the high-$p_\perp$ parton is produced at $\tau=0$, it can initially be either on-shell or off-shell, allowing more than one cut of the diagram to contribute to the energy loss, as detailed in the Appendices. These cuts interfere with each other, contributing to the Landau-Pomeranchuk-Migdal (LPM) effect~\cite{LPM} at high jet energies.

\subsection{HTL Propagators}

The HTL gluon propagator accounts for both transverse and longitudinal components~\cite{Kalash,Klimov,Gyulassy_Selikhov,Rebhan} and is expressed as:
\beq
i D^{\mu\nu}(l)= \frac{P^{\mu \nu }(l)}{l^2 - \Pi_T (l)} + \frac{Q^{\mu \nu }(l)}{l^2 - \Pi_L (l)}\,,
\eeq{dmnMed}
where $l=(l_0,\vec{l})$ is the 4-momentum of the gluon. In the present work we use the standard equilibrium, isotropic HTL self-energies evaluated at the local temperature $T(\tau)$ (local-equilibrium regime). We emphasize that at very early times the medium may be momentum-anisotropic and not locally equilibrated; incorporating such pre-equilibrium anisotropies would require anisotropic hard-loop self-energies/propagators and is beyond the scope of the present work. We also note that non-perturbative studies (including lattice-QCD constraints) indicate deviations from strict weak-coupling HTL expectations at moderate temperatures; a fully non-perturbative treatment at the level of medium propagators is beyond the scope of the present work. Instead, we account for key non-perturbative medium ingredients phenomenologically through finite magnetic screening and running coupling, as described in Subsection~II.D. The transverse and longitudinal projection tensors, $P_{\mu\nu}(l)$ and $Q_{\mu\nu}(l)$, are associated with the transverse ($\Pi_T$) and longitudinal ($\Pi_L$) form factors of the HTL gluon self-energy, given by:
\bea
\label{PiT}
\Pi_T (l) &=& \mu^2 \left[ \frac{y^2}{2} + \frac{y (1 - y^2)}{4} \ln\left(\frac{y+1}{y-1}\right)\right],
\,\,\,
\Pi_L (l) = \mu^2 \left[ 1 - y^2 - \frac{y (1 - y^2)}{2} \ln\left(\frac{y+1}{y-1}\right)\right],
\eea
where $y \equiv l_0 / |\vec{l}|$ and $\mu$ is the Debye screening mass.

To simplify the calculations, we work in the Coulomb gauge, where the non-zero components of the projection tensors reduce to:
\bea
P^{ij} (l) &=& -\delta^{ij} + \frac{l^i  l^j}{\vec{l}^2}\,,
\label{PQmunu}
\qquad
Q^{00}(l) = -\frac{l^2}{\vec{l}^2} = 1 - \frac{l_0^2}{\vec{l}^2} = 1 - y^2.
\eea
The exchanged gluon carries spacelike momentum~\cite{MD_Coll,BT,BT_fermions}, so only the Landau damping contribution ($q_0 \leq |\vec{q}|$) from the cut HTL effective propagator $D(q)$~\cite{MD_Coll,TG,BT} is relevant. Conversely, the radiated gluon has timelike momentum $k = (\omega, \vec{k})$, and its amplitude involves the quasi-particle contribution for $\omega > |\vec{k}|$ from the cut propagator $D(k)$~\cite{DG_TM,Kapusta,Le_Bellac}, which naturally separates these two contributions in momentum space.

\subsection{Gluon radiation spectrum in a locally thermalized, temperature-evolving medium}

We use the same kinematic approximations as in~\cite{MD_PRC,GLV,DG,ASW,AMY}, assuming the validity of the soft gluon ($\omega \ll E$) and soft rescattering ($\omega \gg |\vec{k}| \sim |\vec{q}| \sim q_0, q_z$) limits. In a separate study~\cite{Blagojevic:2018nve}, we relaxed the soft gluon approximation and found that the modifications to energy loss are negligible. Therefore, for the sake of simplicity and transparency, we retain the soft gluon approximation in the present derivation. After evaluating the diagrams detailed in Appendices~\ref{app_M101C}-\ref{app_M12RL}, the interaction rate is given as:
\beqar
\Gamma (E) &=& \frac{1}{N_J} M_{tot} = \frac{1}{N_J} (M_{1,0} + M_{1,1} + M_{1,2}),
\eeqar{Gamma}
where $M_{1,0}$, $M_{1,1}$, and $M_{1,2}$ correspond to contributions with zero, one, or two ends of the exchanged gluon $q$ connected to the radiated gluon $k$. The invariant distribution of high-$p_\perp$ partons $N_J$, determined by the effective jet source current~\cite{GLV}, is:
\beqar
N_J =  D_R \int  \frac{d^3p}{(2\pi)^3 2 E} |J(p)|^2,
\eeqar{NJ}
where $D_R = 3$ accounts for the high-$p_\perp$ parton colors.

The results for $M_{1,0}$, after combining Eqs.~(\ref{M101C_f})-(\ref{M1056RL_f}), are:
\bea
M_{1,0}  &=& \int_0^{L} d\tau \int  \frac{d^3p}{(2\pi)^3 2 E} |J(p)|^2 \int \frac{d^3k}{(2\pi)^3 2 \omega} \int \frac{d^2q}{(2\pi)^2} \,\, 4 g^4 [t_a, t_c][t_c, t_a] T \,\, v(q,T) \nn \\
&\times& \frac{\boldsymbol{k}^2}{(\boldsymbol{k}^2 + \chi(T))^2} \Big(1 - \cos (\xi(T) \tau)\Big)\,,
\eea
or equivalently:
\bea
M_{1,0} &=& D_R \int_0^{L} d\tau \int \frac{d^3p}{(2\pi)^3 2 E} |J(p)|^2 \int \frac{dx}{x} \frac{d^2k}{\pi} \frac{d^2q}{\pi} \,\, \frac{C_R \alpha_s}{\pi} \,\, C_2(G) \alpha_s T \,\, v(q,T) \nn \\
&\times& \frac{\boldsymbol{k}^2}{(\boldsymbol{k}^2 + \chi(T))^2} \Big(1 - \cos (\xi(T) \tau)\Big)\,, \label{M10_f}.
\eea
where $\chi(T) \equiv M^2 x^2 + m_g^2(T)$, $\xi(T) \equiv \frac{\vec{k}^2 + \chi(T)}{x E^+}$, $m_g(T)$ denotes the effective mass of the radiated gluon~\cite{DG_TM} and $v(q,T)$ is the effective potential. Here, we used $[t_a, t_c] [t_c, t_a] = C_2(G) C_R D_R$, with $C_2(G) = 3$, $C_R = 4/3$, and $D_R = 3$. In all expressions in this manuscript, we note that the temperature $T$ is defined as in Subsection IIA, from which we see that it explicitly depends on the proper time $\tau$. 

Similarly, the results for $M_{1,1}$, obtained from Eqs.~(\ref{M1112C_f})-(\ref{M113L4R_f}), are:
\bea
M_{1,1} &=& 2 D_R \int_0^{L} d\tau \int \frac{d^3p}{(2\pi)^3 2 E} |J(p)|^2 \int \frac{dx}{x} \frac{d^2k}{\pi} \frac{d^2q}{\pi} \,\, \frac{C_R \alpha_s}{\pi} \,\, C_2(G) \alpha_s T \,\, v(q,T) \nn \\
&\times& \frac{- \boldsymbol{k} \cdot (\boldsymbol{k} + \boldsymbol{q})}{(\boldsymbol{k}^2 + \chi(T)) ((\boldsymbol{k} + \boldsymbol{q})^2 + \chi(T))} \Big(1 - \cos (\zeta(T) \tau)\Big)\,, \label{M11_f}
\eea
where $\zeta(T) \equiv \frac{(\boldsymbol{k} + \boldsymbol{q})^2 + \chi(T)}{x E^+}$.

Finally, combining Eqs.~(\ref{M12C_f}) and~(\ref{M12RL_f}), $M_{1,2}$ becomes:
\bea
M_{1,2} &=& D_R \int_0^{L} d\tau \int \frac{d^3p}{(2\pi)^3 2 E} |J(p)|^2 \int \frac{dx}{x} \frac{d^2k}{\pi} \frac{d^2q}{\pi} \,\, \frac{C_R \alpha_s}{\pi} \,\, C_2(G) \alpha_s T \,\, v(q,T) \nn \\
&\times& \Bigg[ \frac{2 (\boldsymbol{k} + \boldsymbol{q})^2}{((\boldsymbol{k} + \boldsymbol{q})^2 + \chi(T))^2} \Big(1 - \cos (\zeta(T) \tau)\Big) - \frac{\boldsymbol{k}^2}{(\boldsymbol{k}^2 + \chi(T))^2} \Big(1 - \cos (\xi(T) \tau)\Big) \Bigg]\,. \label{M12_f}
\eea
Using Eqs.~\eqref{NJ}-\eqref{M12_f}, the interaction rate reduces to:
\bea
\Gamma(E) &=& \int_0^{L} d\tau \int \frac{dx}{x} \frac{d^2k}{\pi} \frac{d^2q}{\pi} \,\, \frac{C_R \alpha_s}{\pi}\,\, C_2(G) \alpha_s T \,\, v(q,T) \nn \\
&\times& \frac{2 (\boldsymbol{k} + \boldsymbol{q})}{(\boldsymbol{k} + \boldsymbol{q})^2 + \chi(T)} \left(\frac{\boldsymbol{k} + \boldsymbol{q}}{(\boldsymbol{k} + \boldsymbol{q})^2 + \chi(T)} - \frac{\boldsymbol{k}}{\boldsymbol{k}^2 + \chi(T)}\right) \Big(1 - \cos (\zeta(T) \tau)\Big)\,, \label{Gamma_f}
\eea
leading to the differential gluon radiation spectrum:
\beqar
\frac{d^2N_{g}^{\mathrm{rad}}}{dx d\tau} &=& \int \frac{d^2k}{\pi} \frac{d^2q}{\pi} \,\, \frac{1}{x} \frac{C_R \alpha_s}{\pi}\,\, C_2(G) \alpha_s T \,\, v(q,T) \nn \\
&\times& \frac{2 (\boldsymbol{k} + \boldsymbol{q})}{(\boldsymbol{k} + \boldsymbol{q})^2 + \chi(T)} \left(\frac{\boldsymbol{k} + \boldsymbol{q}}{(\boldsymbol{k} + \boldsymbol{q})^2 + \chi(T)} - \frac{\boldsymbol{k}}{\boldsymbol{k}^2 + \chi(T)}\right) \Big(1 - \cos (\zeta(T) \tau)\Big)\,\Theta\!\left(\frac{|\bk|}{xE}-R\right) \,.\nn\\
\eeqar{DeltaNl}
Here $\Theta(y)$ is the Heaviside step function imposing
$\theta\simeq|\bk|/(xE)>R$. For $R=0$, the $\Theta$ factor equals unity and
Eq.~\eqref{DeltaNl} gives the total radiative spectrum of a single parton. For $R>0$, the
$\Theta$ factor selects gluons emitted outside the cone and therefore gives the finite-$R$
radiative out-of-cone contribution from a jet-initiating parton.

In the present work, ``temperature-evolving medium'' means that the local radiation kernel is
evaluated at the temperature encountered at the scattering point along the high-$p_\perp$
parton trajectory,
$T=T(x_0+\tau\cos\phi,y_0+\tau\sin\phi)\equiv T(\tau)$. Thus the temperature evolution
enters Eq.~\eqref{DeltaNl} through all $T$-dependent medium ingredients, including
$v(\bq,T)$, $m_g(T)$, $\chi(T)$, $\xi(T)$, $\zeta(T)$, and the HTL self-energies
$\Pi_{L,T}(q,T)$, which are evaluated at $T=T(\tau)$. If $T(\tau)$ is replaced by a
constant temperature $T_{\rm const}$, Eq.~\eqref{DeltaNl} has the same local differential form,
but all medium-dependent quantities are evaluated at $T_{\rm const}$; after setting $R=0$
and integrating over $\tau$, this fixed-$T$ limit reproduces the result of~\cite{MD_PRC}.

At the leading local-equilibrium order considered here, the transition from a fixed-temperature
medium to a temperature-evolving medium does not generate additional explicit terms in
Eq.~\eqref{DeltaNl} that would vanish in the homogeneous fixed-temperature limit.
This leading local-equilibrium treatment is complementary to explicit gradient-expansion
corrections~\cite{Sadofyev:2022hhw,Sadofyev:2021,Kuzmin:2023hko}, which address
gradient- and flow-induced effects within GLV/Gyulassy--Wang-type opacity-expansion setups. In the
present dynamical-QCD-medium formalism, the corresponding explicit $\partial_\mu T$ terms belong to the next
order in the gradient expansion and are beyond the scope of this work. More generally, for two-time structures in cut diagrams, the medium is
evaluated at the mid-point of the amplitude and its conjugate, which corresponds to the leading
term of a Wigner/gradient expansion. Physically, the oscillatory factors
$1-\cos(\xi\tau)$ and $1-\cos(\zeta\tau)$ encode LPM interference over the corresponding
formation interval; if the temperature varies appreciably over that interval, higher-order
gradient/non-local terms would become important and are not included here. The corresponding
smooth-profile criterion is given in Appendices~A and~B.

\subsection{Generalization Towards Magnetic Screening and Running Coupling}

As noted in~\cite{MD_PRC}, in the pure HTL approach, each diagram contributing to energy loss in a finite-size dynamical QCD medium is logarithmically divergent as the transverse momentum of the exchanged gluon approaches zero, {\it{i.e.,}} $\boldsymbol{q} \rightarrow 0$. This divergence arises from the contribution of transverse gluon exchange to radiative energy loss~\cite{Wang_Dyn}. While Debye screening renders the longitudinal exchange infrared finite, the transverse exchange leads to a well-known logarithmic singularity due to the lack of magnetic screening~\cite{Le_Bellac} in the HTL approach. However, the infrared divergences are naturally regulated when all diagrams are accounted for, as shown in~\cite{MD_PRC}.

As noted in the introduction, different non-perturbative approaches~\cite{Maezawa:2010vj,Hart:2000ha,Borsanyi:2015yka} suggest a non-zero magnetic mass at RHIC and LHC. Non-perturbative effects are, therefore, important for a realistic description of energy loss interactions suitable for QGP tomography. In~\cite{Djordjevic:2011dd}, we generalized the energy loss formalism to consistently include non-zero magnetic screening, modifying the HTL effective potential $v(q, T)$ from
\beqar
\frac{\mu(T)^2}{\bq^2 \, (\bq^2+\mu(T)^2)} \rightarrow \frac{\mu_E(T)^2}{(\bq^2 + \mu_E(T)^2)\, (\bq^2+\mu_M(T)^2)},
\eeqar{EffectPoten}
where $\mu_E(T)$ and $\mu_M(T)$ represent the electric and magnetic screening masses, respectively, with the magnetic-to-electric mass ratio approximately given by $\mu_M(T)/\mu_E(T) \approx 0.6$~\cite{Borsanyi:2015yka}.

The electric screening (Debye mass) $\mu_E(T)$ is self-consistently determined by solving the expression from~\cite{Peshier:2006ah}:
\beqar
\frac{\mu_E(T)^2}{\Lambda_{\rm QCD}^2} \ln \left(\frac{\mu_E(T)^2}{\Lambda_{\rm QCD}^2}\right)=\frac{1+N_f/6}{11-2/3 \, N_f} \left(\frac{4 \pi T}{\Lambda_{\rm QCD}} \right)^2,
\eeqar{mu}
leading to~\cite{Karmakar:2023ity}
\beqar
\mu_E=\sqrt{\Lambda_{\rm QCD}^2 \frac{\psi(T)}{W(\psi(T))}},
\eeqar{Debye_mass}
where
\beqar
\psi(T)=\frac{1+\frac{n_f}{6}}{11-\frac{2}{3} n_f}
\left(\frac{4 \pi T}{\Lambda_{\rm QCD}}\right)^2 ,
\eeqar{psi_T}
and $W$ is Lambert's $W$ function, while $\Lambda_{\rm QCD}$ is the perturbative QCD scale. This procedure ensures that $\mu_E(T)$ is consistent with lattice QCD~\cite{Peshier:2006ah}.

Although the derivations in the appendices assume constant coupling, we include the running coupling $\alpha_S(Q^2)$, defined as~\cite{Field}:
\beqar
\alpha_S (Q^2)=\frac{4 \pi}{(11-2/3 N_f) \ln (Q^2/\Lambda_{\rm QCD}^2)},
\eeqar{alpha}
where the coupling enters the radiative energy loss expression as $\alpha_S (E \, T) \, \alpha_S (\frac{\bk^2+\chi(T)}{x})$ (see also~\cite{Djordjevic:2013xoa}).

By incorporating these considerations into Eq.~\eqref{DeltaNl}, the gluon radiation spectrum becomes:
\beqar
\frac{d^2N_{g}^{\mathrm{rad}}}{dx\,d\tau}
&=&
\int \frac{d^2k}{\pi}\,\frac{d^2q}{\pi}\;
\frac{2\,C_R C_2(G)\,T}{x}\,
\frac{\alpha_S(E T)\,
      \alpha_S\!\big(\tfrac{\bk^2+\chi(T)}{x}\big)}{\pi}\,
\frac{\mu_E^2(T)-\mu_M^2(T)}
     {(\bq^2+\mu_M^2(T))(\bq^2+\mu_E^2(T))}
\nonumber\\[3pt]
&&\hspace*{-2.2cm}\times
\frac{(\bk\!+\!\bq)}{(\bk\!+\!\bq)^2\!+\!\chi(T)}
\!\left[\frac{(\bk\!+\!\bq)}{(\bk\!+\!\bq)^2\!+\!\chi(T)}\!
       -\!\frac{\bk}{\bk^2\!+\!\chi(T)}\right]\!
\!\left(\!1\!-\!\cos\!\left[\!
   \frac{(\bk\!+\!\bq)^2\!+\!\chi(T)}{xE^+}\,\tau\!\right]\!\right)\!
\Theta\!\left(\!\frac{|\bk|}{xE}\!-\!R\!\right).
\eeqar{DeltaNDynl}

Here $\Theta(y)$ is the Heaviside step function. For $R=0$, the $\Theta$ factor equals unity and Eq.~\eqref{DeltaNDynl} reduces to the total medium-induced radiation spectrum of a single parton. For finite $R$, the $\Theta\!\big(|\bk|/(xE)-R\big)$ term restricts the integration to gluons emitted at angles larger than the cone size, so that the result corresponds to the finite-$R$ radiative out-of-cone contribution from a jet-initiating parton. In this way, Eq.~\eqref{DeltaNDynl} provides an 
expression applicable to both the single-parton limit ($R=0$) and finite-$R$ radiative out-of-cone contributions relevant for future jet applications, and represents the evolution-dependent kernel used to further develop the QGP tomography framework, DREENA~\cite{Zigic:2022xks,Zigic:2021rku}.

\section{Illustrative trends from the derived energy--loss kernel}

This study does not present new full–scale numerical analyses of parton energy loss or radiation spectra, as these have already been implemented and applied within the DREENA framework~\cite{Zigic:2022xks,Zigic:2021rku,Stojku:2020wkh,Stojku:2021yow,Karmakar:2024jak,Karmakar:2023ity}. Our primary goal here is, for the first time, to provide a complete and transparent derivation of the medium–induced gluon radiation spectrum in a temperature–evolving QGP within local thermal equilibrium and the generalized HTL approach. This derivation underpins prior DREENA applications and consolidates the theoretical foundation in a self–contained and accessible form. For comprehensive numerical results and phenomenological implications for hadron observables, we refer the reader to the original DREENA publications.

\begin{figure}[htbp]
  \centering
  \includegraphics[width=\linewidth]{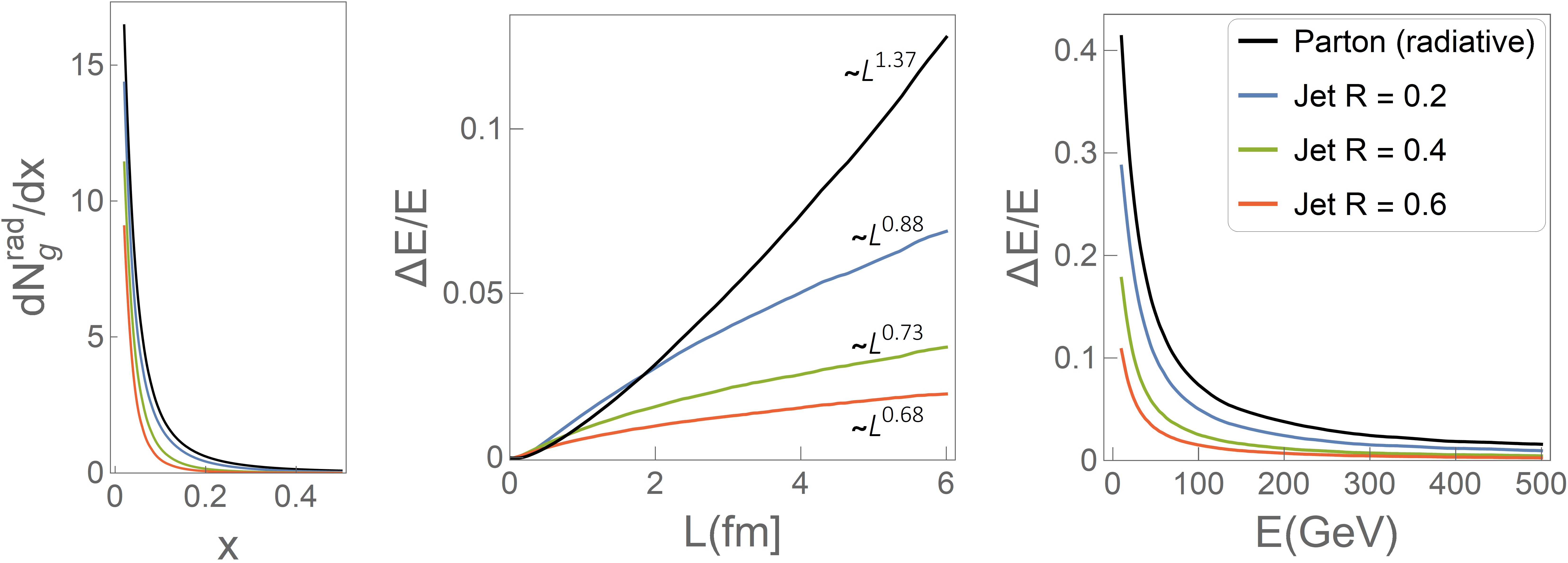}
  \vspace{-0.5em}
  \caption{Illustrative observables from the analytic energy--loss kernel.
(a) Medium--induced gluon radiation spectrum $dN_g/dx$ for an initial parton energy $E_0 = 100~\mathrm{GeV}$,
shown for a parton ($R = 0$) and for finite cone sizes
$R=0.2,0.4,0.6$, corresponding to the finite-$R$ radiative out-of-cone contribution from a
jet-initiating parton (see legend).
Angles are selected via $\theta \simeq k_\perp/(xE)$ so that the factor $\Theta(|\bk|/(xE) - R)$ implements the out-of-cone loss.
(b) Fractional energy loss $\Delta E/E$ as a function of path length $L$ at fixed $E_0 = 100~\mathrm{GeV}$;
slopes shown correspond to log--log fits (see text).
(c) Fractional energy loss $\Delta E/E$ as a function of initial energy $E$ at fixed $L = 4~\mathrm{fm}$.
All panels are computed from Eq.~(17) in a Bjorken 1D background following~\cite{Zigic:2018ovr} (0--10\% central Pb+Pb at $\sqrt{s_{NN}}=5.02~\mathrm{TeV}$),
with initialization at $\tau_0 = 0.6~\mathrm{fm}$ and $T_0 \approx 500~\mathrm{MeV}$;
$\Lambda_{\mathrm{QCD}} = 0.2~\mathrm{GeV}$, $n_f = 3$;
$M = \mu_E/\sqrt{6}$, $m_g = \mu_E/\sqrt{2}$~\cite{DG_TM};
$\mu_E$ defined by Eq.~(14) and $\mu_M/\mu_E = 0.6$~\cite{Borsanyi:2015yka}.}
\label{fig:illustrative}
\end{figure}

\begin{figure}[htbp]
  \centering
  \includegraphics[width=\linewidth]{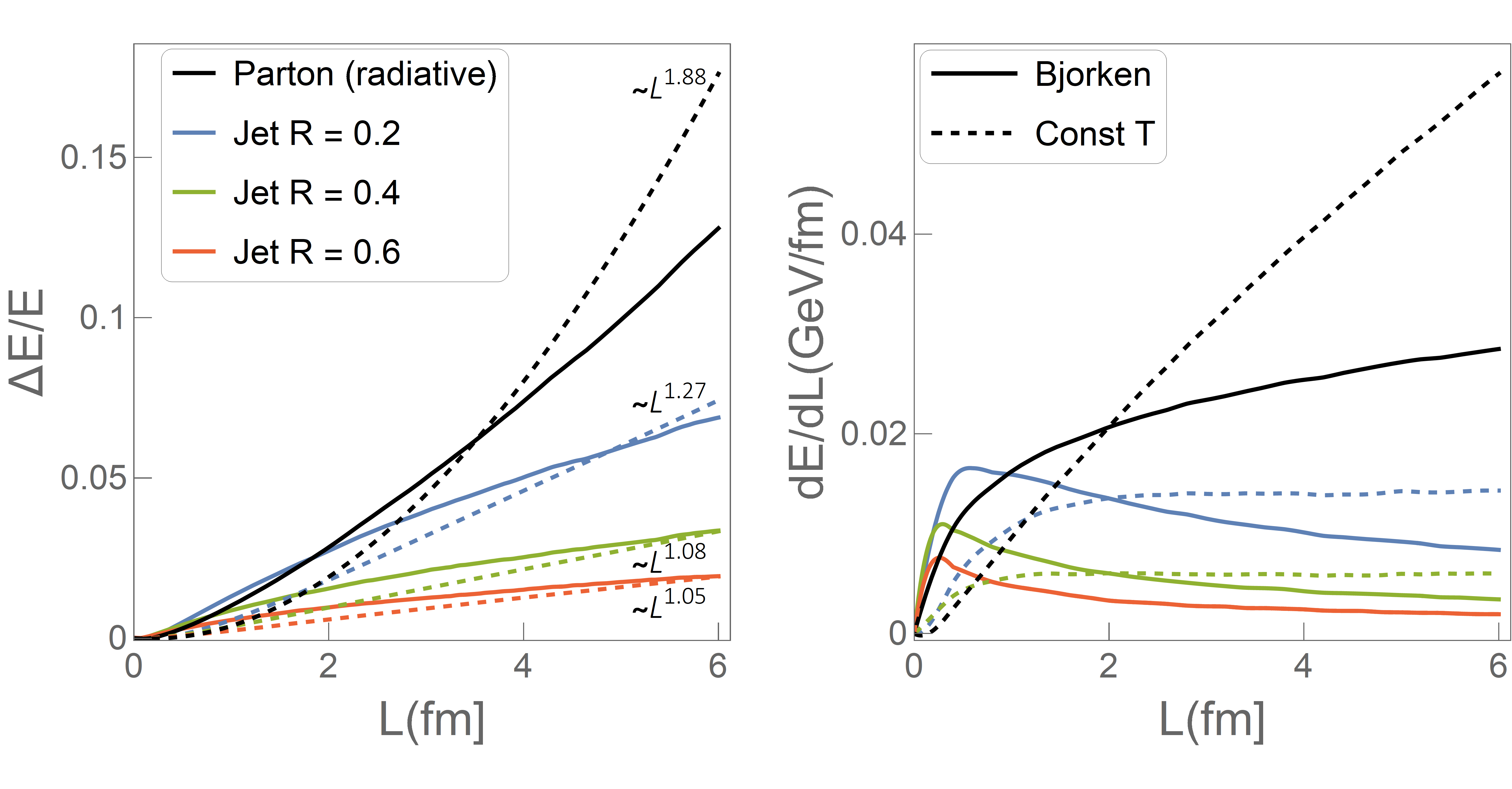}
  \vspace{-0.5em}
\caption{Constant-temperature brick compared to Bjorken 1D expansion.
Solid curves: Bjorken expansion (same setup and parameters as in Fig.~\ref{fig:illustrative}).
Dashed curves: a brick of length $L$ with fixed temperature $T_\mathrm{const}=0.335~\mathrm{GeV}$ 
(all other parameters identical), chosen as a representative effective temperature of the Bjorken profile 
used for $0$--$10\%$ central Pb+Pb collisions at $\sqrt{s_{NN}}=5.02~\mathrm{TeV}$ in~\cite{Zigic:2018ovr,Zigic:2018smz}.
Colors correspond to the partonic case ($R=0$, black) and to finite cone sizes
$R=0.2$ (blue), $R=0.4$ (green), and $R=0.6$ (red).
Left panel: $\Delta E/E$ vs $L$ at fixed $E_0=100~\mathrm{GeV}$; the indicated $\sim L^\beta$ slopes 
illustrate effective log--log fits for the constant-temperature brick over $L\simeq1$--$6~\mathrm{fm}$.
Right panel: the corresponding differential rate $dE/dL$ vs $L$, which highlights the distinct 
path-length build-up in the fixed-temperature case and its reduction in the Bjorken expansion.}
\label{fig:bjorken_vs_brick}
\end{figure}

In Figs.~\ref{fig:illustrative} and~\ref{fig:bjorken_vs_brick}, the finite-$R$ curves denote
the radiative out-of-cone contribution selected by $\Theta(|\bk|/(xE)-R)$ in the single-emitter
approximation, and should not be identified with full reconstructed-jet energy loss. The latter
can also involve elastic energy transfer and its angular redistribution relative to the jet cone,
recoil/medium-response effects, and in-medium shower evolution. The cone-restricted collisional
contribution, including microscopic recoil-sector effects, has been addressed separately in~\cite{Djordjevic:2026ConeColl} and will be combined with the present radiative kernel
in future DREENA jet applications.

From Fig.~\ref{fig:illustrative}, two robust features emerge (stated first for the $R=0$ partonic result, and then for finite-$R$ radiative out-of-cone contributions).

(i) For very short path lengths $L \ll t_f$, formation--time (coherence) effects suppress early small-angle (in-cone) radiation from the parton, keeping the total partonic energy loss small. In this regime, however, the finite-$R$ radiative out-of-cone contribution for a narrow cone
($R\!\sim\!0.2$) can be slightly larger than the total parton loss, because rare large-angle quanta (with shorter formation time) are less suppressed and feed the out-of-cone component, consistent with anti--angular ordering and finite-size coherence arguments~\cite{MehtarTani:2010uam,CasalderreySolana:2010eh}. As $L$ grows, the total parton loss rises faster and the ordering reverses. The same short-$L$ hierarchy reversal is also visible in the constant-$T$ brick (Fig.~\ref{fig:bjorken_vs_brick}), indicating that it is a generic finite-size/coherence effect.

(ii) At fixed $E_0=100~\mathrm{GeV}$ in our 1D Bjorken setup, the partonic energy loss increases with an effective scaling $\Delta E/E \sim L^{\beta}$ with $\beta\!\approx\!1.37$ over $L\simeq1$--$6$~fm (as indicated in Fig.~\ref{fig:illustrative}b). By contrast, the finite-$R$ radiative out-of-cone contribution exhibits a substantially weaker
path-length dependence that decreases with increasing cone size (see Fig.~\ref{fig:illustrative}b). Physically, as $L$ increases in an expanding medium, an increasing fraction of the induced radiation is emitted at smaller angles and is retained inside a larger cone, so the out-of-cone leakage grows more slowly with $L$ than the total partonic loss. This reduction relative to the constant-temperature case is quantified by Fig.~\ref{fig:bjorken_vs_brick}.

In the constant-temperature brick, the partonic loss grows close to quadratically, with an effective $\beta\simeq1.88$, and the corresponding differential rate $dE/dL$ increases with accumulated length (right panel). This is the characteristic path-length build-up of the LPM regime in a uniform medium; in the idealized asymptotic (very high-energy) limit and upon neglecting thermal masses one recovers an exact $L^2$ behavior~\cite{MD_PRC,DH_PRL}, while at finite energies and with temperature-dependent quasiparticle masses the effective exponent remains below $2$, as seen here. In contrast, in the expanding Bjorken medium the differential rate saturates for partons
and decreases for the finite-$R$ radiative out-of-cone contributions at large $L$, reflecting
the suppression of late-time contributions due to cooling and the progressive shift of induced
radiation toward smaller angles that remain inside the cone. Quantitatively, this leads to a
markedly reduced $L$-dependence when going from the constant-temperature brick to Bjorken
expansion, most prominently for the finite-$R$ curves: e.g., for $R=0.2,0.4,0.6$ the effective
exponents decrease from $\beta\simeq1.27,1.08,1.05$ (constant $T$) to
$\beta\simeq0.88,0.73,0.68$ (Bjorken), respectively (cf.\ Fig.~\ref{fig:illustrative}b).

These qualitative behaviors are consistent with recent studies performed within different frameworks.
Analyses based on universal scaling of high-$p_T$ hadron suppression indicate a path-length
dependence for partons between linear and quadratic~\cite{Djordjevic:2018ita,Arleo:2022universal,
Arleo:2024scaling}, matching the $R=0$ trend in our illustrations. For the finite-$R$ radiative
out-of-cone contribution, the sublinear exponent ($\beta_{R=0.4}\!\approx\!0.7$) is obtained
directly from our microscopic radiative kernel, without using jet data as input to the calculation. 
Interestingly, this value agrees with an independent Bayesian extraction of the effective
path-length exponent for jets ($\sim\!0.6$ for $R=0.4$)~\cite{Wu:2023PRC}, with the understanding
that the Bayesian extraction constrains a full jet-energy-loss framework, whereas the present
calculation isolates the radiative-only finite-$R$ contribution.
Other jet studies, however, report diverse effective $L$-dependences, ranging from near-linear
behavior~\cite{Arleo:2024scaling} to stronger scaling ($\sim\!1.5$--$2$)~\cite{Ogrodnik:2024qug}.
Although the precise finite-$R$ exponent may change once elastic/recoil effects, medium
response, shower evolution, or more realistic medium evolution are included, the qualitative
ordering---partonic radiative loss increasing faster with $L$ than the finite-$R$ radiative
out-of-cone contribution---is robust within the present radiative calculation. This ordering is physically
intuitive: radiation emitted at smaller angles remains inside the cone, so the out-of-cone
component grows more slowly with $L$ than the total partonic loss. To our knowledge, explicit
sublinear $L$-scaling for this finite-$R$ radiative contribution has not been reported before;
thus, our findings present a distinct, testable scenario. Finally, since in the small energy loss regime
$1-R_{AA}\!\approx\!\langle\Delta E\rangle/E$ (see, e.g.,~\cite{Djordjevic:2018ita}), the exponents
observed here for $\Delta E/E$ are expected to be reflected in future $R_{AA}$ predictions based
on this radiative kernel.

\section{Conclusion}

In this work, we have extended our previously developed dynamical energy loss formalism to include the space-time temperature evolution of QGP. This extension naturally incorporates temperature evolution within the dynamical energy loss framework, which is essential for QGP tomography. By allowing for general spatial and temporal temperature profiles within a local-equilibrium (slowly varying) approximation, the presented formalism offers a more realistic treatment of the QGP medium as encountered in heavy-ion collisions. Importantly, the resulting first-order-in-opacity radiative kernel applies to single-parton energy loss ($R=0$) and, through an out-of-cone selection, to the finite-$R$ radiative out-of-cone contribution from a jet-initiating parton. It therefore provides a unified starting point for hadron quenching and future jet applications within the DREENA framework.

The derived energy loss expression has been successfully integrated into our DREENA framework, designed for precision QGP tomography. DREENA allowed us to constrain bulk QGP parameters by comparing theoretical predictions for low- and high-$p_\perp$ hadron observables with experimental data, providing new insights into the properties of this extreme state of matter.
The ability of the framework to account for temperature evolution within this approximation significantly enhances its accuracy and applicability to experimental hadron data from RHIC and LHC, marking a substantial advancement in our understanding of high-$p_\perp$ parton–medium interactions and the properties of the QGP. In Section~III, we further showed that the same kernel suggests a robust ordering of path-length dependences, with the total partonic radiative loss increasing faster with $L$ than the finite-$R$ radiative out-of-cone contribution; the latter exhibits a sublinear effective scaling and thus provides a concrete, testable prediction for future comparisons within the present radiative single-emitter framework. The explicit constant-temperature brick vs Bjorken comparison illustrates how medium expansion weakens the effective path-length build-up
relative to a constant-temperature medium.

While the present formalism incorporates temperature evolution under local-equilibrium
assumptions, explicit gradient- and flow-induced corrections~\cite{Sadofyev:2022hhw,Sadofyev:2021,Kuzmin:2023hko} represent a higher-order extension of the present leading
local-equilibrium treatment and remain an important direction for future work. Other
refinements, such as non-equilibrium/anisotropic medium effects, also remain beyond the
scope of the present work~\cite{Romatschke:2003ms,Romatschke:2003vc}.

In addition, while the dynamical energy loss derived in this study is implemented within the DREENA framework as a precision tool for constraining QGP properties from hadron data, we also recognize the growing role of full jet observables in QGP tomography. In the broader literature, several complementary directions are being actively pursued to describe jet--medium interactions, including multi-stage modeling frameworks (e.g., JETSCAPE~\cite{JETSCAPE:2022jer}), coherence-based approaches~\cite{Casalderrey-Solana:2012evi}, the Jet-Med framework~\cite{Caucal:2018dla}, and hybrid models~\cite{Casalderrey-Solana:2018wrw}. At present, however, using jet measurements in the intermediate $p_\perp$ range (8--35~GeV) for precision tomography remains challenging, whereas hadron observables in this region exhibit strong sensitivity to medium properties; this motivates our focus on hadrons as a controlled and effective avenue for precision QGP tomography. Within the DREENA framework, a systematic extension to incorporate jet observables---thereby enabling a unified description of both hadron and jet quenching within a single, self-contained framework---constitutes a key next step toward next-generation precision QGP tomography.

\begin{acknowledgments}
We thank Igor Salom for valuable discussions. This work is supported by the European Research Council, grant ERC-2016-COG: 725741, and by the Ministry of Science and Technological Development of the Republic of Serbia. BK was also supported by the program Excellence Initiative–Research University of the University of Wroclaw of the Ministry of Education and Science, and Czech Science Foundation under project No. 25-16877S.
\end{acknowledgments}

\appendix

\section{Notation, assumptions and the propagators}
\label{appa}
In the following appendices, we perform the calculations using light-cone coordinates~\cite{Kogut}. The light-cone space-time coordinates $[x^+, x^-, \bx]$ and momentum coordinates $[p^+, p^-, \bp]$ are related to the laboratory frame coordinates $(t, z, \bx)$ and $(E, p_z, \bp)$ as follows:
\bea
x^+ &=& (t + z), \, \, x^- = (t - z), \label{x+-} \\
p^+ &=& (E + p_z), \, \, p^- = (E - p_z), \label{p+-}
\eea
where $\bx$ and $\bp$ are the transverse coordinates.

We consider a high-$p_\perp$ quark of finite mass $M$ and large spatial momentum $p' \gg M$, produced inside the medium at the point $x_0$. For simplicity, we choose the initial momentum of the quark to be aligned along the $z$-axis:
\bea
p'= [E'^+,p'^{-},\mathbf{0}]. \label{pprime0}
\eea
We consider the momentum of the virtual exchanged gluon and the radiated real gluon to be, respectively:
\bea
q &=& [q^+,q^-,\bq] = (q_0,\vec{\mathbf{q}}) = (q_0, q_z, \bq), \,\,\,\,\, q_0 \leq |\vec{\mathbf{q}}|, \label{q} \\
k &=& [k^+,k^-,\bk] = (k_0,\vec{\mathbf{k}}) = (k_0, k_z, \bk), \,\,\,\,\, k_0 \geq |\vec{\mathbf{k}}|. \label{k}
\eea
We assume the validity of the soft gluon ($\omega \ll E$) and soft rescattering ($|\bq| \sim |\bk| \ll k_z$) approximations~\cite{MD_PRC,GLV,DG,ASW,AMY}. Together with the conservation of energy and momentum ($p' = p + k + q$), one obtains:
\bea
p = \Bigl[E^+,\, p^-=\frac{\bp^2+M^2}{E^+},\,\bp\Bigr]
\; .
\eea

For the calculation of the Feynman diagrams in Appendices \ref{app_M101C}-\ref{app_M12RL}, we require the light-cone propagators $G_{++}^+(x)$, $G_{--}^-(x)$, and $G_{-+}(x)$ (the latter not being in light-cone coordinates) for the quark $p$, the radiated gluon $k$, and the exchanged gluon $q$. The relevant propagators are derived in~\cite{MD_PRC}. The propagators for the exchanged gluon are given by:
\beqar
D_{++}^{+ \mu \nu} (x_i-x_j) = D_{--}^{- \mu \nu} (x_i-x_j) \approx
D_{+-}^{\mu \nu} (x_i-x_j) \approx \int \frac{d^4 q} {(2 \pi)^4} D^>_{\mu\nu}
\big(q, T(\bar x_{ij})\big)\, e^{ -i q (x_i-x_j)} \, ,
\eeqar{delta_virtual_gluon}
where $\bar x_{ij}\equiv (x_i+x_j)/2$. In a homogeneous medium, Eq.~\eqref{delta_virtual_gluon} is exact. In an evolving medium, translational invariance is lost and the exact two-point function depends separately on $x_i$ and $x_j$; in our local-equilibrium (slowly varying) treatment we retain the leading term of a Wigner/gradient expansion by evaluating the medium-dependent quantities at the mid-point $\bar x_{ij}$ (equivalently, at the local proper time along the high-$p_\perp$ parton trajectory). Gradient corrections are suppressed by the small parameter $|x_i-x_j|/\tau_T$, where $\tau_T\equiv T/|dT/d\tau|$. For notational simplicity, here and below we use $D^{>}(q,T)$ as shorthand for $D^{>}(q,T(\bar{x}_{ij}))$ as defined in Eq.~(A7), i.e.\ the local temperature evaluated at the mid-point of the corresponding two time arguments. The quantity $D^{>}(q,T)$ below is the effective single-HTL cut gluon propagator for the exchanged gluon~\cite{DH_Inf}:

\bea
D^>_{\mu\nu} (q,T) = \theta\left(1-\frac{q_0^2}{\vq^2}\right) (1+f(q_0,T))\;
2 \, {\rm Im} \left(
\frac{P_{\mu \nu} (q)}{q^2{-}\Pi_{T}(q,T)} +
\frac{Q_{\mu \nu} (q)}{q^2{-}\Pi_{L}(q,T)} \right) \; , \label{exchanged_cut}
\eea
where $f(q_0,T)=(e^{q_0/T}{-}1)^{-1}$. $\Pi_T(q,T)$ and $\Pi_L(q,T)$ (see Eq.~(\ref{PiT})) are the transverse and longitudinal gluon self-energies, respectively. Additionally, $P_{\mu \nu}$ and $Q_{\mu \nu}$ represent the transverse and longitudinal projectors of the HTL gluon self-energy, respectively. 

Since the exchanged gluons are virtual, with momentum given in Eq.~\eqref{q}, only the Landau damping contribution from the gluon spectral function contributes to the above propagator.

In a finite-temperature QCD medium, the relevant radiative gluon propagator can be simplified as~\cite{DG_TM}:
\bea
D_{++}^{+\mu \nu} (x_i-x_j)&=&D_{--}^{-\mu \nu} (x_i-x_j)=
\int \frac{dk^+ d^2 k} {(2 \pi)^3 2 k^+}
\theta(k^+) P^{\mu \nu} (k) e^{-i k(x_i-x_j)} \label{delta+++gluon}\\
&&D_{-+}^{\mu \nu} (x_i-x_j)=\int \frac{d^3 k} {(2 \pi)^3 2 \omega}
P^{\mu \nu} (k) e^{-i k(x_i-x_j)} ,\label{delta-+gluon}
\eea
where $\omega(T) \approx \sqrt{\vk^2 + m_g^2(T)}$ and $k^+ = \omega + k_z$. The gluon mass is given by $m_g(T) \approx \mu_(T)/\sqrt{2}$~\cite{DG_TM}, with $\mu(T)$ being the Debye mass. The four-momentum $k$ can then be expressed as:
\bea
k = \Bigl[k^+,\, k^-=\frac{\bk^2{+}m_g^2(T)}{k^+},\,\bk\Bigr].
\eea

The relevant propagators for the high-$p_\perp$ quark are given by:
\beqar
\Delta_{++}^{+} (x_i-x_j)&=&\Delta_{--}^{-} (x_i-x_j)=
\int \frac{dp^+ d^2 p} {(2 \pi)^3 2 p^+} \theta(p^+) e^{-i p(x_i-x_j)} \label{delta+++jet} \\
&&\Delta_{-+} (x_i-x_j)=\int \frac{d^3 p} {(2 \pi)^3 2 E} e^{-i p(x_i-x_j)}. \label{delta-+jet}
\eea

Furthermore, in~\cite{DH_Inf}, it was shown that it is reasonable to assume $q_z \sim |\bq|$. Since $|\bk| \ll k_z$ and $q_z{\,\sim\,}|\bq|{\,\sim\,}|\bk|$, we then have $q_z{\,\ll\,}k_z$. Therefore, we get $k_z{+}q_z \approx k_z$ and $p_z{+}k_z{+}q_z \approx p_z{+}k_z \approx p_z{+}q_z \approx p_z$. Now, we define:
\beqar
  x \equiv \frac{k^+}{E^+} \approx \frac{(k+q)^+}{(E+q)^+} \; .
\eeqar{x}
We further define $\chi(T)$, $\xi(T)$, and $\zeta(T)$ as:
\beqar
\chi(T) &\equiv& M^2 x^2 + m_g^2(T) \nonumber \\
\xi(T) &\equiv& \frac{\bk^2+\chi(T)}{x E^{+}}  \nonumber \\
\zeta(T) &\equiv& \frac{(\bk{+}\bq)^{2}+\chi(T)}{x E^{+}} \nonumber \\
\zeta(T) - \xi(T) &=& \frac{(\bk{+}\bq)^{2} - \bk^2}{x E^{+}} \, .
\eeqar{xi-zeta}

Finally, since $\xi(T), \zeta(T) \ll |\bk| {\,\sim\,} |\bq|$, and $q_z{\,\sim\,}|\bq|{\,\sim\,}|\bk|$, we obtain $\xi(T), \zeta(T) \ll q_z$, which leads to the following relations:
\beqar
q^- + \xi(T) &=& q^0 - (q_z-\xi(T)) \approx  q^0 - q_z \nonumber \\
q^- + \zeta(T) &=& q^0 - (q_z-\zeta(T)) \approx  q^0 - q_z \, \nonumber \\
q^- \pm (\zeta - \xi(T)) &=& q^0 - (q_z \pm (\xi(T) - \zeta(T))) \approx  q^0 - q_z \,. \label{coll_approx}
\eeqar{q-xi-zeta}
For clarity in presenting the cumbersome expressions in Appendices~\ref{app_M101C}-\ref{app_M12RL}, we will not always explicitly indicate the $T$ dependence in $\omega,\, m_g,\, \mu, \,\chi,\, \xi, \,\zeta,\, D^>(q)$; however, it is understood that these variables depend on $T$. We assume that the QGP is in local equilibrium, meaning the temperature changes gradually with position and time. This study analytically derives the energy loss along the path of the high-$p_\perp$ parton, focusing on  proper time ($\tau$) dependence of the temperature profile. Since the position along the high-$p_\perp$ parton's trajectory is directly proportional to $\tau$, the temperature $T$ in this study depends only on $\tau$, as described in Subsection IIA.

To simplify the comparison between constant $T$ and evolving $T$ medium, in the following appendices, we keep the same notation and structure of the Feynman diagrams as in~\cite{MD_PRC}. 
In all Appendices, the diagrams are labeled as follows: In $M_{1, i, j, C}$, the number $1$ indicates contributions to the energy loss at first order in opacity; $i$ specifies how many ends of the virtual gluon $q$ are connected to the radiated gluon $k$; and $j$ identifies the specific diagram within that class. The letter $C$ indicates the central cut of the diagram, while in subsequent sections, the letters $R$ and $L$ will denote right and left cuts, respectively. Also, similar to~\cite{MD_PRC}, the calculations in Appendices~\ref{app_M101C}-\ref{app_M12RL} are performed under the standard equilibrium (isotropic) HTL setup for a locally equilibrated QGP; non-perturbative medium effects are subsequently incorporated phenomenologically through magnetic screening and running coupling as described in Subsection~II.D.
. 

\section{Calculation of Diagram $M_{1,0,1, C}$}
\label{app_M101C}
\begin{center}
\begin{figure}[htbp]
\includegraphics[scale=0.7]{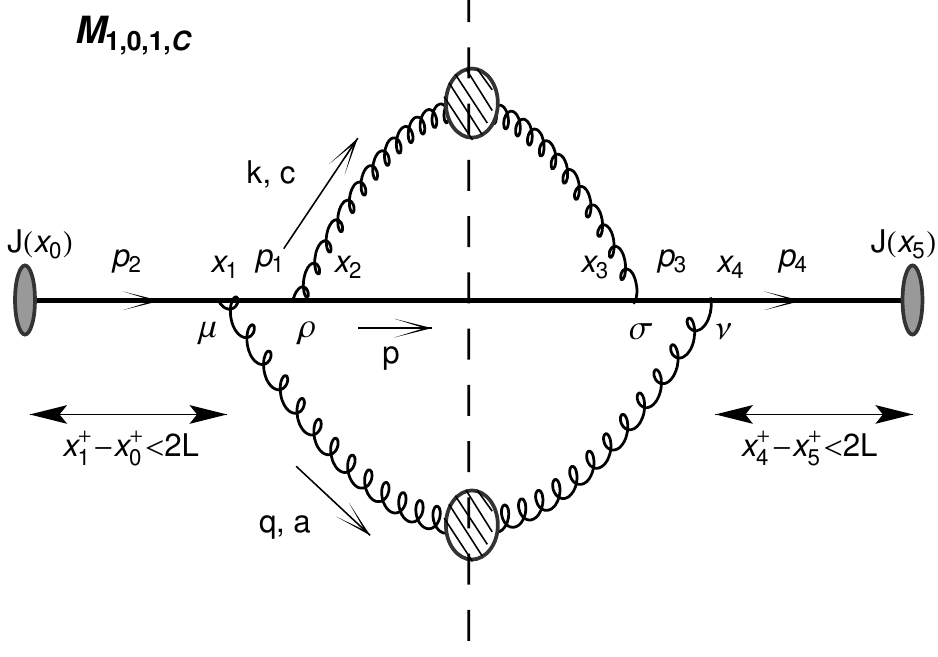}
\caption{Feynman diagram $M_{1,0,1, C}$, contributing to radiative energy loss at first order in opacity. The left (right) gray ellipse represents the source $J$, which produces a high-$p_\perp$ parton with momentum $p_2$ ($p_5$) at time $x_0$ ($x_5$). Large dashed circles denote effective HTL gluon propagators \cite{DG_TM}. The cut gluon propagators with momenta $k$ and $q$ correspond to the radiated gluon and a collisional interaction with a parton in the medium, respectively. Specific time points are denoted by $x_i$. Calculations are performed in the light cone coordinate system, with conditions $x_1^+-x_0^+ < 2L$ and $x_5^+-x_4^+ < 2L$, ensuring that the collisional interaction occurs before the high-$p_\perp$ parton exits the QGP. The total path-length $L$ traversed by the high-$p_\perp$ parton represents the distance traveled from its creation point $(x_0, y_0)$ at $\tau_0 = 0$ until the condition $T(x_0 + \tau \cos\phi, y_0 + \tau \sin\phi) < T_c$ is satisfied.}
\label{fig_M101C}
\end{figure}
\end{center}

In Appendices~\ref{app_M101C}-\ref{app_M105R6L}, we compute the amplitudes of diagrams where both ends of the exchanged gluon $q$ are attached to the high-$p_\perp$ parton, {\it{i.e.,}} neither end is connected to the radiated gluon $k$, and no three-gluon vertex is involved. This appendix presents the calculation of the diagram shown in Fig.~\ref{fig_M101C}.

\bea
M_{1,0,1, C}&= & \int \prod_{i=0}^{5} d^{4} x_{i} J\left(x_{0}\right) \Delta_{++}^{+}\left(x_{1}-x_{0}\right) v_{\mu}^{+}\left(x_{1}\right) D_{-+}^{\mu \nu}\left(x_{4}-x_{1}\right) \Delta_{++}^{+}\left(x_{2}-x_{1}\right) v_{\rho}^{+}\left(x_{2}\right) \nn\\
&&\times D_{-+}^{\rho \sigma}\left(x_{3}-x_{2}\right)  \Delta_{-+}\left(x_{3}-x_{2}\right) v_{\sigma}^{-}\left(x_{3}\right) \Delta_{--}^{-}\left(x_{4}-x_{3}\right) v_{\nu}^{-}\left(x_{4}\right) \Delta_{--}^{-}\left(x_{5}-x_{4}\right) J\left(x_{5}\right) \nn \\
& &\times \theta\left(x_{1}^{+}-x_{0}^{+}\right) \theta\left(x_{2}^{+}-x_{1}^{+}\right) \theta\left(2 L-\left(x_{1}^{+}-x_{0}^{+}\right)\right) \theta\left(x_{4}^{+}-x_{5}^{+}\right) \theta\left(x_{3}^{+}-x_{4}^{+}\right) \nn\\
&&\times \theta\left(2 L-\left(x_{4}^{+}-x_{5}^{+}\right)\right). \label{M101C}
\eea
Here $J$ represents the source of the high-$p_\perp$ parton, $\Delta$ corresponds to the high-$p_\perp$ parton propagator, $D$ denotes the gluon propagators, and $v$ represents the vertices. Using the expressions for the propagators provided in the previous section, we obtain:
\bea
M_{1,0,1, C}&= & \int \prod_{i=0}^{5} d^{4} x_{i} J\left(x_{0}\right) \int_{-\infty}^{\infty} \int_{0}^{\infty} \frac{d p_{2}^{+} d^{2} p_{2}}{(2 \pi)^{3} 2 p_{2}^{+}} e^{-i p_{2} \cdot\left(x_{1}-x_{0}\right)} \int \frac{d^{4} q}{(2 \pi)^{4}} D_{\mu \nu}^{>}(q,T) e^{-i q\left(x_{4}-x_{1}\right)} \nn\\
& &\times \int_{-\infty}^{\infty} \int_{0}^{\infty} \frac{d p_{1}^{+} d^{2} p_{1}}{(2 \pi)^{3} 2 p_{1}^{+}} e^{-i p_{1} \cdot\left(x_{2}-x_{1}\right)}\left(-i g\left(p_{2}+p_{1}\right)^{\mu} t_{a}\right) \nn \\
& &\times(-1) \int \frac{d^{3} k}{(2 \pi)^{3} 2 \omega} P_{\rho \sigma}(k) e^{-i k \cdot\left(x_{3}-x_{2}\right)} \int \frac{d^{3} p}{(2 \pi)^{3} 2 E} e^{-i p \cdot\left(x_{3}-x_{2}\right)}\left(-i g\left(p_{1}+p\right)^{\rho} t_{c}\right) \nn \\
&& \times \int_{-\infty}^{\infty} \int_{0}^{\infty} \frac{d p_{3}^{+} d^{2} p_{3}}{(2 \pi)^{3} 2 p_{3}^{+}} e^{-i p_{3} \cdot\left(x_{4}-x_{3}\right)}\left(i g\left(p+p_{3}\right)^{\sigma} t_{c}\right) \int_{-\infty}^{\infty} \int_{0}^{\infty} \frac{d p_{4}^{+} d^{2} p_{4}}{(2 \pi)^{3} 2 p_{4}^{+}} e^{-i p_{4} \cdot\left(x_{5}-x_{4}\right)}\nn\\
&&\times \left(i g\left(p_{3}+p_{4}\right)^{\nu} t_{a}\right)  J\left(x_{5}\right) \theta\left(x_{1}^{+}-x_{0}^{+}\right) \theta\left(x_{2}^{+}-x_{1}^{+}\right) \theta\left(2 L-\left(x_{1}^{+}-x_{0}^{+}\right)\right) \theta\left(x_{4}^{+}-x_{5}^{+}\right)\nn\\
&&\times \theta\left(x_{3}^{+}-x_{4}^{+}\right) \theta\left(2 L-\left(x_{4}^{+}-x_{5}^{+}\right)\right) \nn \\
&= & \int_{-\infty}^{\infty} \int_{0}^{\infty} \prod_{i=1}^{4} \frac{d p_{i}^{+} d^{2} p_{i}}{(2 \pi)^{3} 2 p_{i}^{+}} \int \frac{d^{3} k}{(2 \pi)^{3} 2 \omega} \int \frac{d^{3} p}{(2 \pi)^{3} 2 E} \int \frac{d^{4} q}{(2 \pi)^{4}} g^{4} t_{a} t_{c} t_{c} t_{a} \,I \label{M101C_b}
\eea
where
\bea
I&= & \int \prod_{i=0}^{5} d^{4} x_{i} \theta\left(x_{1}^{+}-x_{0}^{+}\right) \theta\left(x_{2}^{+}-x_{1}^{+}\right) \theta\left(2 L-\left(x_{1}^{+}-x_{0}^{+}\right)\right) \theta\left(x_{4}^{+}-x_{5}^{+}\right) \theta\left(x_{3}^{+}-x_{4}^{+}\right) \nn\\
&&\times \theta\left(2 L-\left(x_{4}^{+}-x_{5}^{+}\right)\right) e^{-i p_{2} \cdot\left(x_{1}-x_{0}\right)} e^{-i q \cdot\left(x_{4}-x_{1}\right)} e^{-i p_{1} \cdot\left(x_{2}-x_{1}\right)} e^{-i(p+k) \cdot\left(x_{3}-x_{2}\right)} e^{-i p_{3} \cdot\left(x_{4}-x_{3}\right)} \nn\\
&&\times e^{-i p_{4} \cdot\left(x_{5}-x_{4}\right)} J\left(x_{0}\right) J\left(x_{5}\right) \left(p_{1}+p_{2}\right)^{\mu} D_{\mu \nu}^{>}(q,T)\left(p_{3}+p_{4}\right)^{\nu}\left(p+p_{1}\right)^{\rho} P_{\rho \sigma}(k)\left(p+p_{3}\right)^{\sigma} \nn\\
&= & |J(p)|^{2}(2 \pi)^{3} \delta\left(\left(p_{2}-p-k-q\right)^{+}\right) \delta^{2}\left(\boldsymbol{p}_{2}-\boldsymbol{p}-\boldsymbol{k}-\boldsymbol{q}\right)(2 \pi)^{3} \delta\left(\left(p_{1}-p-k\right)^{+}\right) \delta^{2}\left(\boldsymbol{p}_{1}-\boldsymbol{p}-\boldsymbol{k}\right) \nn \\
&& \times(2 \pi)^{3} \delta\left(\left(p_{3}-p-k\right)^{+}\right) \delta^{2}\left(\boldsymbol{p}_{3}-\boldsymbol{p}-\boldsymbol{k}\right)(2 \pi)^{3} \delta\left(\left(p_{4}-p-k-q\right)^{+}\right) \delta^{2}\left(\boldsymbol{p}_{4}-\boldsymbol{p}-\boldsymbol{k}-\boldsymbol{q}\right) I_{1}, \nn\\
\label{M101C_I}
\eea
where
\bea
I_{1}&=&\int_{0}^{\infty} d x_{2}^{\prime+} e^{-\frac{i}{2}\left(p_{1}-p-k\right)^{-} x_{2}^{\prime+}} \int_{0}^{2 L} d x_{1}^{\prime+} e^{-\frac{i}{2}\left(p_{2}-p-k-q\right)^{-} x_{1}^{\prime+}} \int_{0}^{\infty} d x_{3}^{\prime+} e^{\frac{i}{2}\left(p_{3}-p-k\right)^{-} x_{3}^{\prime+}} \nn\\
&&\times \int_{0}^{2 L} d x_{4}^{\prime+} e^{\frac{i}{2}\left(p_{4}-p-k-q\right)^{-} x_{4}^{\prime+}} \left(p_{1}+p_{2}\right)^{\mu} D_{\mu \nu}^{>}(q,T)\left(p_{3}+p_{4}\right)^{\nu}\left(p+p_{1}\right)^{\rho} P_{\rho \sigma}(k)\left(p+p_{3}\right)^{\sigma} .\nn\\
\label{M101C_I1a}
\eea
Here $x_{1}^{\prime}=x_{1}-x_{0}, x_{2}^{\prime}=x_{2}-x_{1}, x_{3}^{\prime}=x_{3}-x_{4}, x_{4}^{\prime}=x_{4}-x_{5}$.

By applying the $\delta$ functions from Eq.~\eqref{M101C_I} and using
\bea
&&p_{i}^{-} =\frac{\boldsymbol{p}_{i}^{2}+M^{2}}{p_{i}^{+}}, \label{M101C_approx1pi}\\
&&k^{-}  =\frac{\boldsymbol{k}^{2}+m_{g}(T)^{2}}{k^{+}}, \label{M101C_approx1}
\eea
we obtain (noting that $\boldsymbol{p}+\boldsymbol{k}+\boldsymbol{q}=0 \rightarrow \boldsymbol{p}+\boldsymbol{k}=-\boldsymbol{q}$):
\bea
p_{1}^{-}=p_{3}^{-}=\frac{\boldsymbol{q}^{2}+M^{2}}{(p+k)^{+}}. \label{M101C_approx2}
\eea

In the soft gluon and soft rescattering approximation, we find (where $x \equiv \frac{k^{+}}{E^{+}}$, and $\xi(T)$ and $\zeta(T)$ are defined in Eq.~\eqref{xi-zeta}):
\bea
&& \left(p_{1}-p-k\right)^{-}=\left(p_{3}-p-k\right)^{-}=\frac{\boldsymbol{k}^{2}+\chi(T)}{x E^{+}}=-\xi(T), \nn \\
&& \left(p_{2}-p-k\right)^{-}=\frac{(\boldsymbol{k}+\boldsymbol{q})^{2}+\chi(T)}{x E^{+}}=-\zeta(T). \label{M101C_approx3}
\eea

Similarly, for highly energetic partons~\cite{MD_PRC},
\bea
\left(p+p_{1}\right)^{\rho} P_{\rho \sigma}(k)\left(p+p_{3}\right)^{\sigma} & \approx&- \frac{4\boldsymbol{k}^{2}}{x^{2}}\nn  \\
\left(p_{1}+p_{2}\right)^{\mu} P_{\mu \nu}(q)\left(p_{3}+p_{4}\right)^{\nu} & \approx& -\left(p_{1}+p_{2}\right)^{\mu} Q_{\mu \nu}(q)\left(p_{3}+p_{4}\right)^{\nu} \approx-E^{+2} \frac{\boldsymbol{q}^{2}}{\overrightarrow{\mathbf{q}}^{2}}, \label{M101C_reln1}
\eea

By using Eq.~\eqref{M101C_reln1} and Eq.~\eqref{exchanged_cut}, $\left(p_{1}+p_{2}\right)^{\mu} D_{\mu \nu}^{>}(q,T)\left(p_{3}+p_{4}\right)^{\nu}$ becomes
\bea
\left(p_{1}+p_{2}\right)^{\mu} D_{\mu \nu}^{>}(q,T)\left(p_{3}+p_{4}\right)^{\nu} & \approx&\left(p_{1}+p_{2}\right)^{\mu} 2 \operatorname{Im}\left(\frac{P_{\mu \nu}(q)}{q^{2}-\Pi_{T}(q,T)}+\frac{Q_{\mu \nu}(q)}{q^{2}-\Pi_{L}(q,T)}\right)\left(p_{3}+p_{4}\right)^{\nu}\nn \\
& \approx& \theta\left(1-\frac{q_{0}^{2}}{\overrightarrow{\mathbf{q}}^{2}}\right) f\left(q_{0},T\right) E^{+2} \frac{\boldsymbol{q}^{2}}{\overrightarrow{\mathbf{q}}^{2}} 2 \operatorname{Im}\left(\frac{1}{q^{2}-\Pi_{L}(q,T)}-\frac{1}{q^{2}-\Pi_{T}(q,T)}\right)\nn\\
&\equiv& F(q_0,q_z,\boldsymbol{q},T), \label{M101C_reln2}
\eea
where $f(q_0,T)=(e^{q_0/T}-1)^{-1}$. For small $q_0$, $f(q_0,T)$ can be expanded to~\cite{DH_Inf}
\bea
f(q_0,T)\sim T/q_0 . \label{M101C_approx4}
\eea

By using Eqs.~\eqref{M101C_approx1pi}-\eqref{M101C_approx4}, Eq.~\eqref{M101C_I1a} reduces to
\bea
I_{1}&=& 4\int_0^{\infty} dx_2'^{+}\, e^{-\frac{i}{2}\xi(T)  x_2'^{+}}\int_0^{\infty}  dx_3'^{+}\, e^{\frac{i }{2}\xi(T) x_3'^{+}} \,\frac{-4\boldsymbol{k}^2}{x^2} \int_0^L dl_1 \, e^{i(\zeta(T) + q)l_1} \int_0^{L}dl_4 \,e^{-i(\zeta(T) + q^-)l_4}\nn\\
&&\times\theta\left(1-\frac{q_{0}^{2}}{\overrightarrow{\mathbf{q}}^{2}}\right)
 \frac{T}{q_0}   E^{+2} \frac{\boldsymbol{q}^{2}}{\overrightarrow{\mathbf{q}}^{2}} 2 \operatorname{Im}\left(\frac{1}{q^{2}-\Pi_{L}(q,T)}-\frac{1}{q^{2}-\Pi_{T}(q,T)}\right) , \label{M101C_I1b}
\eea
where $l_1=x_1'^{+}/2$ and $l_4=x_4'^{+}/2$.

In an evolving medium, $\xi(T)$ (through its dependence on $m_g(T)$ and hence on $\chi(T)=M^2x^2+m_g^2(T)$) may, in principle, vary along the light-cone separations $x_2'^+$ and $x_3'^+$. Within our local-equilibrium (slowly varying) approximation we evaluate $\xi$ at the local temperature corresponding to the scattering time (mid-point prescription) and treat it as constant over the support of these integrals. This is justified because the integrals are dominated by $|x'^+|\sim 1/|\xi|$; hence the relative correction from temperature variation is parametrically of order $(1/|\xi|)/\tau_T\sim t_f/\tau_T$, where $\tau_T\equiv T/|dT/d\tau|$ is the temperature-variation scale and
$t_f\sim 1/|\xi(T)| = xE^+/(\boldsymbol{k}^2+\chi(T))$ is the corresponding formation interval. Finite parton and thermal gluon masses entering $\chi(T)$ therefore regulate the collinear limit and typically shorten $t_f$ relative to the schematic massless ($\chi\!\to\!0$) estimate $t_f\sim xE^+/\boldsymbol{k}^2 \approx \omega/\boldsymbol{k}^2$. The same smoothness criterion underlies the controlled treatment of the $l_1$--$l_4$ integration below.

Further, by using soft gluon, soft rescattering approximation, we have $\zeta(T) + q^- \approx q^-$~\cite{DH_Inf}. Therefore,
\bea
I_1 &=&  \frac{-16}{\xi(T)^2}\frac{4\boldsymbol{k}^2}{x^2} \int_0^L dl_1 \, e^{i q^-l_1} \int_0^{L}dl_4 \,e^{-i q^-l_4}\theta\left(1-\frac{q_{0}^{2}}{\overrightarrow{\mathbf{q}}^{2}}\right) E^{+2} \frac{\boldsymbol{q}^{2}}{\overrightarrow{\mathbf{q}}^{2}} \nn\\
&\times &  \frac{ T}{q_0}2\operatorname{Im}\left(\frac{1}{q^{2}-\Pi_{L}(q,T)}-\frac{1}{q^{2}-\Pi_{T}(q,T)}\right) ,\label{M101C_I1c}
\eea
where the second line of Eq.~\eqref{M101C_I1c} depends on the temperature $T$, which also affects $\xi(T), \Pi_{L}(q,T)$, and $\Pi_{T}(q,T)$. In a non-uniform medium this dependence is, in principle, associated with both $l_1$ and $l_4$ (i.e., with the two time arguments of the amplitude and its conjugate). To make the local-equilibrium assumption explicit and controlled, we introduce average and relative coordinates
\bea
l \equiv \frac{l_1+l_4}{2}, \qquad d \equiv l_1-l_4, \label{ld_def}
\eea
so that $l_1=l+d/2$ and $l_4=l-d/2$. Under local equilibrium, the medium properties vary slowly on the scale relevant for the collisional interaction, and we may evaluate the temperature-dependent prefactor at the mid-point $l$, up to gradient corrections. Equivalently, to leading order in a gradient expansion in $d$ we write
\bea
G\!\left(q,T\!\left(l\pm \frac{d}{2}\right)\right) &=& G\!\left(q,T(l)\right) \pm \frac{d}{2}\,\frac{dT}{dl}\,\partial_T G\!\left(q,T(l)\right)
\\
&&\hspace{-1.3em}+\;\frac{d^{2}}{8}\left[\frac{d^{2}T}{dl^{2}}\,\partial_T G\!\left(q,T\right)+\left(\frac{dT}{dl}\right)^{2}\,\partial_T^{2}G\!\left(q,T\right)\right]_{T=T(l)}+{\cal O}(d^{3}),
\label{G_gradexp}
\eea
so that odd powers of $d$ drop out after the (symmetric) integration over $d$ in a cut diagram, and the first non-vanishing corrections are ${\cal O}(d^2)$.
\bea
I_1 &=& \int_0^L dl_1 \, e^{i q^- l_1} \int_0^{L} dl_4 \, e^{-i q^- l_4}\,
G\!\left(q,\,T\!\left(\frac{l_1+l_4}{2}\right)\right) ,\label{M101C_I1d}
\eea
where
\bea
G(q,T) \!=\!\frac{-16}{\xi(T)^2}\!\frac{4\boldsymbol{k}^2}{x^2}
\theta\!\left(1-\frac{q_{0}^{2}}{\overrightarrow{\mathbf{q}}^{2}}\right)
\frac{T}{q_0} \,E^{+2}\,\frac{\boldsymbol{q}^{2}}{\overrightarrow{\mathbf{q}}^{2}}\,
2\! \operatorname{Im}\!\left(\!\frac{1}{q^{2}-\Pi_{L}(q,T)}\!-\!\frac{1}{q^{2}-\Pi_{T}(q,T)}\!\right) .
\label{G_def}
\eea
In Eq.~\eqref{M101C_I1d}, the temperature entering the collisional prefactor is evaluated at the
mid-point $l=(l_1+l_4)/2$ (leading term of the gradient/Wigner expansion in $d=l_1-l_4$).
The first non-vanishing corrections arise at ${\cal O}(d^2)$ and involve $\partial_l T$
(and higher derivatives); they are parametrically suppressed provided the temperature varies on a scale
$\tau_T\equiv T/|dT/dl|$ (equivalently, when expressed in proper time along the trajectory, $\tau_T=T/|dT/d\tau|$)
that is much larger than the typical relative separation $|d|$ contributing to the collisional kernel, i.e.\ $|d|/\tau_T\ll 1$.

To calculate the integrals in Eq.~\eqref{M101C_I1d}, we note that this part corresponds to the collisional interaction between the high-$p_\perp$ parton and a medium parton~\cite{MD_Coll}. For a constant temperature ($T=\text{const}$), $G(q, T)$ can be extracted from the integrals over $l_1$ and $l_4$, simplifying Eq.~\eqref{M101C_I1d} to:
\bea
I_1^{\text{const}} &=&  G(q,T) \int_0^L dl_1\, e^{iq^- l_1} \int_0^L dl_4 \, e^{-i q^- l_4}\nn\\
&=&  G(q,T)  \frac{e^{iq^- L}-1}{i q^-}\frac{e^{-i q^- L}-1}{-iq^-}\nn\\
&=&  G(q,T)  \frac{2(1-\cos{(q^-L)})}{(q^-)^2}\nn\\
&=&  G(q,T)  \frac{4 \sin^2{(\frac{q^-L}{2})}}{(q^-)^2}. \label{M101C_I2const}
\eea
We previously showed in~\cite{MD_Coll} that the finite size effects for collisional interactions are negligible, further reducing this expression to:
\bea
\lim_{L \to \infty}  \left(\frac{\sin\left(\frac{q^-L}{2}\right)}{\frac{q^-}{2}}\right)^2 = 2\pi L \,\delta (q^-). \label{M101C_Tconst}
\eea
To generalize our calculations from $T=\text{const}$ to the local equilibrium case, we again exploit that finite-size effects in the collisional part are negligible~\cite{MD_Coll}. Using Eq.~\eqref{ld_def} in Eq.~\eqref{M101C_I1d} and keeping the leading (mid-point) term in the gradient expansion, we obtain
\bea
I_1 &\simeq& \int_{0}^{L} dl\,G\!\left(q,T(l)\right)\int_{-\Delta(l)}^{\Delta(l)} dd\,e^{i q^- d},\qquad \Delta(l)\equiv 2\min(l,L-l). \label{I1_ld}
\eea
Neglecting boundary corrections (finite-size effects) in the relative coordinate $d$, we may extend the $d$-integration to $(-\infty,\infty)$, which yields
\bea
I_1  &\simeq&  2 \pi \,\delta(q^-) \int_0^L dl\, G\left(q, T(l)\right) . \label{M101C_I1g}
\eea
For notational economy in subsequent appendices, we will use the shorthand
\bea
\int_0^L dl\,e^{i q^- l} \Rightarrow 2\pi\,\delta(q^-), \label{M101_approx5}
\eea
with the understanding that it implements the above relative-coordinate ($d$) integration in the local-equilibrium/large-$L$ limit.

Therefore, $M_{1,0,1, C}$ reduces to:
\bea
M_{1,0,1,C}&=& \int_0^L d\tau  \int \frac{d^3p}{(2\pi)^3 2E} |J(p)|^2 \int \frac{d^3k}{(2\pi)^3 2\omega} 4g^4 t_a t_c t_c t_a \frac{\boldsymbol{k}^2}{(\boldsymbol{k}^2+\chi(T))^2} I_q(T), \label{M101C_c}
\eea
where
\bea
I_q (T)&=& T \int \frac{d^4q}{(2\pi)^4}2\pi \delta(q_0-q_z) \frac{1}{q_0} \frac{\boldsymbol{q}^{2}}{\overrightarrow{\mathbf{q}}^{2}}  2 \operatorname{Im}\left(\frac{1}{q^{2}-\Pi_{L}(q,T)}-\frac{1}{q^{2}-\Pi_{T}(q,T)}\right).  \label{M101C_Iq}
\eea
Here, the variable $l_1$ is replaced with the directly proportional proper time $\tau$, explicitly assuming that the temperature $T$ depends on $\tau$.

In the high-temperature limit, $I_q(T)$ reduces to (see Appendix C in~\cite{DH_Inf} for more details):
\bea
I_q(T)=T \int \frac{d^2q}{(2\pi)^2}v(q,T), \label{M101C_Iq_highT}
\eea
where $v(q,T)=\frac{\mu^2(T)}{\boldsymbol{q}^2(\boldsymbol{q}^2+\mu^2(T))}$ is the effective potential. Finally, Eq.~\eqref{M101C_c} becomes:
\bea
M_{1,0,1,C} = 4t_a t_c t_c t_a \int_0^L d\tau \int \frac{d^3p}{(2\pi)^3 2E} |J(p)|^2 \int \frac{d^3k}{(2\pi)^3  2\omega} \int \frac{d^2q}{(2\pi)^2}\,\, g^4 \, T \, v(q, T) \frac{\boldsymbol{k}^2}{(\boldsymbol{k}^2+\chi(T))^2}. \nn\\ \label{M101C_f}
\eea

\section{Calculation of Diagrams $M_{1,0,2,C}$, $M_{1,0,2,R}$, and $M_{1,0,2,L}$}
\label{app_M102}
\begin{center}
\begin{figure}[htbp]
\includegraphics[scale=0.67]{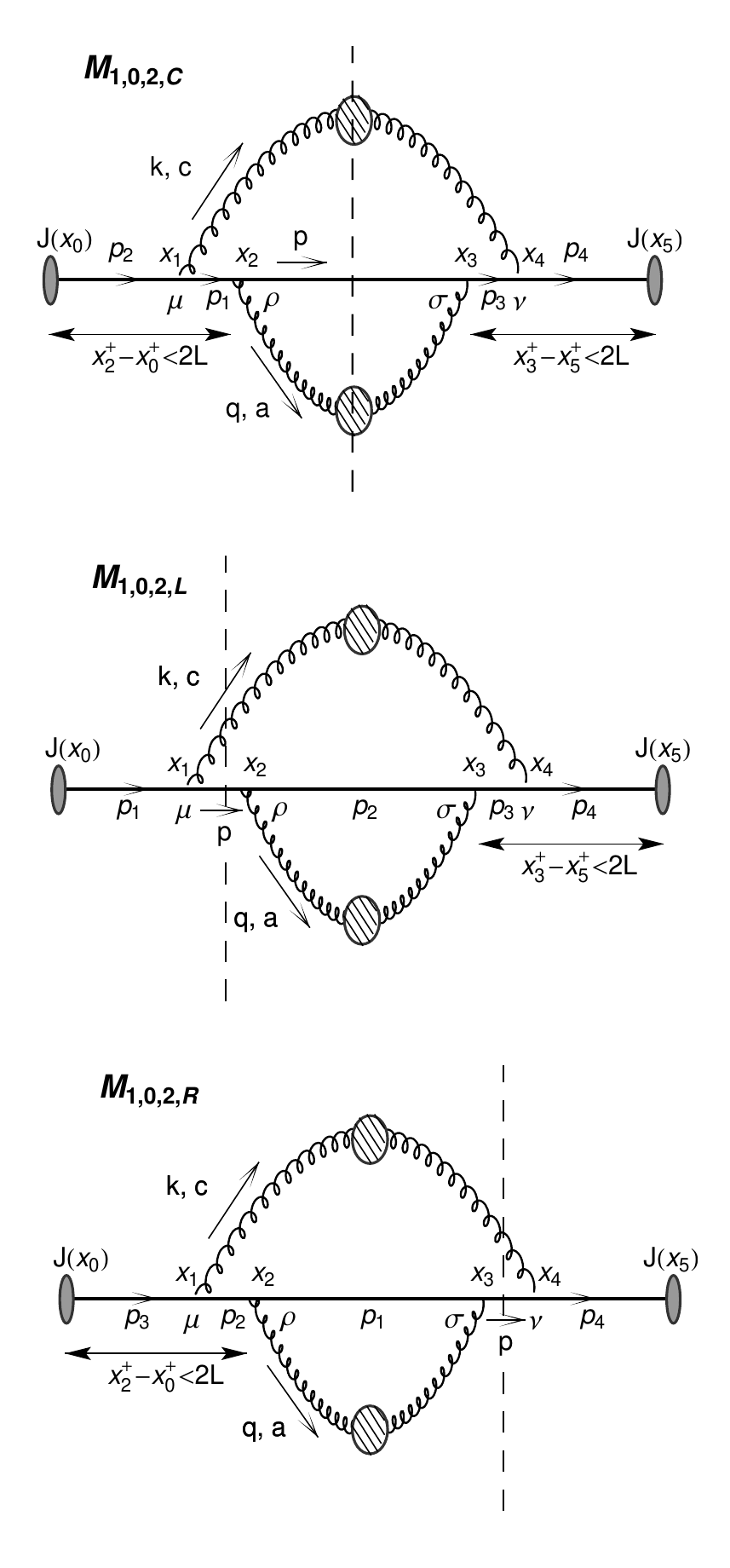}
\caption{Feynman diagrams $M_{1,0,2,C}$, $M_{1,0,2,L}$, and $M_{1,0,2,R}$, labeled equivalently to Fig.~\ref{fig_M101C}. The top, middle, and bottom panels depict three possible cuts (central, left, and right, respectively) of the same 2-HTL Feynman diagram, all contributing to the first-order in opacity radiative energy loss.}
\label{fig_M102}
\end{figure}
\end{center}

We will first calculate the cut diagram $M_{1,0,2,C}$, shown in the left panel of Fig.~\ref{fig_M102}. 
\bea
M_{1,0,2, C}&= & \int \prod_{i=0}^{5} d^{4} x_{i} J\left(x_{0}\right) \Delta_{++}^{+}\left(x_{1}-x_{0}\right) v_{\mu}^{+}\left(x_{1}\right) D_{-+}^{\mu \nu}\left(x_{4}-x_{1}\right) \Delta_{++}^{+}\left(x_{2}-x_{1}\right) v_{\rho}^{+}\left(x_{2}\right) D_{-+}^{\rho \sigma}\left(x_{3}-x_{2}\right)\nn  \\
&& \times \Delta_{-+}\left(x_{3}-x_{2}\right) v_{\sigma}^{-}\left(x_{3}\right) \Delta_{--}^{-}\left(x_{4}-x_{3}\right) v_{\nu}^{-}\left(x_{4}\right) \Delta_{--}^{-}\left(x_{5}-x_{4}\right) J\left(x_{5}\right) \nn \\
&& \times \theta\left(x_{1}^{+}-x_{0}^{+}\right) \theta\left(x_{2}^{+}-x_{1}^{+}\right) \theta\left(2 L-\left(x_{2}^{+}-x_{0}^{+}\right)\right) \theta\left(x_{4}^{+}-x_{5}^{+}\right) \theta\left(x_{3}^{+}-x_{4}^{+}\right) \theta\left(2 L-\left(x_{3}^{+}-x_{5}^{+}\right)\right)\nn \\
&= & -\int_{-\infty}^{\infty} \int_{0}^{\infty} \prod_{i=1}^{4} \frac{d p_{i}^{+} d^{2} p_{i}}{(2 \pi)^{3} 2 p_{i}^{+}} \int \frac{d^{3} k}{(2 \pi)^{3} 2 \omega} \int \frac{d^{3} p}{(2 \pi)^{3} 2 E} \int \frac{d^{4} q}{(2 \pi)^{4}} g^{4} t_{c} t_{a} t_{a} t_{c} I \label{M102C}
\eea
where
\bea
I&= & \int \prod_{i=0}^{5} d^{4} x_{i} \theta\left(x_{1}^{+}-x_{0}^{+}\right) \theta\left(x_{2}^{+}-x_{1}^{+}\right) \theta\left(2 L-\left(x_{2}^{+}-x_{0}^{+}\right)\right) \theta\left(x_{4}^{+}-x_{5}^{+}\right) \theta\left(x_{3}^{+}-x_{4}^{+}\right) \nn\\
&&\times \theta\left(2 L-\left(x_{3}^{+}-x_{5}^{+}\right)\right) e^{-i p_{2} \cdot\left(x_{1}-x_{0}\right)} e^{-i k \cdot\left(x_{4}-x_{1}\right)} e^{-i p_{1} \cdot\left(x_{2}-x_{1}\right)} e^{-i(p+q) \cdot\left(x_{3}-x_{2}\right)} e^{-i p_{3} \cdot\left(x_{4}-x_{3}\right)} \nn\\
&&\times e^{-i p_{4} \cdot\left(x_{5}-x_{4}\right)} J\left(x_{0}\right) J\left(x_{5}\right)\left(p_{1}+p_{2}\right)^{\mu} P_{\mu \nu}(k)\left(p_{3}+p_{4}\right)^{\nu}\left(p+p_{1}\right)^{\rho} D_{\rho \sigma}^{>}(q)\left(p+p_{3}\right)^{\sigma} \nn \\
&= & |J(p)|^{2}(2 \pi)^{3} \delta\left(\left(p_{2}-p_{1}-k\right)^{+}\right) \delta^{2}\left(\boldsymbol{p}_{2}-\boldsymbol{p}_{1}-\boldsymbol{k}\right)(2 \pi)^{3} \delta\left(\left(p_{1}-p-q\right)^{+}\right) \delta^{2}\left(\boldsymbol{p}_{1}-\boldsymbol{p}-\boldsymbol{q}\right) \nn \\
&& \times(2 \pi)^{3} \delta\left(\left(p_{3}-p-q\right)^{+}\right) \delta^{2}\left(\boldsymbol{p}_{3}-\boldsymbol{p}-\boldsymbol{q}\right)(2 \pi)^{3} \delta\left(\left(p_{4}-p_{3}-k\right)^{+}\right) \delta^{2}\left(\boldsymbol{p}_{4}-\boldsymbol{p}_{3}-\boldsymbol{k}\right) I_{1}, \label{M102C_I}
\eea
and where
\bea
I_{1}&=&\int_{0}^{2 L} d x_{2}^{\prime+} e^{-\frac{i}{2}\left(p_{1}-p-q\right)^{-} x_{2}^{\prime+}} \int_{0}^{x_{2}^{\prime+}} d x_{1}^{\prime+} e^{-\frac{i}{2}\left(p_{2}-p_{1}-k\right)^{-} x_{1}^{\prime+}} \int_{0}^{2 L} d x_{3}^{\prime+} e^{\frac{i}{2}\left(p_{3}-p-q\right)^{-} x_{3}^{\prime+}} \nn\\
&&\times \int_{0}^{x_{3}^{\prime+}} d x_{4}^{\prime+} e^{\frac{i}{2}\left(p_{4}-p_{3}-k\right)^{-} x_{4}^{\prime+}} \left(p_{1}+p_{2}\right)^{\mu} P_{\mu \nu}(k)\left(p_{3}+p_{4}\right)^{\nu}\left(p+p_{1}\right)^{\rho} D_{\rho \sigma}^{>}(q)\left(p+p_{3}\right)^{\sigma}. \label{M102C_I1_0}
\eea
Here, we defined 
$x_{1}^{\prime}=x_{1}-x_{0}$, $x_{2}^{\prime}=x_{2}-x_{0}$, $x_{3}^{\prime}=x_{3}-x_{5}$, and $x_{4}^{\prime}=x_{4}-x_{5}$.

Now, by applying the $\delta$ functions from Eq.~\eqref{M102C_I}, and using Eqs.~\eqref{M101C_approx1pi}, \eqref{M101C_approx1}, we obtain $p_{1}^{-}=p_{3}^{-}$ and $p_{2}^{-}=p_{4}^{-}$.

Using Eqs.~\eqref{M101C_reln1} and \eqref{M101C_reln2}, $I_1$ becomes:
\bea
I_1 &=& 4 \int_{0}^{ L} d l_{2}\, e^{-i\left(p_{1}-p-q\right)^{-} l_{2}} \int_{0}^{2l_{2}} d x_{1}^{\prime+} e^{-\frac{i}{2}\left(p_{2}-p_{1}-k\right)^{-} x_{1}^{\prime+}} \int_{0}^{L} d l_{3}\, e^{i\left(p_{3}-p-q\right)^{-} l_{3}} \nn\\
&& \times \int_{0}^{2l_{3}} d x_{4}^{\prime+} e^{\frac{i}{2}\left(p_{4}-p_{3}-k\right)^{-} x_{4}^{\prime+}} \frac{-4\boldsymbol{k}^{2}}{x^{2}} \theta\left(1-\frac{q_{0}^{2}}{\overrightarrow{\mathbf{q}}^{2}}\right) \frac{T}{q_{0}} E^{+2} \frac{\boldsymbol{q}^{2}}{\overrightarrow{\mathbf{q}}^{2}} \nn\\
&& \times 2 \operatorname{Im}\left(\frac{1}{q^{2}-\Pi_{L}(q,T)}-\frac{1}{q^{2}-\Pi_{T}(q,T)}\right), \label{M102C_I1}
\eea
where $l_2 = x_2'^+/2$ and $l_3 = x_3'^+/2$. Similar to Appendix~\ref{app_M101C}, the second line of Eq.~\eqref{M102C_I1} depends on the temperature $T$ through $\xi(T)$, $\Pi_L(q,T)$, and $\Pi_T(q,T)$. In a non-uniform medium this dependence is, in principle, associated with both time arguments ($l_2$ and $l_3$), i.e., with the amplitude and its conjugate. Within our local-equilibrium (slowly varying) approximation we evaluate these medium-dependent quantities at the mid-point of the two time arguments (leading term of a gradient/Wigner expansion), which reduces them to a single local temperature; the suppression of gradient corrections is discussed in Appendix~\ref{app_M101C}, Eqs.~\eqref{ld_def}--\eqref{M101C_I1g}. For notational simplicity we denote this local temperature by $T(l_2)$ below.

In soft gluon, soft rescattering approximation
\bea
&&\left(p_{1}-p\right)^{-} \approx \frac{\boldsymbol{k}^{2}-(\boldsymbol{k}+\boldsymbol{q})^{2}}{E^{+}} \nn \\
&& \left(p_{1}+k-p_{2}\right)^{-} \approx \xi(T) . \label{M102C_approx1}
\eea

Now, we perform the $x_1'^+$ and $x_4'^+$ integrations in Eq.~\eqref{M102C_I1}:
\bea
I_1 = \int_0^{L} dl_3 \, e^{-i\left(p_1-p-q\right)^- l_3} \left( 1-e^{i(p_1+k-p_2)^- l_3}\right) \int_0^{L} dl_2 \, e^{i(p_3-p-q)^- l_2} \left(1-e^{-i(p_1+k-p_2)^- l_2} \right) G\left(q,T\right), \nn\\
\label{M102C_I1_2_0}
\eea
where $G(q,T)$ is defined in Eq.~\eqref{G_def}.

Here, we note that the spacetime integrations over $l_2$ and $l_3$ correspond to the collisional interaction. We perform the $l_3$ integration using Eq.~\eqref{M101_approx5}, leading to:
\bea
I_1 &=& 2\pi \int_0^{L} dl_2 \Big[ \delta\big((p_2-p-k-q)^-\big) \Big( 1-e^{i(p_1+k-p_2)^- l_2}\Big) + \delta\big((p_1-p-q)^-\big) \nn\\
&&\times \Big( 1-e^{-i(p_1+k-p_2)^- l_2} \Big)\Big] G\left(q,T(l_2)\right). \label{M102C_I1_2}
\eea
Note that $\left(p_{1}-p\right)^{-} \ll \xi(T) \ll |\boldsymbol{k}|, |\boldsymbol{q}|, q_{z}$, leading to:
\bea
&& \left(\delta\left(\left(p_{2}-p-k-q\right)^{-}\right)+\delta\left(\left(p_{1}-p-q\right)^{-}\right)\right) \approx 2 \delta\left(\left(p_{2}-p-k-q\right)^{-}\right) \approx 2 \delta\left(q_{0}-q_{z}\right),\nn \\
&& \left(\delta\left(\left(p_{2}-p-k-q\right)^{-}\right)-\delta\left(\left(p_{1}-p-q\right)^{-}\right)\right) \approx \delta\left(q_{0}-q_{z}+\xi(T)\right)-\delta\left(q_{0}-q_{z}\right) .\label{M102C_approx2}
\eea
By using Eq.~\eqref{M102C_approx2}, Eq.~\eqref{M102C_I1_2} finally reduces to:
\bea
I_1 &=& \frac{\pi}{2}  \int_0^{L} d\tau \Big[ 2\delta(q_0-q_z) \Big( 1-\frac{e^{-i\xi(T) \tau}}{2}-\frac{e^{i\xi(T) \tau}}{2} \Big) + (\delta(q_0-q_z+\xi(T))-\delta(q_0-q_z))\nn\\
&&\times \Big( \frac{e^{-i\xi(T) \tau}}{2} -\frac{e^{i\xi(T) \tau}}{2}\Big)\Big]G\left(q,T\right),\label{M102C_I1f}
\eea
where, as in the previous section, the variable $l_2$ is here replaced with the directly proportional proper time $\tau$, and we explicitly assume that the temperature $T$ depends on $\tau$. 

By using Eqs.~\eqref{M102C_I} and \eqref{M102C_I1f}, and after performing integrations over $p_{1},\, p_{2},\, p_{3}$, and $p_{4}$, Eq.~\eqref{M102C} becomes:
\bea
M_{1,0,2, C}&= &  \int_0^{L} d\tau \int \frac{d^{3} p}{(2 \pi)^{3} 2 E}|J(p)|^{2} \int \frac{d^{3} k}{(2 \pi)^{3} 2 \omega} 4  g^{4} t_{c} t_{a} t_{a} t_{c}  \frac{\boldsymbol{k}^{2}}{\left(\boldsymbol{k}^{2}+\chi(T)\right)^{2}} \nn \\
&& \times \Big[2I_q(T) \Big( 1-\frac{e^{-i\xi(T) \tau}}{2}-\frac{e^{i\xi(T) \tau}}{2} \Big)  + J_q(T) \Big( \frac{e^{-i\xi(T) \tau}}{2} -\frac{e^{i\xi(T) \tau}}{2}\Big)   \Big] ,\label{M102C_2}
\eea
where $I_{q}(T)$ is given by Eq.~\eqref{M101C_Iq_highT}, and
\bea
J_{q}(T)&=&\int \frac{d^{4} q}{(2 \pi)^{4}}2\pi\left(\delta\left(q_{0}-q_{z}+\xi(T)\right)-\delta\left(q_{0}-q_{z}\right)\right) \frac{T}{q_{0}}\frac{\boldsymbol{q}^{2}}{\overrightarrow{\mathbf{q}}^{2}} 2 \operatorname{Im}\left(\frac{1}{q^{2}-\Pi_{L}(q,T)}-\frac{1}{q^{2}-\Pi_{T}(q,T)}\right)\nn\\
&=&\left.\xi(T) \int \frac{d^{3} q}{(2 \pi)^{3}}\frac{2\pi}{E^{+2}} \frac{d F\left(q_0,q_z,\boldsymbol{q},T\right)}{d q_{0}}\right|_{q_{0}=q_{z}}=0, \label{M102C_reln2}
\eea
as $\left.\frac{d F\left(q_0,q_z,\boldsymbol{q},T\right)}{d q_{0}}\right|_{q_{0}=q_{z}}$ is an odd function of $q_{z}$.

Finally, by using Eqs.~\eqref{M101C_Iq_highT} and \eqref{M102C_reln2}, Eq.~\eqref{M102C_2} reduces to:
\bea
M_{1,0,2, C}
&=& 8  t_{c} t_{a} t_{a} t_{c}\int_0^{L} d\tau\int \frac{d^{3} p}{(2 \pi)^{3} 2 E}|J(p)|^{2} \int \frac{d^{3} k}{(2 \pi)^{3} 2 \omega} \int \frac{d^{2} q}{(2 \pi)^{2}} \, g^{4}\,T\,  v(q,T) \nn\\
&& \times \frac{\boldsymbol{k}^{2}}{\left(\boldsymbol{k}^{2}+\chi(T)\right)^{2}} \Big(1-\cos{\left( \xi(T) \tau\right)}\Big). \label{M102C_f}
\eea

\bigskip

Now we proceed with the cut diagrams $M_{1,0,2, L}$ and $M_{1,0,2, R}$, shown in the central and right panels of Fig.~\ref{fig_M102}, respectively. We begin with $M_{1,0,2, R}$:
\bea
M_{1,0,2, R}&= & \int \prod_{i=0}^{5} d^{4} x_{i} J\left(x_{0}\right) \Delta_{++}^{+}\left(x_{1}-x_{0}\right) v_{\mu}^{+}\left(x_{1}\right) D_{-+}^{\mu \nu}\left(x_{4}-x_{1}\right) \Delta_{++}^{+}\left(x_{2}-x_{1}\right) v_{\rho}^{+}\left(x_{2}\right) \nn\\
&&\times D_{++}^{+\rho \sigma}\left(x_{3}-x_{2}\right)  \Delta_{++}^{+}\left(x_{3}-x_{2}\right) v_{\sigma}^{+}\left(x_{3}\right) \Delta_{-+}\left(x_{4}-x_{3}\right) v_{\nu}^{-}\left(x_{4}\right) \Delta_{--}^{-}\left(x_{5}-x_{4}\right) J\left(x_{5}\right) \nn \\
& &\times \theta\left(x_{1}^{+}-x_{0}^{+}\right) \theta\left(x_{2}^{+}-x_{1}^{+}\right) \theta\left(x_{3}^{+}-x_{2}^{+}\right) \theta\left(2 L-\left(x_{2}^{+}-x_{0}^{+}\right)\right) \theta\left(x_{4}^{+}-x_{5}^{+}\right) \nn \\
&= & \int_{-\infty}^{\infty} \int_{0}^{\infty} \prod_{i=1}^{4} \frac{d p_{i}^{+} d^{2} p_{i}}{(2 \pi)^{3} 2 p_{i}^{+}} \int \frac{d^{3} k}{(2 \pi)^{3} 2 \omega} \int \frac{d^{3} p}{(2 \pi)^{3} 2 E} \int \frac{d^{4} q}{(2 \pi)^{4}} g^{4} t_{c} t_{a} t_{a} t_{c} I \label{M102R}
\eea
where
\bea
I&= & \int \prod_{i=0}^{5} d^{4} x_{i} \theta\left(x_{1}^{+}-x_{0}^{+}\right) \theta\left(x_{2}^{+}-x_{1}^{+}\right) \theta\left(x_{3}^{+}-x_{2}^{+}\right) \theta\left(2 L-\left(x_{2}^{+}-x_{0}^{+}\right)\right) \theta\left(x_{4}^{+}-x_{5}^{+}\right) \nn \\
&& \times e^{-i p_{3} \cdot\left(x_{1}-x_{0}\right)} e^{-i k \cdot\left(x_{4}-x_{1}\right)} e^{-i p_{2} \cdot\left(x_{2}-x_{1}\right)} e^{-i\left(p_{1}+q\right) \cdot\left(x_{3}-x_{2}\right)} e^{-i p \cdot\left(x_{4}-x_{3}\right)} e^{-i p_{4} \cdot\left(x_{5}-x_{4}\right)} J\left(x_{0}\right) J\left(x_{5}\right) \nn \\
&&\times \left(p_{2}+p_{3}\right)^{\mu} P_{\mu \nu}(k)\left(p_{+} p_{4}\right)^{\nu}\left(p_{1}+p_{2}\right)^{\rho} D_{\rho \sigma}^{>}(q)\left(p+p_{1}\right)^{\sigma}\nn\\
&= & |J(p)|^{2}(2 \pi)^{3} \delta\left(\left(p_{3}-p_{2}-k\right)^{+}\right) \delta^{2}\left(\boldsymbol{p}_{3}-\boldsymbol{p}_{2}-\boldsymbol{k}\right)(2 \pi)^{3} \delta\left(\left(p_{2}-p\right)^{+}\right) \delta^{2}\left(\boldsymbol{p}_{2}-\boldsymbol{p}\right) \nn \\
&& \times(2 \pi)^{3} \delta\left(\left(p_{1}-p+q\right)^{+}\right) \delta^{2}\left(\boldsymbol{p}_{1}-\boldsymbol{p}+\boldsymbol{q}\right)(2 \pi)^{3} \delta\left(\left(p_{4}-p-k\right)^{+}\right) \delta^{2}\left(\boldsymbol{p}_{4}-\boldsymbol{p}-\boldsymbol{k}\right) I_{1}, \label{M102R_I}
\eea
and where
\bea
I_{1}&=&\int_{0}^{2 L} d x_{2}^{\prime+} e^{-\frac{i}{2}\left(p_{2}-p\right)^{-} x_{2}^{\prime+}} \int_{0}^{x_{2}^{\prime+}} d x_{1}^{\prime+} e^{-\frac{i}{2}\left(p_{3}-p_{2}-k\right)^{-} x_{1}^{\prime+}} \int_{0}^{\infty} d x_{3}^{\prime+} e^{-\frac{i}{2}\left(p_{1}-p+q\right)^{-} x_{3}^{\prime+}}\nn\\
&&\times \int_{0}^{\infty} d x_{4}^{\prime+} e^{\frac{i}{2}\left(p_{4}-p-k\right)^{-} x_{4}^{\prime+}}\left(p_{2}+p_{3}\right)^{\mu} P_{\mu \nu}(k)\left(p_{+} p_{4}\right)^{\nu}\left(p_{1}+p_{2}\right)^{\rho} D_{\rho \sigma}^{>}(q)\left(p+p_{1}\right)^{\sigma}. \label{M102R_I1}
\eea
Here we defined $x_{1}^{\prime}=x_{1}-x_{0}, x_{2}^{\prime}=x_{2}-x_{0}, x_{3}^{\prime}=x_{3}-x_{2}, x_{4}^{\prime}=x_{4}-x_{5}$.

By using Eq.~\eqref{M101C_approx1pi} and by applying the $\delta$ functions from Eq.~\eqref{M102R_I}, we obtain the following relations:
\bea
&& p_{2}^{-}=p^{-} \rightarrow\left(p_{2}-p\right)^{-}=0, \nn\\
& &p_{3}^{-}=p_{4}^{-}=\frac{(\boldsymbol{p}+\boldsymbol{k})^{2}+M^{2}}{(p+k)^{+}} \approx \frac{M^{2}}{E^{+}}, \nn\\
&&\left(p+k-p_{3}\right)^{-}=\xi(T) ,\nn\\
&&\left(p_{1}-p+q\right)^{-} \approx q^{-} . \label{M102R_approx1}
\eea

Using Eqs.~\eqref{M101C_reln1} and \eqref{M101C_reln2}, Eq.~\eqref{M102R_I1} can be written as 
\bea
I_{1} & =&4\int_{0}^{L} d l_{2} \int_{0}^{2l_{2}} d x_{1}^{\prime+} e^{\frac{i}{2} \xi(T) x_{1}^{\prime+}} \int_{0}^{\infty} d l_{3} e^{-i\left(p_{1}-p+q\right)^{-} l_{3}} \int_{0}^{\infty} d x_{4}^{\prime+} e^{-\frac{i}{2} \xi(T) x_{4}^{\prime+}} \Big(\frac{-4\boldsymbol{k}^{2}}{x^{2}}\Big)\theta\left(1-\frac{q_{0}^{2}}{\overrightarrow{\mathbf{q}}^{2}}\right)\nn\\
&&\times \frac{T}{q_0} E^{+2} \frac{\boldsymbol{q}^{2}}{\overrightarrow{\mathbf{q}}^{2}} 2 \operatorname{Im}\left(\frac{1}{q^{2}-\Pi_{L}(q,T)}-\frac{1}{q^{2}-\Pi_{T}(q,T)}\right)\nn \\
& =&\int_{0}^{L} d l_{2} \,\Big( 1-e^{i\xi(T) l_2} \Big)\int_{0}^{\infty} d l_{3} \,e^{-i q^{-} l_{3}}\,G(q,T(l_2)), \label{M102R_I1f}
\eea
where $G(q,T)$ is defined in Eq.~\eqref{G_def}.

By using Eqs. \eqref{M102R}, \eqref{M102R_I} and \eqref{M102R_I1f}, $M_{1,0,2, R}$ becomes
\bea
M_{1,0,2, R}&= & \int_{0}^{L} d l_{2} \Big( 1-e^{i\xi(T) l_2} \Big) \int_{0}^{\infty} d l_3 e^{-i q^{-} l_3}\int \frac{d^{3} p}{(2 \pi)^{3} 2 E}|J(p)|^{2} \int \frac{d^{3} k}{(2 \pi)^{3} 2 \omega}(-4) g^{4} t_{c} t_{a} t_{a} t_{c}  \nn\\
&&\times \frac{\boldsymbol{k}^{2}}{\left(\boldsymbol{k}^{2}+\chi(T)\right)^{2}}  \frac{1}{E^{+2}} \int \frac{d^{4} q}{(2 \pi)^{4}} \,F(q_0,q_z,\boldsymbol{q},T). \label{M102R_f}
\eea
Similarly, $M_{1,0,2, L}$ can be computed as
\bea
M_{1,0,2, L}&= &  \int_{0}^{L} d l_{2}\, \Big( 1-e^{-i\xi(T) l_2} \Big)\int_{0}^{\infty} d l_3\, e^{i q^{-} l_3}\int \frac{d^{3} p}{(2 \pi)^{3} 2 E}|J(p)|^{2} \int \frac{d^{3} k}{(2 \pi)^{3} 2 \omega}(-4) g^{4} t_{c} t_{a} t_{a} t_{c} \nn\\
&&\times \frac{\boldsymbol{k}^{2}}{\left(\boldsymbol{k}^{2}+\chi(T)\right)^{2}} \frac{1}{E^{+2}} \int \frac{d^{4} q}{(2 \pi)^{4}}\,F(q_0,q_z,\boldsymbol{q},T)  . \label{M102L_f}
\eea
Therefore, $M_{1,0,2, R}+M_{1,0,2, L}$ becomes
\bea
M_{1,0,2, R}+M_{1,0,2, L}=  -4 g^{4} t_{c} t_{a} t_{a} t_{c} \int \frac{d^{3} p}{(2 \pi)^{3} 2 E}|J(p)|^{2} \int \frac{d^{3} k}{(2 \pi)^{3} 2 \omega}   \frac{1}{E^{+2}} \, I_{2}, \label{M102R_M102L}
\eea
where
\bea
I_2 &=& \left(\int_0^{L} dl_2\, \Big( 1-e^{i \xi(T) l_2}\Big) \int_0^\infty dl_3\,e^{-i q^- l_3} +\int_0^{L} dl_2\, \Big( 1-e^{-i \xi(T) l_2}\Big) \int_0^\infty dl_3\,e^{i q^- l_3} \right)  \frac{\boldsymbol{k}^{2}}{\left(\boldsymbol{k}^{2}+\chi(T)\right)^{2}}\nn\\
&&\times \int \frac{d^{4} q}{(2 \pi)^{4}}\, F(q_0,q_z,\boldsymbol{q},T)\nn\\
&=& \Bigg(\frac{1}{2}\int_0^{L} dl_2 \Big( 2-e^{i\xi(T) l_2} - e^{-i\xi(T) l_2}\Big)  \int_{-\infty}^\infty dl_3\, e^{-iq^-l_3} + i\int_0^{L} dl_2 \Big( e^{i\xi(T) l_2} - e^{-i\xi(T) l_2}\Big)\nn\\
&&\times \int_0^\infty dl_3\, \sin(q^-l_3) \Bigg) \frac{\boldsymbol{k}^{2}}{\left(\boldsymbol{k}^{2}+\chi(T)\right)^{2}}\int \frac{d^{4} q}{(2 \pi)^{4}}\, F(q_0,q_z,\boldsymbol{q},T)\nn\\
&=& \Bigg(\frac{1}{2}\int_0^{L}  dl_2 \Big( 2-e^{i\xi(T) l_2} - e^{-i\xi(T) l_2}\Big) 2\pi \delta(q^-) +i \int_0^{L} dl_2 \Big( e^{i\xi(T) l_2} - e^{-i\xi(T) l_2}\Big) \int_0^\infty dl_3\, \sin(q^-l_3)\Bigg)\nn\\
&&\times\frac{\boldsymbol{k}^{2}}{\left(\boldsymbol{k}^{2}+\chi(T)\right)^{2}}\int \frac{d^{4} q}{(2 \pi)^{4}}\, F(q_0,q_z,\boldsymbol{q},T).
\label{M102R_I2}
\eea
Finally, by using Eqs.~\eqref{M102R_I2}, Eq. \eqref{M102R_M102L} becomes
\bea
&&M_{1,0,2, R}+M_{1,0,2, L} = \nn\\
&=&  - \int_0^{L}  dl_2 \Big( 2-e^{i\xi(T) l_2} - e^{-i\xi(T) l_2}\Big)\int \frac{d^{3} p}{(2 \pi)^{3} 2 E}|J(p)|^{2} \int \frac{d^{3} k}{(2 \pi)^{3} 2 \omega} 2g^{4} t_{c} t_{a} t_{a} t_{c}  \frac{\boldsymbol{k}^{2}}{\left(\boldsymbol{k}^{2}+\chi(T)\right)^{2}} I_{q}(T)\nn \\
&-i&  \int_0^{L} dl_2 \Big( e^{i\xi(T) l_2} - e^{-i\xi(T) l_2}\Big)\int \frac{d^{3} p}{(2 \pi)^{3} 2 E}|J(p)|^{2} \int \frac{d^{3} k}{(2 \pi)^{3} 2 \omega} 4 g^{4} t_{c} t_{a} t_{a} t_{c} \frac{\boldsymbol{k}^{2}}{\left(\boldsymbol{k}^{2}+\chi(T)\right)^{2}} K_{q}(T), \label{M102R_M102L_2}
\eea
where $I_{q}(T)$ is given in Eq. \eqref{M101C_Iq_highT}, and $K_{q}(T)$ is given by
\bea
K_{q}(T)=\int_{0}^{\infty} d l_3 \int \frac{d q_{0} d q_{z} d^{2} q}{(2 \pi)^{4}}\left(\sin \left(q_{0} l_3\right) \cos \left(q_{z} l_3\right)-\cos \left(q_{0} l_3\right) \sin \left(q_{z} l_3\right)\right) F\left(q_0,q_z,\boldsymbol{q},T\right),\nn\\
\label{M102R_Jq}
\eea
$F\left(q_{0}, q_{z}, \boldsymbol{q},T \right)$ is even function of both $q_{0}$ and $q_{z}$, leading to
\bea
& &\int_{-\infty}^{\infty} d q_{0} \sin \left(q_{0} l_3\right) F\left(q_0,q_z,\boldsymbol{q},T\right)=0, \nn \\
&& \int_{-\infty}^{\infty} d q_{z} \sin \left(q_{z} l_3\right) F\left(q_0,q_z,\boldsymbol{q},T\right)=0. \label{M102R_reln2}
\eea
Therefore, we have
\bea
K_{q}(T)=0. \label{M102R_Jqf}
\eea
Finally, by using Eqs. \eqref{M101C_Iq_highT} and \eqref{M102R_Jqf}, Eq. \eqref{M102R_M102L_2} becomes
\bea
M_{1,0,2, R}+M_{1,0,2, L}
&= & -4 t_{c} t_{a} t_{a} t_{c}\int_0^{L}  d\tau  \int \frac{d^{3} p}{(2 \pi)^{3} 2 E}|J(p)|^{2} \int \frac{d^{3} k}{(2 \pi)^{3} 2 \omega} \int  \frac{d^{2} q}{(2 \pi)^{2}}  \, g^{4}\,T \,v(q,T)\,\nn\\
&& \times \frac{\boldsymbol{k}^{2}}{\left(\boldsymbol{k}^{2}+\chi(T)\right)^{2}}
\Big( 1-\cos{(\xi(T) \tau)}\Big). \label{M102RL_f}
\eea

\section{Calculation of Diagrams $M_{1,0,3, C}$ and $M_{1,0,4, C}$}
\label{app_M1034C}
\begin{center}
\begin{figure}[htbp]
\includegraphics[scale=0.7]{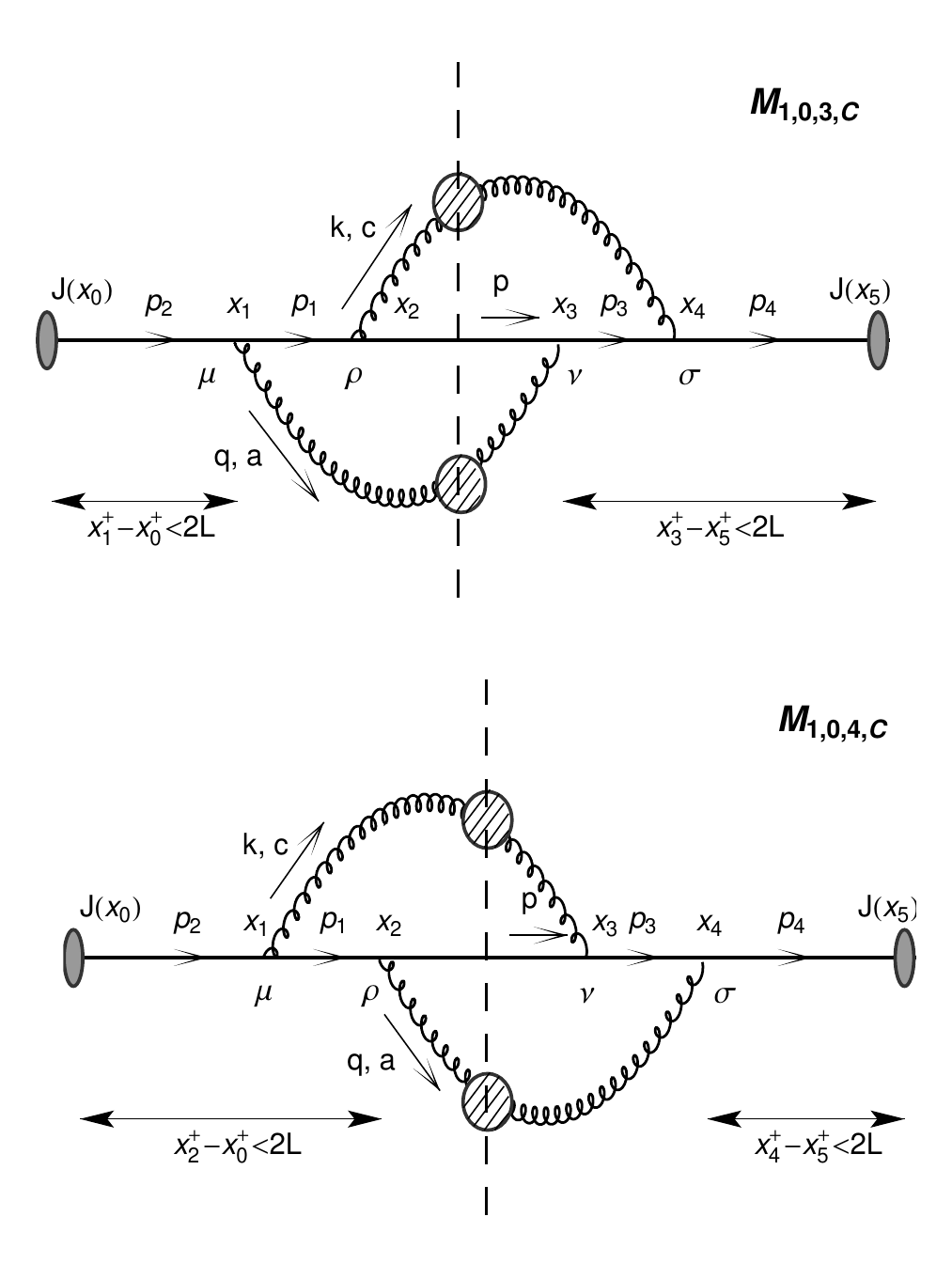}
\caption{Feynman diagrams $M_{1,0,3, C}$ and $M_{1,0,4, C}$, labeled in the same way as Fig.~\ref{fig_M101C}.}
\label{fig_M1034C}
\end{figure}
\end{center}
In this Appendix, we compute the cut diagrams $M_{1,0,3, C}$ and $M_{1,0,4, C}$, shown in Fig.~\ref{fig_M1034C}. We start with $M_{1,0,3, C}$:
\bea
M_{1,0,3, C}&= & \int \prod_{i=0}^{5} d^{4} x_{i} J\left(x_{0}\right) \Delta_{++}^{+}\left(x_{1}-x_{0}\right) v_{\mu}^{+}\left(x_{1}\right) D_{-+}^{\mu \nu}\left(x_{3}-x_{1}\right) \Delta_{++}^{+}\left(x_{2}-x_{1}\right) v_{\rho}^{+}\left(x_{2}\right) D_{-+}^{\rho \sigma}\left(x_{4}-x_{2}\right) \nn \\
&& \times \Delta_{-+}\left(x_{3}-x_{2}\right) v_{\nu}^{-}\left(x_{3}\right) \Delta_{--}^{-}\left(x_{4}-x_{3}\right) v_{\sigma}^{-}\left(x_{4}\right) \Delta_{--}^{-}\left(x_{5}-x_{4}\right) J\left(x_{5}\right) \nn \\
& &\times \theta\left(x_{1}^{+}-x_{0}^{+}\right) \theta\left(x_{2}^{+}-x_{1}^{+}\right) \theta\left(2 L-\left(x_{1}-x_{0}\right)^{+}\right) \theta\left(x_{4}^{+}-x_{5}^{+}\right) \theta\left(x_{3}^{+}-x_{4}^{+}\right) \theta\left(2 L-\left(x_{3}-x_{5}\right)^{+}\right).\nn\\ \label{M103C}
\eea
Following Appendix~\ref{app_M101C} and the first part of Appendix~\ref{app_M102}, Eq.~\eqref{M103C} can be calculated as
\bea
M_{1,0,3, C}&= -2 t_{a} t_{c} t_{a} t_{c} & \int_0^{L} d\tau\, \Big( 1-e^{i\xi(T) \tau}\Big) \int \frac{d^{3} p}{(2 \pi)^{3} 2 E}|J(p)|^{2} \int \frac{d^{3} k}{(2 \pi)^{3} 2 \omega}  \int \frac{d^{2} q}{(2 \pi)^{2}} g^{4}\,T\,v(q,T) \nn\\
&&\times\frac{\boldsymbol{k}^{2}}{\left(\boldsymbol{k}^{2}+\chi(T)\right)^{2}}.
\eea
Since $M_{1,0,4, C}$ is a complex conjugate of $M_{1,0,3, C}$, we obtain
\bea
M_{1,0,3, C}+M_{1,0,4, C}
= &  -4  t_{a} t_{c} t_{a} t_{c}&\,\int_0^{L} d\tau  \int \frac{d^{3} p}{(2 \pi)^{3} 2 E}|J(p)|^{2}  \int \frac{d^{3} k}{(2 \pi)^{3} 2 \omega} \int \frac{d^{2} q}{(2 \pi)^{2}} g^{4}\,T\,v(q,T) \nn\\
&& \times \frac{\boldsymbol{k}^{2}}{\left(\boldsymbol{k}^{2}+\chi(T)\right)^{2}}\Big( 1-\cos{(\xi(T) \tau)}\Big). \label{M1034C_f}
\eea
%
\section{Calculation of Diagrams $M_{1,0,3, R}$ and $M_{1,0,4, L}$}
\label{app_M103R4L}
\begin{center}
\begin{figure}[htbp]
\includegraphics[scale=0.7]{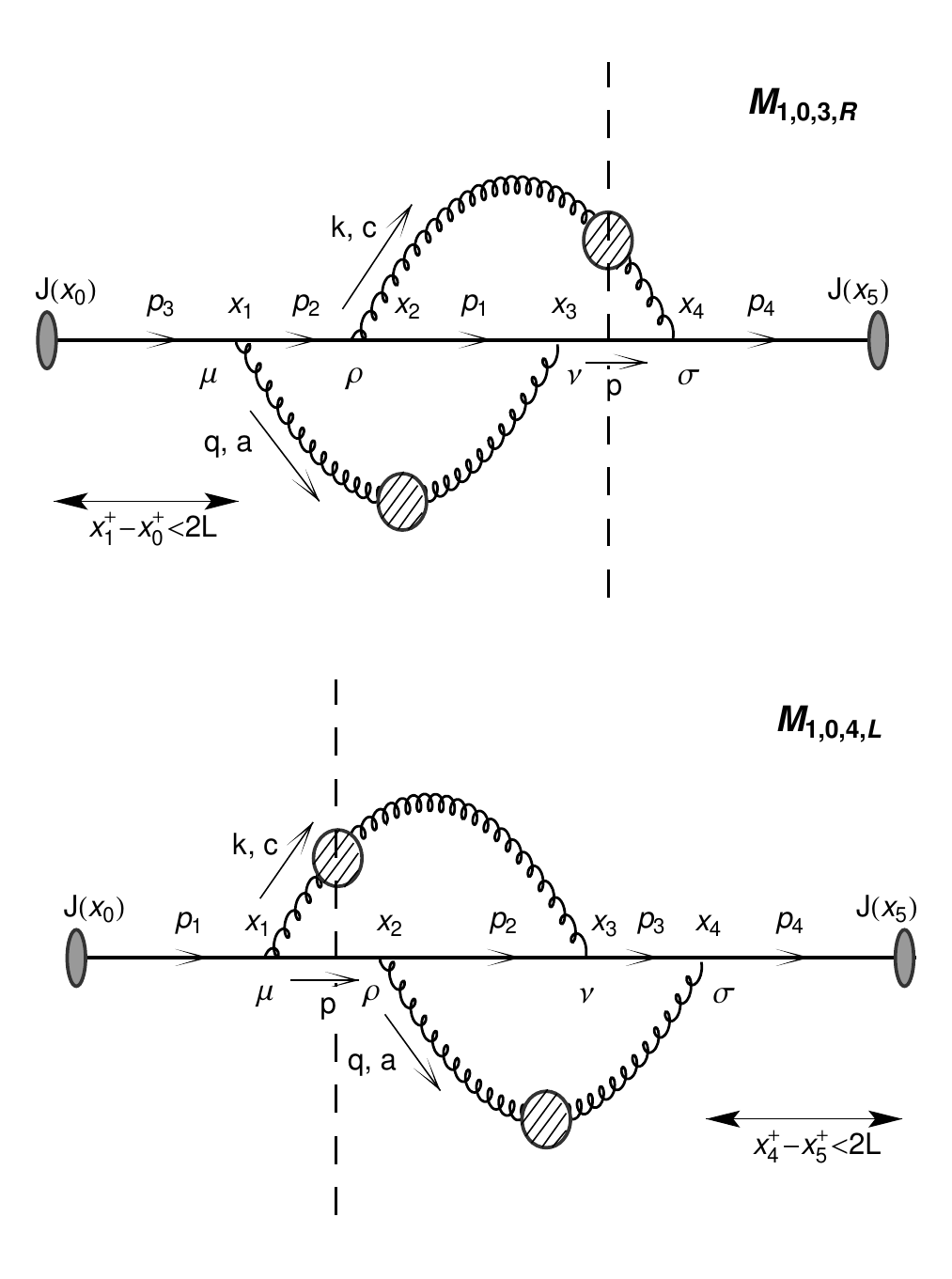}
\caption{Feynman diagrams $M_{1,0,3, R}$ and $M_{1,0,4, L}$, labeled in the same way as Fig.~\ref{fig_M101C}.}
\label{fig_M103R4L}
\end{figure}
\end{center}

The cut diagram $M_{1,0,3, R}$ shown in the left panel of Fig.~\ref{fig_M103R4L} can be calculated as
\bea
M_{1,0,3, R}&= & \int \prod_{i=0}^{5} d^{4} x_{i} J\left(x_{0}\right) \Delta_{++}^{+}\left(x_{1}-x_{0}\right) v_{\mu}^{+}\left(x_{1}\right) D_{++}^{+\mu \nu}\left(x_{3}-x_{1}\right) \Delta_{++}^{+}\left(x_{2}-x_{1}\right) v_{\rho}^{+}\left(x_{2}\right) D_{-+}^{\rho \sigma}\left(x_{4}-x_{2}\right) \nn\\
& \times &\Delta_{++}^{+}\left(x_{3}-x_{2}\right) v_{\lambda}^{+}\left(x_{3}\right) \Delta_{-+}\left(x_{4}-x_{3}\right) v_{\sigma}^{-}\left(x_{4}\right) \Delta_{--}^{-}\left(x_{5}-x_{4}\right) J\left(x_{5}\right) \nn\\
& \times &\prod_{i=0}^{3} \theta\left(x_{i+1}^{+}-x_{i}^{+}\right) \theta\left(x_{4}^{+}-x_{5}^{+}\right) \theta\left(2 L-\left(x_{1}-x_{0}\right)^{+}\right) \label{M103R4L}
\eea
Following Appendix F of~\cite{MD_PRC}, it can be shown that
\bea
M_{1,0,3, R}+M_{1,0,4, L} \approx 0 . \label{M103R4L_f}
\eea
\section{Calculation of Diagrams $M_{1,0,5, R}$ and $M_{1,0,6, L}$}
\label{app_M105R6L}
\begin{center}
\begin{figure}[htbp]
\includegraphics[scale=0.7]{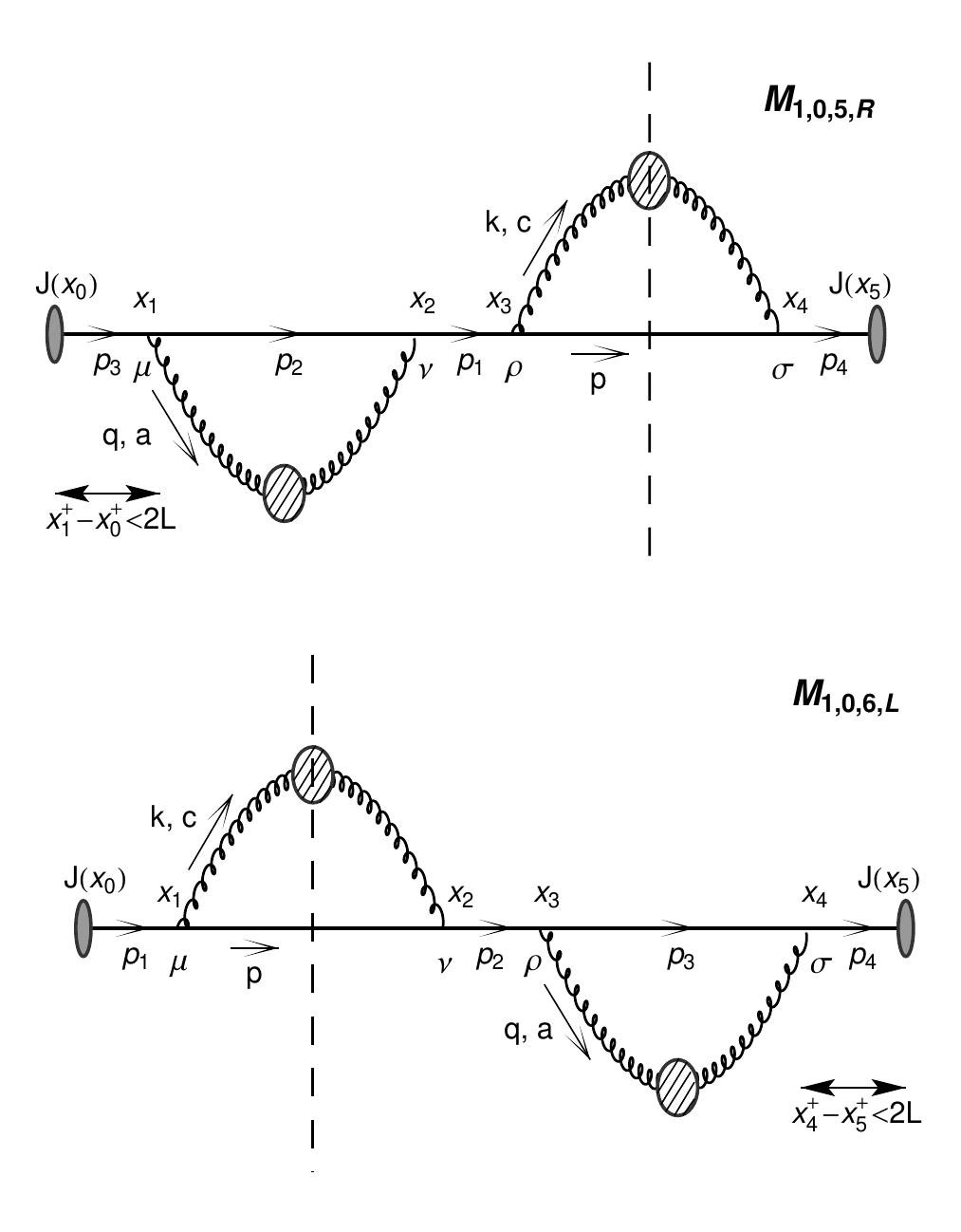}
\caption{Feynman diagrams $M_{1,0,5, R}$ and $M_{1,0,6, L}$, labeled in the same way as Fig.~\ref{fig_M101C}.}
\label{fig_M105R6L}
\end{figure}
\end{center}
In this Appendix, we calculate the cut diagrams $M_{1,0,5, R}$ and $M_{1,0,6, L}$, shown in Fig.~\ref{fig_M105R6L}. We begin with $M_{1,0,5, R}$:
\bea 
M_{1,0,5, R}&= & \int \prod_{i=0}^{5} d^{4} x_{i} J\left(x_{0}\right) \Delta_{++}^{+}\left(x_{1}-x_{0}\right) v_{\mu}^{+}\left(x_{1}\right) D_{++}^{+\mu \nu}\left(x_{2}-x_{1}\right) \Delta_{++}^{+}\left(x_{2}-x_{1}\right) v_{\nu}^{+}\left(x_{2}\right) \nn\\
& \times &\Delta_{++}^{+}\left(x_{3}-x_{2}\right)  v_{\rho}^{+}\left(x_{3}\right) D_{-+}^{\rho \sigma}\left(x_{4}-x_{3}\right) \Delta_{-+}\left(x_{4}-x_{3}\right) v_{\sigma}^{-}\left(x_{4}\right) \Delta_{--}^{-}\left(x_{5}-x_{4}\right) J\left(x_{5}\right) \nn \\
& \times & \theta\left(x_{1}^{+}-x_{0}^{+}\right) \theta\left(x_{2}^{+}-x_{1}^{+}\right) \theta\left(x_{3}^{+}-x_{2}^{+}\right) \theta\left(x_{5}^{+}-x_{4}^{+}\right) \theta\left(2 L-\left(x_{1}-x_{0}\right)^{+}\right) .\label{M105R}
\eea 
Following the computation of $M_{1,0,2,R}$ in Appendix~\ref{app_M102}, Eq.~\eqref{M105R} becomes
\bea 
M_{1,0,5, R}&= & -4  t_{a} t_{a} t_{c} t_{c}\int_0^{L} dl_1 \, e^{i\xi(T) l_1} \int_{0}^{\infty} d l_2 \, e^{-i  q^{-}l_2}\int \frac{d^{3} p}{(2 \pi)^{3} 2 E}|J(p)|^{2} \int \frac{d^{3} k}{(2 \pi)^{3} 2 \omega}  \nn \\
&\times&   \frac{\boldsymbol{k}^{2}}{\left(\boldsymbol{k}^{2}+\chi(T)\right)^{2}} \int \frac{d^4q}{(2\pi)^4}\theta\left(1-\frac{q_{0}^{2}}{\overrightarrow{\mathbf{q}}^{2}}\right)
g^{4}\,\frac{T}{q_0} \frac{\boldsymbol{q}^{2}}{\overrightarrow{\mathbf{q}}^{2}} 2 \operatorname{Im}\left(\frac{1}{q^{2}-\Pi_{L}(q,T)}-\frac{1}{q^{2}-\Pi_{T}(q,T)}\right) .
\eea
Since $M_{1,0,6, L}$ is the complex conjugate of $M_{1,0,5, R}$, we obtain
\bea 
M_{1,0,5, R}+M_{1,0,6, L}
= &- 4 t_{a} t_{a} t_{c} t_{c}&\int_0^{L} d\tau \int \frac{d^{3} p}{(2 \pi)^{3} 2 E}|J(p)|^{2} \int \frac{d^{3} k}{(2 \pi)^{3} 2 \omega}  \int \frac{d^{2} q}{(2 \pi)^{2}}\,g^{4} \,T\,v(q,T) \nn\\
&&\times \frac{\boldsymbol{k}^{2}}{\left(\boldsymbol{k}^{2}+\chi(T)\right)^{2}} \cos{( \xi(T) \tau)} .  \label{M1056RL_f}
\eea 

\section{Calculation of Diagrams $M_{1,1,1, C}$ and $M_{1,1,2, C}$}
\label{app_M1112C}
\begin{center}
\begin{figure}[htbp]
\includegraphics[scale=0.7]{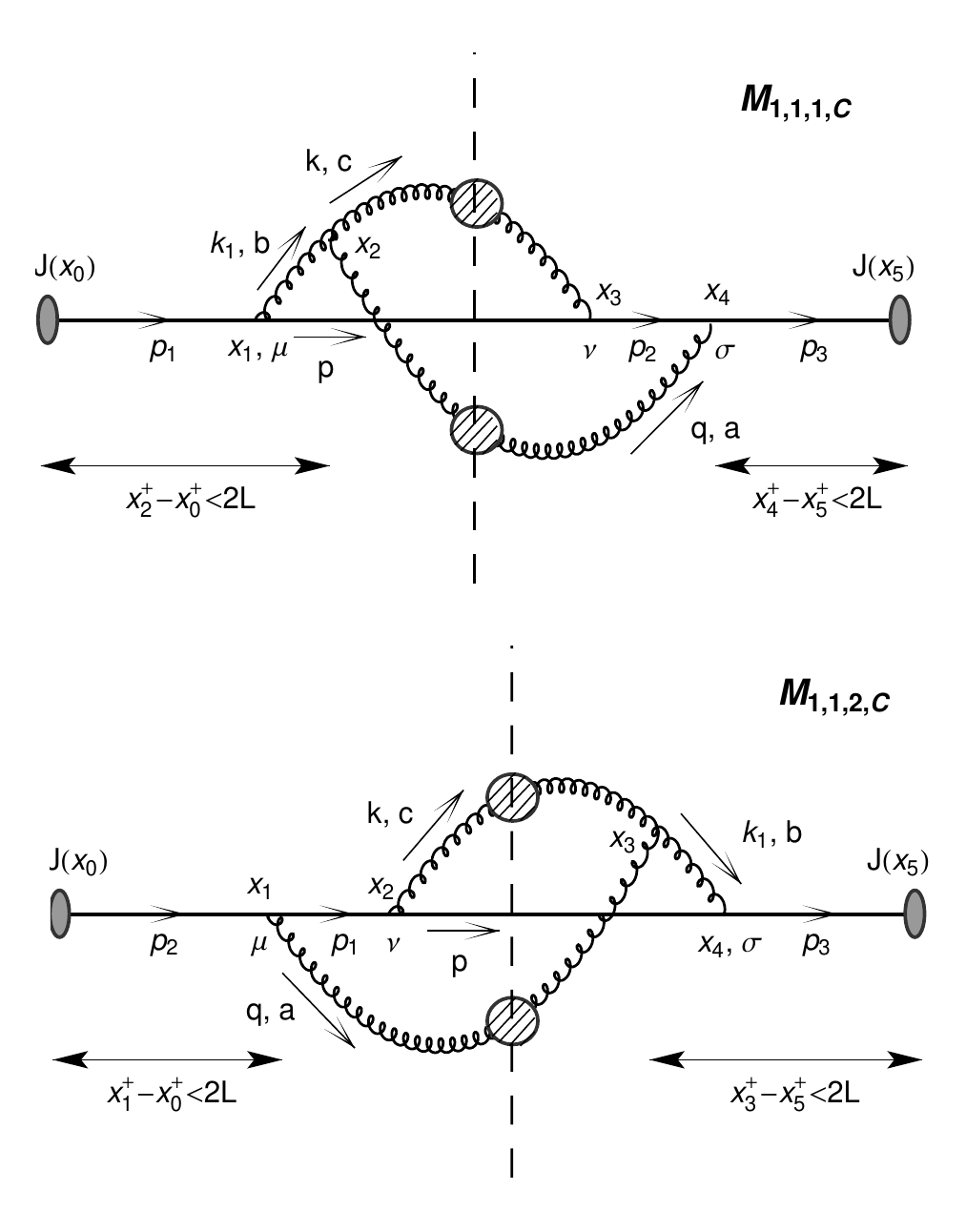}
\caption{Feynman diagrams $M_{1,1,1, C}$ and $M_{1,1,2, C}$, labeled in the same way as Fig.~\ref{fig_M101C}.}
\label{fig_M1112C}
\end{figure}
\end{center}
In Appendices \ref{app_M1112C} - \ref{app_M113L4R}, we calculate the diagrams where one end of the exchanged gluon $q$ is attached to the high-$p_\perp$ parton, while the other end is connected to the radiated gluon $k$. Consequently, a 3-gluon vertex is involved in the diagrams. Here, we present the calculation of the diagrams shown in Fig.~\ref{fig_M1112C}, starting with the cut diagram $M_{1,1,1, C}$:
\bea
M_{1,1,1, C}&= & \int \prod_{i=0}^{5} d x_{i} J\left(x_{0}\right) \Delta_{++}^{+}\left(x_{1}-x_{0}\right) v_{\mu}^{+}\left(x_{1}\right) D_{++}^{+\mu \rho_{1}}\left(x_{2}-x_{1}\right) v_{\rho_{1} \rho_{2} \rho_{3}}^{+}\left(x_{2}\right) \Delta_{-+}\left(x_{3}-x_{1}\right)\nn \\
&& \times D_{-+}^{\rho_{3} \nu}\left(x_{3}-x_{2}\right) v_{\nu}^{-}\left(x_{3}\right) D_{-+}^{\rho_{2} \sigma}\left(x_{4}-x_{2}\right) v_{\sigma}^{-}\left(x_{4}\right) \Delta_{--}^{-}\left(x_{4}-x_{3}\right) \Delta_{--}^{-}\left(x_{5}-x_{4}\right) J\left(x_{5}\right) \nn\\
&& \times \theta\left(x_{1}^{+}-x_{0}^{+}\right) \theta\left(x_{2}^{+}-x_{1}^{+}\right) \theta\left(2 L-\left(x_{2}-x_{0}\right)^{+}\right) \theta\left(x_{3}^{+}-x_{4}^{+}\right) \theta\left(x_{4}^{+}-x_{5}^{+}\right) \nn\\
&&\times \theta\left(2 L-\left(x_{4}-x_{5}\right)^{+}\right) \nn\\
&= & -\int_{-\infty}^{\infty} \int_{0}^{\infty} \prod_{i=1}^{3} \frac{d p_{i}^{+} d^{2} p_{i}}{(2 \pi)^{3} 2 p_{i}^{+}} \frac{d k_{1}^{+} d^{2} k_{1}}{(2 \pi)^{3} 2 k_{1}^{+}} \int \frac{d^{3} k}{(2 \pi)^{3} 2 \omega} \frac{d^{3} p}{(2 \pi)^{3} 2 E} \frac{d^{4} q}{(2 \pi)^{4}}(-i) g^{4} f^{c b a} t_{c} t_{b} t_{a}\, I, \label{M111C}
\eea
where
\bea
I&= & \int \prod_{i=0}^{5} d x_{i} \theta\left(x_{1}^{+}-x_{0}^{+}\right) \theta\left(x_{2}^{+}-x_{1}^{+}\right) \theta\left(2 L-\left(x_{2}-x_{0}\right)^{+}\right) \theta\left(x_{3}^{+}-x_{4}^{+}\right) \theta\left(x_{4}^{+}-x_{5}^{+}\right) \nn\\
&&\times \theta\left(2 L-\left(x_{4}-x_{5}\right)^{+}\right)  e^{-i p_{1}\left(x_{1}-x_{0}\right)} e^{-i k_{1}\left(x_{2}-x_{1}\right)} e^{-i p\left(x_{3}-x_{1}\right)} e^{-i k\left(x_{3}-x_{2}\right)} e^{-i q\left(x_{4}-x_{2}\right)}\nn\\
&&\times \, e^{-i p_{2}\left(x_{4}-x_{3}\right)} e^{-i p_{3}\left(x_{5}-x_{4}\right)} J\left(x_{0}\right) J\left(x_{5}\right) \left(p+p_{1}\right)^{\mu} P_{\mu \rho_{1}}\left(k_{1}\right)\Big(g^{\rho_{1} \rho_{3}}\left(k_{1}+k\right)^{\rho_{2}}\nn\\
&& +g^{\rho_{2} \rho_{3}}(q-k)^{\rho_{1}}+g^{\rho_{1} \rho_{2}}\left(-k_{1}-q\right)^{\rho_{3}}\Big) P_{\rho_{3} \nu}(k) D_{\rho_{2} \sigma}^{>}(q)\left(p+p_{2}\right)^{\nu}\left(p_{2}+p_{3}\right)^{\sigma}\nn \\
&= & |J(p)|^{2}(2 \pi)^{3} \delta\left(\left(p_{1}-p-k_{1}\right)^{+}\right) \delta^{2}\left(\boldsymbol{p}_{1}-\boldsymbol{p}-\boldsymbol{k}_{1}\right)(2 \pi)^{3} \delta\left(\left(k_{1}-k-q\right)^{+}\right) \delta^{2}\left(\boldsymbol{k}_{1}-\boldsymbol{k}-\boldsymbol{q}\right) \nn \\
&& \times(2 \pi)^{3} \delta\left(\left(p+k-p_{2}\right)^{+}\right) \delta^{2}\left(\boldsymbol{p}+\boldsymbol{k}-\boldsymbol{p}_{2}\right)(2 \pi)^{3} \delta\left(\left(p+k+q-p_{1}\right)^{+}\right) \delta^{2}\left(\boldsymbol{p}+\boldsymbol{k}+\boldsymbol{q}-\boldsymbol{p}_{1}\right) I_{1}\nn\\ \label{M111C_I}
\eea
and where
\bea
I_{1}&= & \int_{0}^{2 L} d x_{2}^{\prime+} e^{-\frac{i}{2}\left(k_{1}-k-q\right)^{-} x_{2}^{\prime+}} \int_{0}^{x_{2}^{\prime+}} d x_{1}^{\prime+} e^{-\frac{i}{2}\left(p_{1}-p-k_{1}\right)^{-} x_{1}^{\prime+}} \int_{0}^{\infty} d x_{3}^{\prime+} e^{-\frac{i}{2}\left(p+k-p_{2}\right)^{-} x_{3}^{\prime+}} \nn\\
&& \times \int_{0}^{2 L} d x_{4}^{\prime+} e^{-\frac{i}{2}\left(p+k+q-p_{3}\right)^{-} x_{4}^{\prime+}} \left(p+p_{1}\right)^{\mu} P_{\mu \rho_{1}}\left(k_{1}\right)\Big(g^{\rho_{1} \rho_{3}}\left(k_{1}+k\right)^{\rho_{2}}+g^{\rho_{2} \rho_{3}}(q-k)^{\rho_{1}}\nn\\
&&+g^{\rho_{1} \rho_{2}}\left(-k_{1}-q\right)^{\rho_{3}}\Big) P_{\rho_{3} \nu}(k) D_{\rho_{2} \sigma}^{>}(q)\left(p+p_{2}\right)^{\nu}\left(p_{2}+p_{3}\right)^{\sigma}\nn \\
&=& \frac{4}{(p_2-p-k)^-(p_1-p-k_1)^-} \int_0^{2L} dx_2'^+ e^{-\frac{i}{2}(k_1-k-q)^- x_2'^+} \Big( 1-e^{-\frac{i}{2}(p_1-k_1-p)^- x_2'^+}\Big) \nn\\
&& \times \int_0^{2L} dx_4'^+ e^{-\frac{i}{2}(p+k+q-p_3)^-x_4'^+}\left(p+p_{1}\right)^{\mu} P_{\mu \rho_{1}}\left(k_{1}\right)\Big(g^{\rho_{1} \rho_{3}}\left(k_{1}+k\right)^{\rho_{2}}+g^{\rho_{2} \rho_{3}}(q-k)^{\rho_{1}}\nn\\
&&+g^{\rho_{1} \rho_{2}}\left(-k_{1}-q\right)^{\rho_{3}}\Big) P_{\rho_{3} \nu}(k) D_{\rho_{2} \sigma}^{>}(q)\left(p+p_{2}\right)^{\nu}\left(p_{2}+p_{3}\right)^{\sigma}. \label{M111C_I1}
\eea
Now we compute
\bea
&&\left(p_{1}+p_{2}\right)^{\mu}\left(p+p_{2}\right)^{\nu}\left(p_{2}+p_{3}\right)^{\sigma} P_{\mu \rho_{1}}\left(k_{1}\right) P_{\rho_{3} \nu}(k) D_{\rho_{2} \sigma}^{>}(q)\Big(g^{\rho_{1} \rho_{3}}\left(k+k_{1}\right)^{\rho_{2}}+g^{\rho_{2} \rho_{3}}(q-k)^{\rho_{1}}\nn\\
&&+g^{\rho_{1} \rho_{2}}\left(-k_{1}-q\right)^{\rho_{3}}\Big ) \nn \\
&&\approx\left\{\left(p_{1}+p_{2}\right)^{\mu} P_{\mu \rho_{1}}\left(k_{1}\right) P_{\nu}^{\rho_{1}}(k)\left(p+p_{2}\right)^{\nu}\right\}\left\{\left(k+k_{1}\right)^{\rho} D_{\rho \sigma}^{>}(q)\left(p_{2}+p_{3}\right)^{\sigma}\right\} \nn\\
&&\approx\left(-4 \frac{\boldsymbol{k} \cdot \boldsymbol{k}_{1}}{x^{2}}\right)\theta\left(1-\frac{q_{0}^{2}}{\overrightarrow{\mathbf{q}}^{2}}\right) \frac{T}{q_0} E^{+} k^{+} \frac{\boldsymbol{q}^{2}}{\overrightarrow{\mathbf{q}}^{2}} 2 \operatorname{Im}\left(\frac{1}{q^{2}-\Pi_{L}(q,T)}-\frac{1}{q^{2}-\Pi_{T}(q,T)}\right). \label{M111C_reln}
\eea
Therefore, $I_1$ can be written as
\bea
I_1 &=& \frac{16}{(p_2-p-k)^-(p_1-p-k_1)^-} \int_0^{L} dl_2\, e^{-i(k_1-k-q)^- l_2} \Big( 1-e^{-i(p_1-k_1-p)^- l_2}\Big) \nn\\
&& \times \int_0^{L} dl_4\, e^{-i(p+k+q-p_3)^-l_4} \left(-4 \frac{\boldsymbol{k} \cdot \boldsymbol{k}_{1}}{x^{2}}\right)\theta\left(1-\frac{q_{0}^{2}}{\overrightarrow{\mathbf{q}}^{2}}\right) \frac{T}{q_0} E^{+} k^{+} \frac{\boldsymbol{q}^{2}}{\overrightarrow{\mathbf{q}}^{2}}\nn\\
&&\times 2 \operatorname{Im}\left(\frac{1}{q^{2}-\Pi_{L}(q,T)}-\frac{1}{q^{2}-\Pi_{T}(q,T)}\right)\nn\\
&=&\!  \int_0^{L} \!dl_2\, e^{-i(k_1-k-q)^- l_2} \Big( \!1-e^{-i(p_1-k_1-p)^- l_2}\!\Big)\!
\int_0^{L} \!dl_4\, e^{-i(p+k+q-p_3)^-l_4}\,
H\!\left(\!q,\,T\!\left(\!\frac{l_2\!+\!l_4}{2}\!\right)\!\right),  \label{M111C_I1b}
\eea
where, in a non-uniform medium, the temperature dependence is in principle associated with both time arguments
($l_2$ and $l_4$), i.e.\ with the amplitude and its conjugate. Consistently with Appendix~\ref{app_M101C},
we evaluate medium-dependent quantities at the mid-point $l=(l_2+l_4)/2$ (leading term of a gradient/Wigner expansion),
with gradient corrections suppressed as discussed around Eqs.~\eqref{ld_def}--\eqref{M101C_I1g}.
\bea
H(q,T) &=&\frac{16}{(p_2-p-k)^-(p_1-p-k_1)^-} \left(-4 \frac{\boldsymbol{k} \cdot \boldsymbol{k}_{1}}{x^{2}}\right)
\theta\left(1-\frac{q_{0}^{2}}{\overrightarrow{\mathbf{q}}^{2}}\right) \frac{T}{q_0} E^{+} k^{+}
\frac{\boldsymbol{q}^{2}}{\overrightarrow{\mathbf{q}}^{2}}\nn\\
&&\times 2 \operatorname{Im}\left(\frac{1}{q^{2}-\Pi_{L}(q,T)}-\frac{1}{q^{2}-\Pi_{T}(q,T)}\right). \label{H_def}
\eea
By using Eq.~\eqref{M101C_approx1pi} and applying the $\delta$ functions from Eq.~\eqref{M111C_I}, we obtain
\bea
&& p_{1}^{-}=p_{3}^{-}=\frac{M^{2}}{p_{1}^{+}}=\frac{M^{2}}{(p+k+q)^{+}} ,\\
&& p^{-}=\frac{(\boldsymbol{k}+\boldsymbol{q})^{2}+M^{2}}{p^{+}}, \quad k^{-}=\frac{\boldsymbol{k}^{2}+m_{g}(T)^{2}}{k^{+}}, \quad k_{1}^{-}=\frac{(\boldsymbol{k}+\boldsymbol{q})^{2}+m_{g}(T)^{2}}{(k+q)^{+}} \quad p_{2}^{-}=\frac{\boldsymbol{q}^{2}+M^{2}}{(p+k)^{+}},
\eea
leading to
\bea
\left(p+k-p_{2}\right)^{-}&=&\frac{k^{2}+\chi}{ x E} \equiv \xi(T), \label{M111C_approx1}\\
\left(p_{1}-k_{1}-p\right)^{-} &\approx&-\frac{(\boldsymbol{k}+\boldsymbol{q})^{2}+\chi(T)}{ x E} \equiv-\zeta(T), \label{M111C_approx2}\\
\left( p+k+q-p_1\right)^- &\approx& q^-, \label{M111C_approx3} \\
\left( k_1-k-q \right)^- &\approx& -q^-. \label{M111C_approx4}
\eea
Using these approximations, we obtain
\bea
I_1 &=& \int_0^{L} dl_2\, e^{-i(k_1-k-q)^- l_2} \Big( 1-e^{-i(p_1-k_1-p)^- l_2}\Big)   2 \pi \delta\left(q_{0}-q_{z}\right) H(q,T(l_2)). \label{M111C_I1f}
\eea
Additionally, we note that
\bea
-i f^{a b c} t_{a} t_{b} t_{c}=\frac{1}{2}\left[t_{a}, t_{c}\right]\left[t_{c}, t_{a}\right] .\label{M111C_reln2}
\eea
Finally, by using Eqs.~\eqref{M101C_Iq}, \eqref{M111C_I1f}, and \eqref{M111C_reln2}, Eq.~\eqref{M111C} reduces to
\bea
M_{1,1,1, C}&= & -  2\left[t_{a}, t_{c}\right]\left[t_{c}, t_{a}\right]\int_0^{L} dl_2\,  \Big( 1-e^{i\zeta(T) l_2} \Big) \int \frac{d^{3} p}{(2 \pi)^{3} 2 E}|J(p)|^{2} \int \frac{d^{3} k}{(2 \pi)^{3} 2 \omega}  \int \frac{d^{2} q}{(2 \pi)^{2}}  \nn\\
&&\times \,g^{4}\, T v(q,T)\,
 \frac{\boldsymbol{k} \cdot(\boldsymbol{k}+\boldsymbol{q})}{\left(\boldsymbol{k}^{2}+\chi(T)\right)\left((\boldsymbol{k}+\boldsymbol{q})^{2}+\chi(T)\right)}.
\eea
As $M_{1,1,2, C}$ is the complex conjugate of $M_{1,1,1, C}$, one finally obtains
\bea
M_{1,1,1, C}+M_{1,1,2, C}
&=&-4 \left[t_{a}, t_{c}\right]\left[t_{c}, t_{a}\right]\int_0^{L} d\tau\,   \int \frac{d^{3} p}{(2 \pi)^{3} 2 E}|J(p)|^{2} \int \frac{d^{3} k}{(2 \pi)^{3} 2 \omega} \int \frac{d^{2} q}{(2 \pi)^{2}}\, g^{4} \, T \,v(q,T) \nn\\
&& \times \frac{\boldsymbol{k} \cdot(\boldsymbol{k}+\boldsymbol{q})}{\left(\boldsymbol{k}^{2}+\chi(T)\right)\left((\boldsymbol{k}+\boldsymbol{q})^{2}+\chi(T)\right)}\Big( 1-\cos(\zeta(T) \tau) \Big). \label{M1112C_f}
\eea
\section{Calculation of Diagrams $M_{1,1,2, R}$ and $M_{1,1,1, L}$}
\label{app_M111L2R}
\begin{center}
\begin{figure}[htbp]
\includegraphics[scale=0.7]{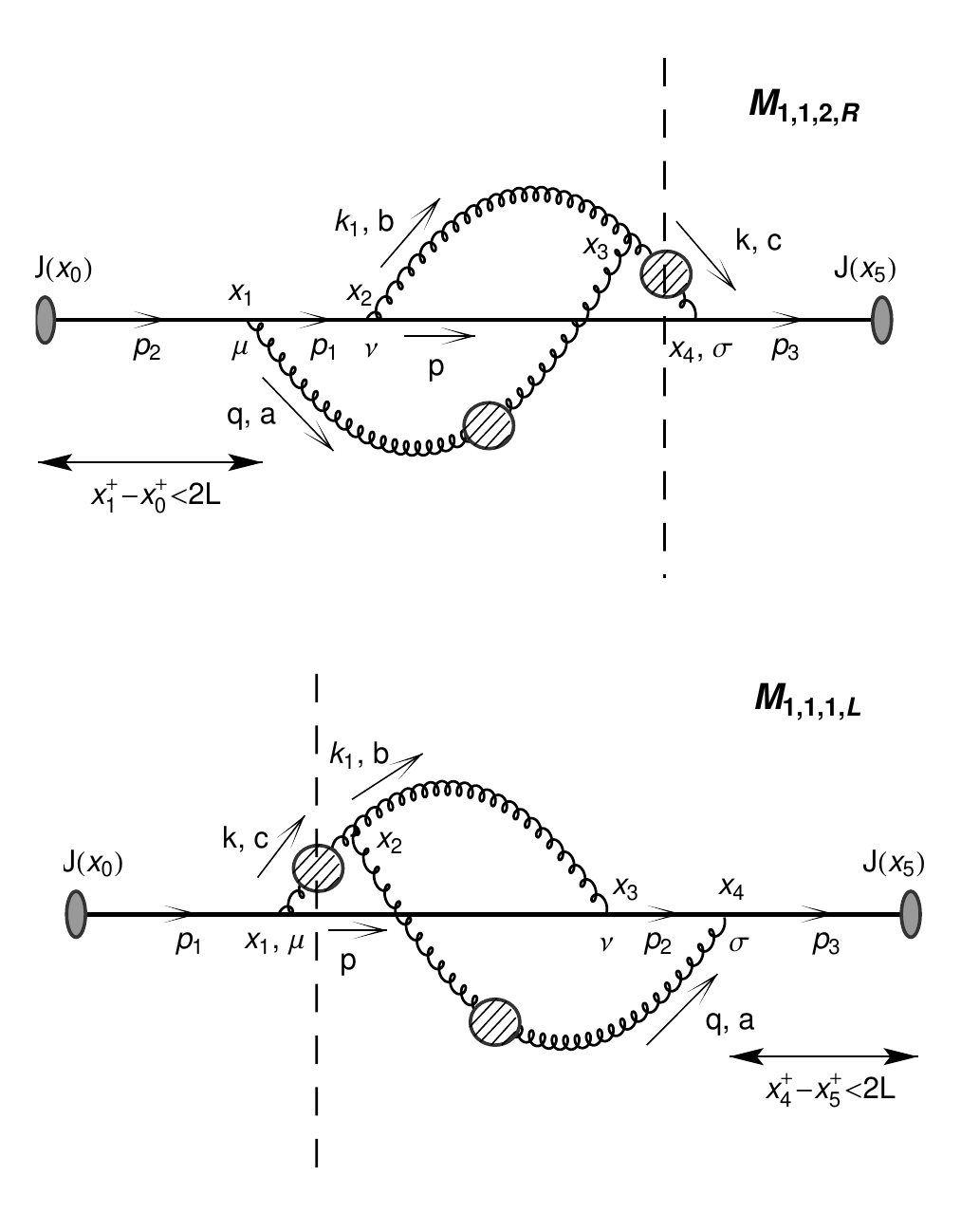}
\caption{Feynman diagrams $M_{1,1,2, R}$ and $M_{1,1,1, L}$, labeled in the same way as Fig.~\ref{fig_M101C}.}
\label{fig_M111L2R}
\end{figure}
\end{center}
The cut diagram $M_{1,1,2, R}$, shown in the left panel of Fig.~\ref{fig_M111L2R}, can be calculated as
\bea
M_{1,1,2, R}&= & \int \prod_{i=0}^{5} d x_{i} J\left(x_{0}\right) \Delta_{++}^{+}\left(x_{1}-x_{0}\right) v_{\mu}^{+}\left(x_{1}\right) D_{++}^{+\mu \rho_{2}}\left(x_{3}-x_{1}\right) v_{\rho_{1} \rho_{2} \rho_{3}}^{+}\left(x_{3}\right) \Delta_{++}^{+}\left(x_{2}-x_{1}\right) \nn\\
& &\times D_{++}^{\rho_{1} \nu}\left(x_{3}-x_{2}\right) v_{\nu}^{+}\left(x_{2}\right) D_{-+}^{\rho_{3} \sigma}\left(x_{4}-x_{2}\right) v_{\sigma}^{-}\left(x_{4}\right) \Delta_{-+}\left(x_{4}-x_{2}\right) v_{--}^{-}\left(x_{5}-x_{4}\right) J\left(x_{5}\right) \nn\\
& &\times \theta\left(x_{1}^{+}-x_{0}^{+}\right) \theta\left(x_{2}^{+}-x_{1}^{+}\right) \theta\left(2 L-\left(x_{1}-x_{0}\right)^{+}\right) \theta\left(x_{3}^{+}-x_{2}^{+}\right) \theta\left(x_{4}^{+}-x_{5}^{+}\right) . \label{M112R}
\eea
By applying the same procedure as in Appendix~I of~\cite{MD_PRC}, we obtain
\bea
M_{1,1,1, L}+M_{1,1,2, R}=0. \label{M1112RL_f}
\eea
\section{Calculation of Diagrams $M_{1,1,3, C}$ and $M_{1,1,4, C}$}
\label{app_M1134C}
\begin{center}
\begin{figure}[htbp]
\includegraphics[scale=0.7]{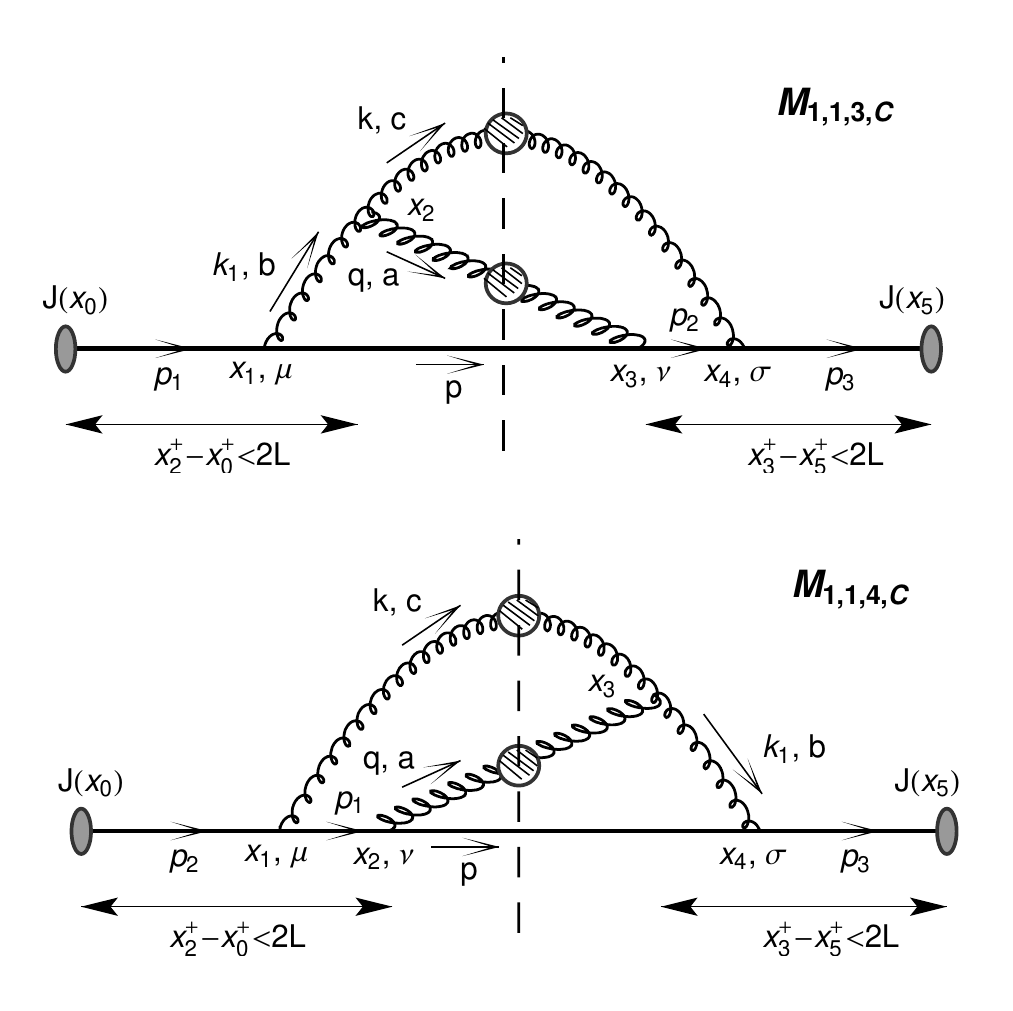}
\caption{Feynman diagrams $M_{1,1,3, C}$ and $M_{1,1,4, C}$, labeled in the same way as Fig.~\ref{fig_M101C}.}
\label{fig_M1134C}
\end{figure}
\end{center}
We here calculate the cut diagrams $M_{1,1,3, C}$ and $M_{1,1,4, C}$, shown in Fig.~\ref{fig_M1134C}. $M_{1,1,3, C}$ can be computed as
\bea
M_{1,1,3, C}&= & \int \prod_{i=0}^{5} d x_{i} J\left(x_{0}\right) \Delta_{++}^{+}\left(x_{1}-x_{0}\right) v_{\mu}^{+}\left(x_{1}\right) D_{++}^{+\mu \rho_{1}}\left(x_{2}-x_{1}\right) v_{\rho_{1} \rho_{2} \rho_{3}}^{+}\left(x_{2}\right) \Delta_{-+}\left(x_{3}-x_{1}\right) \nn \\
& \times& D_{-+}^{\rho_{2} \lambda}\left(x_{3}-x_{2}\right) v_{\lambda}^{-}\left(x_{3}\right) \Delta_{--}^{-}\left(x_{4}-x_{3}\right) D_{-+}^{\rho_{3} \sigma}\left(x_{4}-x_{2}\right) v_{\sigma}^{-}\left(x_{4}\right) \Delta_{--}^{-}\left(x_{5}-x_{4}\right) J\left(x_{5}\right) \nn \\
& \times& \int_{-\infty}^{\infty} \int_{0}^{\infty} \frac{d p_{3}^{+} d^{2} p_{3}}{(2 \pi)^{3} 2 p_{3}^{+}} e^{-i p_{3}\left(x_{5}-x_{4}\right)}\left(i g\left(p_{2}+p_{3}\right)^{\sigma} t_{b}\right) J\left(x_{5}\right)  \theta\left(x_{1}^{+}-x_{0}^{+}\right) \theta\left(x_{2}^{+}-x_{1}^{+}\right) \nn\\
&\times &\theta\left(2 L-\left(x_{2}-x_{0}\right)^{+}\right) \theta\left(x_{3}^{+}-x_{4}^{+}\right) \theta\left(x_{4}^{+}-x_{5}^{+}\right) \theta\left(2 L-\left(x_{3}-x_{5}\right)^{+}\right) . \label{M113C}
\eea
Similar to Appendix~\ref{app_M1112C}, Eq.~\eqref{M113C} can be calculated as
\bea
M_{1,1,3, C}&= &   -2\left[t_{a}, t_{c}\right]\left[t_{c}, t_{a}\right]\,\int_0^{L}d\tau\,\left(1-e^{i\zeta(T) \tau}-e^{-i\xi(T) \tau}- e^{i(\zeta(T)-\xi(T))\tau}\right) \int \frac{d^{3} p}{(2 \pi)^{3} 2 E}|J(p)|^{2} \nn\\
&&\times \int \frac{d^{3} k}{(2 \pi)^{3} 2 \omega}\int  \frac{d^{2} q}{(2 \pi)^{2}}  \,g^{4}\,T\, v(q,T)
 \frac{\boldsymbol{k} \cdot(\boldsymbol{k}+\boldsymbol{q})}{\left(\boldsymbol{k}^{2}+\chi(T)\right)\left((\boldsymbol{k}+\boldsymbol{q})^{2}+\chi(T)\right)}.
\eea
Since $M_{1,1,4, C}$ is the complex conjugate of $M_{1,1,3, C}$, one obtains
\bea
&&M_{1,1,3, C}+M_{1,1,4,C}\nn\\
&= &  -4\left[t_{a}, t_{c}\right]\left[t_{c}, t_{a}\right] \int_0^{L}d\tau  \int \frac{d^{3} p}{(2 \pi)^{3} 2 E}|J(p)|^{2}\int \frac{d^{3} k}{(2 \pi)^{3} 2 \omega} \int \frac{d^{2} q}{(2 \pi)^{2}} \,g^{4}\, T\, v(q,T)\nn\\
&\times & 
 \frac{\boldsymbol{k} \cdot(\boldsymbol{k}+\boldsymbol{q})}{\left(\boldsymbol{k}^{2}+\chi(T)\right)\left((\boldsymbol{k}+\boldsymbol{q})^{2}+\chi(T)\right)}\left[1-\cos\left(\zeta(T) \tau\right)-\cos\left(\xi(T) \tau\right)+ \cos\left((\zeta(T)-\xi(T)) \tau\right)\right].\nn\\
 \label{M1134C_f}
\eea

\section{Calculation of Diagrams $M_{1,1,3, R},\, M_{1,1,3, L},\, M_{1,1,4, L}$, and $M_{1,1,4, R}$}
\label{app_M113L4R}
\begin{center}
\begin{figure}[htbp]
\includegraphics[scale=0.7]{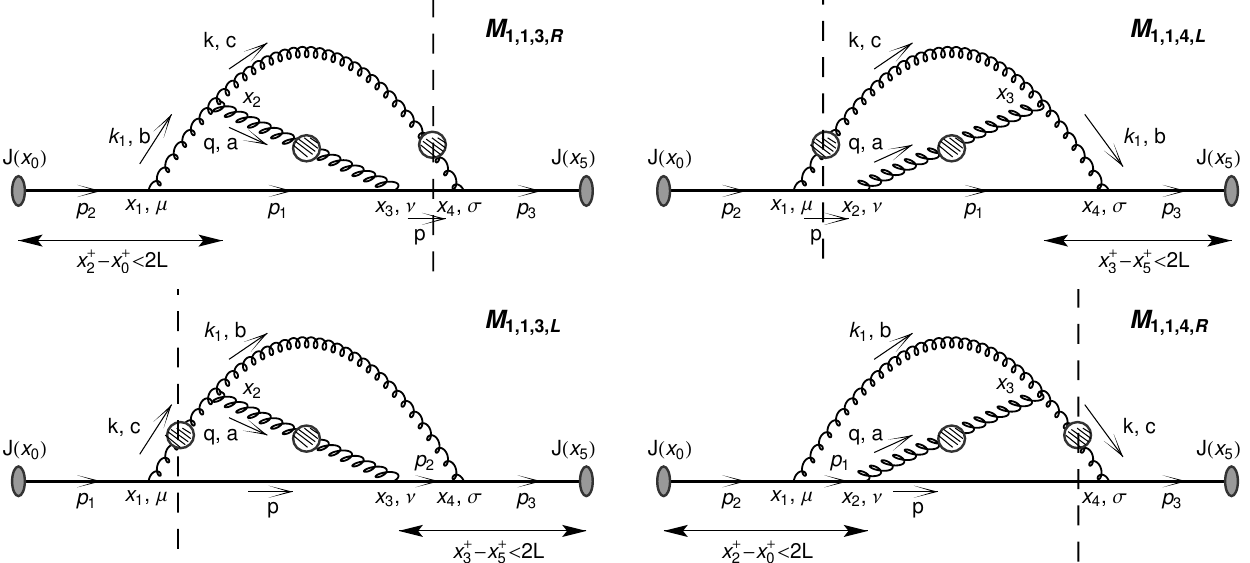}
\caption{Feynman diagrams $M_{1,1,3, R},\, M_{1,1,3, L},\, M_{1,1,4, R}$ and $M_{1,1,4, L}$, labeled in the same way as Fig.~\ref{fig_M101C}.}
\label{fig_M113L4R}
\end{figure}
\end{center}

In this Appendix, we calculate the cut diagrams $M_{1,1,3, R}$, $M_{1,1,3, L}$, $M_{1,1,4, R}$, and $M_{1,1,4, L}$ shown in Fig.~\ref{fig_M113L4R}. $M_{1,1,3, R}$ can be calculated as
\bea
M_{1,1,3, R}&= & \int \prod_{i=0}^{5} d x_{i} J\left(x_{0}\right) \Delta_{++}^{+}\left(x_{1}-x_{0}\right) v_{\mu}^{+}\left(x_{1}\right) D_{++}^{+\mu \rho_{1}}\left(x_{2}-x_{1}\right) v_{\rho_{1} \rho_{2} \rho_{3}}^{+}\left(x_{2}\right) \Delta_{++}^{+}\left(x_{3}-x_{1}\right) \nn\\
& \times & D_{++}^{+\rho_{2} \nu}\left(x_{3}-x_{2}\right) v_{\nu}^{+}\left(x_{3}\right) D_{-+}^{\rho_{3} \sigma}\left(x_{4}-x_{2}\right) v_{\sigma}^{-}\left(x_{4}\right) \Delta_{-+}\left(x_{4}-x_{3}\right) \Delta_{--}^{-}\left(x_{5}-x_{4}\right) J\left(x_{5}\right) \nn\\
& \times &\theta\left(x_{1}^{+}-x_{0}^{+}\right) \theta\left(x_{2}^{+}-x_{1}^{+}\right) \theta\left(2 L-\left(x_{2}-x_{0}\right)^{+}\right) \theta\left(x_{3}^{+}-x_{2}^{+}\right) \theta\left(x_{4}^{+}-x_{5}^{+}\right) \nn \\
&\approx & -\int_{-\infty}^{\infty} \int_{0}^{\infty} \prod_{i=1}^{3} \frac{d p_{i}^{+} d^{2} p_{i}}{(2 \pi)^{3} 2 p_{i}^{+}} \frac{d k_{1}^{+} d^{2} k_{1}}{(2 \pi)^{3} 2 k_{1}^{+}} \int \frac{d^{3} k}{(2 \pi)^{3} 2 \omega} \frac{d^{3} p}{(2 \pi)^{3} 2 E} \frac{d^{4} q}{(2 \pi)^{4}}(-i) g^{4} f^{c b a} t_{c} t_{b} t_{a} \, I, \label{M113R}
\eea
where
\bea
I&= & \int \prod_{i=0}^{5} d x_{i} \theta\left(x_{1}^{+}-x_{0}^{+}\right) \theta\left(x_{2}^{+}-x_{1}^{+}\right) \theta\left(2 L-\left(x_{2}-x_{0}\right)^{+}\right) \theta\left(x_{3}^{+}-x_{2}^{+}\right) \theta\left(x_{4}^{+}-x_{5}^{+}\right) \nn \\
& \times& e^{-i p_{2}\left(x_{1}-x_{0}\right)} e^{-i k_{1}\left(x_{2}-x_{1}\right)} e^{-i p_{1}\left(x_{3}-x_{1}\right)} e^{-i q\left(x_{3}-x_{2}\right)} e^{-i k\left(x_{4}-x_{2}\right)} e^{-i p\left(x_{4}-x_{3}\right)} e^{-i p_{3}\left(x_{5}-x_{4}\right)}\nn\\
&\times& J\left(x_{0}\right) J\left(x_{5}\right) \left(p_{1}+p_{2}\right)^{\mu}\left(p+p_{1}\right)^{\nu}\left(p+p_{3}\right)^{\sigma} P_{\mu \rho_{1}}\left(k_{1}\right) D_{\rho_{2} \nu}^{>}(q) P_{\rho_{3} \sigma}(k)\Big (g^{\rho_{1} \rho_{3}}\left(k_{1}+k\right)^{\rho_{2}}\nn\\
&+&g^{\rho_{2} \rho_{3}}(q-k)^{\rho_{1}}+g^{\rho_{1} \rho_{2}}\left(-k_{1}-q\right)^{\rho_{3}}\Big)\nn \\
&= & |J(p)|^{2}(2 \pi)^{3} \delta\left(\left(p_{2}-p_{1}-k_{1}\right)^{+}\right) \delta^{2}\left(\boldsymbol{p}_{2}-\boldsymbol{p}_{1}-\boldsymbol{k}_{1}\right)(2 \pi)^{3} \delta\left(\left(k_{1}-k+p_{1}-p\right)^{+}\right)\nn\\
&&\times \delta^{2}\left(\boldsymbol{k}_{1}-\boldsymbol{k}+\boldsymbol{p}_{1}-\boldsymbol{p}\right) (2 \pi)^{3} \delta\left(\left(p_{1}+q-p\right)^{+}\right) \delta^{2}\left(\boldsymbol{p}_{1}+\boldsymbol{q}-\boldsymbol{p}\right)(2 \pi)^{3} \delta\left(\left(p_{3}-p-k\right)^{+}\right) \nn\\
&&\times \delta^{2}\left(\boldsymbol{p}_{3}-\boldsymbol{p}-\boldsymbol{k}\right) I_{1}, \label{M113R_I}
\eea
where
\bea
I_{1}&= & \int_{0}^{2 L} d x_{2}^{\prime+} e^{-\frac{i}{2}\left(k_{1}-k+p_{1}-p\right)^{-} x_{2}^{\prime+}} \int_{0}^{x_{2}^{\prime+}} d x_{1}^{\prime+} e^{-\frac{i}{2}\left(p_{2}-p_{1}-k_{1}\right)^{-} x_{1}^{\prime+}}  \int_{0}^{\infty} d x_{3}^{\prime+} e^{-\frac{i}{2}\left(p_{1}+q-p\right)^{-} x_{3}^{\prime+}} \nn\\
&\times &\int_{0}^{\infty} d x_{4}^{\prime+} e^{\frac{i}{2}\left(p_{3}-p-k\right)^{-} x_{4}^{\prime+}}\left(p_{1}+p_{2}\right)^{\mu}\left(p+p_{1}\right)^{\nu}\left(p+p_{3}\right)^{\sigma} P_{\mu \rho_{1}}\left(k_{1}\right) D_{\rho_{2} \nu}^{>}(q) P_{\rho_{3} \sigma}(k)\nn\\
&\times &\Big (g^{\rho_{1} \rho_{3}}\left(k_{1}+k\right)^{\rho_{2}}+g^{\rho_{2} \rho_{3}}(q-k)^{\rho_{1}}+g^{\rho_{1} \rho_{2}}\left(-k_{1}-q\right)^{\rho_{3}}\Big). \nn\\ \label{M113R_I1}
\eea
Using the following relation
\bea
&&\left(p_{1}+p_{2}\right)^{\mu}\left(p+p_{1}\right)^{\nu}\left(p+p_{3}\right)^{\sigma} P_{\mu \rho_{1}}\left(k_{1}\right) D_{\rho_{2} \nu}^{>}(q) P_{\rho_{3} \sigma}(k)\Big (g^{\rho_{1} \rho_{3}}\left(k_{1}+k\right)^{\rho_{2}}+g^{\rho_{2} \rho_{3}}(q-k)^{\rho_{1}}\nn\\
&&+g^{\rho_{1} \rho_{2}}\left(-k_{1}-q\right)^{\rho_{3}}\Big) \nn \\
&&\approx\left\{\left(p_{1}+p_{2}\right)^{\mu} P_{\mu \rho_{1}}\left(k_{1}\right) P_{\sigma}^{\rho_{1}}(k)\left(p+p_{3}\right)^{\sigma}\right\}\left\{\left(k+k_{1}\right)^{\rho_{2}} D_{\rho_{2} \nu}^{>}(q)\left(p+p_{1}\right)^{\nu}\right\}\nn\\
&& \approx- \frac{4\boldsymbol{k} \cdot(\boldsymbol{k}+\boldsymbol{q})}{x^{2}} \theta\left(1-\frac{q_{0}^{2}}{\overrightarrow{\mathbf{q}}^{2}}\right) \frac{T}{q_0} E^{+} k^{+} \frac{\boldsymbol{q}^{2}}{\overrightarrow{\boldsymbol{q}}^{2}} 2 \operatorname{Im}\left(\frac{1}{q^{2}-\Pi_{L}(q,T)}-\frac{1}{q^{2}-\Pi_{T}(q,T)}\right) \label{M113R_approx2}
\eea
in Eq.~\eqref{M113R_I1}, one obtains
\bea
I_1&=& \int_{0}^{2 L} d x_{2}^{\prime+} e^{-\frac{i}{2}\left(k_{1}-k+p_{1}-p\right)^{-} x_{2}^{\prime+}} \int_{0}^{x_{2}^{\prime+}} d x_{1}^{\prime+} e^{-\frac{i}{2}\left(p_{2}-p_{1}-k_{1}\right)^{-} x_{1}^{\prime+}}  \int_{0}^{\infty} d x_{3}^{\prime+} e^{-\frac{i}{2}\left(p_{1}+q-p\right)^{-} x_{3}^{\prime+}} \nn\\
&\times &\int_{0}^{\infty} d x_{4}^{\prime+} e^{\frac{i}{2}\left(p_{3}-p-k\right)^{-} x_{4}^{\prime+}} \Big(- \frac{4\boldsymbol{k} \cdot(\boldsymbol{k}+\boldsymbol{q})}{x^{2}} \Big)\theta\left(1-\frac{q_{0}^{2}}{\overrightarrow{\mathbf{q}}^{2}}\right) \frac{ T}{q_0} E^{+} k^{+} \frac{\boldsymbol{q}^{2}}{\overrightarrow{\boldsymbol{q}}^{2}} \nn\\
&&\times 2 \operatorname{Im}\left(\frac{1}{q^{2}-\Pi_{L}(q,T)}-\frac{1}{q^{2}-\Pi_{T}(q,T)}\right) .\nn\\
\label{M113R_I1b}
\eea

By using Eq.~\eqref{M101C_approx1} and the $\delta$ functions from Eq.~\eqref{M113R_I}, we obtain, under the soft gluon and soft rescattering approximation,
\bea
& & p_{2}^{-}=p_{3}^{-}=\frac{M^{2}}{E^{+}}, \nn\\
&& \left(p_{2}-p_{1}-k_{1}\right)^{-}=-\zeta(T), \nn \\
&& \left(p_{3}-p-k\right)^{-}=\left(p_{2}-p-k\right)^{-}=-\xi(T), \nn \\
&& \left(k_{1}-k+p_{1}-p\right)^{-}=\zeta(T)-\xi(T), \nn \\
&& \left(q+p_{1}-p\right)^{-} \approx q_{0}-q_{z} . \label{M113R_approx1}
\eea
Using Eq. \eqref{M113R_approx1}, Eq. \eqref{M113R_I1b} becomes
\bea
I_1 &=&\int_0^{L} dl_2 \left( e^{-i\left(\zeta(T)-\xi(T) \right)l_2}-e^{i\xi(T) l_2}\right) \int_{0}^{\infty} dl_3\, e^{-i q^{-} l_3} \, H(q,T(l_2)), \label{M113R_I1f}
\eea
where $H(q,T(l_2))$ is defined in Eq.~\eqref{H_def}.

Finally, by using Eqs.~\eqref{H_def} and~\eqref{M113R_I1f}, Eq. \eqref{M113R} reduces to
\bea
M_{1,1,3, R}&= & 2\left[t_{a}, t_{c}\right]\left[t_{c}, t_{a}\right]\int_0^{L} dl_2 \left( e^{-i\left(\zeta(T)-\xi(T) \right)l_2}-e^{i\xi(T) l_2}\right)\int_{0}^{\infty} dl_3\,e^{-i q^{-} l_3} \int \frac{d^{3} p}{(2 \pi)^{3} 2 E}|J(p)|^{2} \nn\\
&&\times \int \frac{d^{3} k}{(2 \pi)^{3} 2 \omega} \int \frac{d^{4} q}{(2 \pi)^{4}}   \frac{\boldsymbol{k} \cdot(\boldsymbol{k}+\boldsymbol{q})}{\left((\boldsymbol{k}+\boldsymbol{q})^{2}+\chi(T)\right)\left(\left(\boldsymbol{k}^{2}+\chi(T)\right)\right.}  \theta\left(1-\frac{q_{0}^{2}}{\overrightarrow{\mathbf{q}}^{2}}\right) g^{4}\frac{T}{q_0} \frac{\boldsymbol{q}^{2}}{\overrightarrow{\mathbf{q}}^{2}} \nn\\
&&\times  2 \operatorname{Im}\left(\frac{1}{q^{2}-\Pi_{L}(q,T)}-\frac{1}{q^{2}-\Pi_{T}(q,T)}\right) .
\eea
$M_{1,1,4, L}$ is the complex conjugate of $M_{1,1,3, R}$. Moreover, the relationship $M_{1,1,3, L} + M_{1,1,4, R} = M_{1,1,3, R} + M_{1,1,4, L}$ holds, leading to the final result.
\bea
M_{1,1,3, L} &+&  M_{1,1,3, R} + M_{1,1,4, R}+M_{1,1,4, L} =\nn\\
&=& 4 \left[t_{a}, t_{c}\right]\left[t_{c}, t_{a}\right] \int_0^{L} d\tau   \int \frac{d^{3} p}{(2 \pi)^{3} 2 E}|J(p)|^{2} \int \frac{d^{3} k}{(2 \pi)^{3} 2 \omega} \int \frac{d^{2} q}{(2 \pi)^{2}} \, g^{4}\,T\, v(q,T) \nn\\
&& \times \frac{\boldsymbol{k} \cdot(\boldsymbol{k}+\boldsymbol{q})}{\left((\boldsymbol{k}+\boldsymbol{q})^{2}+\chi(T)\right)\left(\boldsymbol{k}^{2}+\chi(T)\right)}\Big ( \cos{\big((\xi(T)-\zeta(T))\tau \big )}-\cos{\left(\xi(T) \tau \right)}\Big).
\label{M113L4R_f}
\eea

\section{Calculation of Diagram $M_{1,2, C}$}
\label{app_M12C}
\begin{center}
\begin{figure}[htbp]
\includegraphics[scale=0.7]{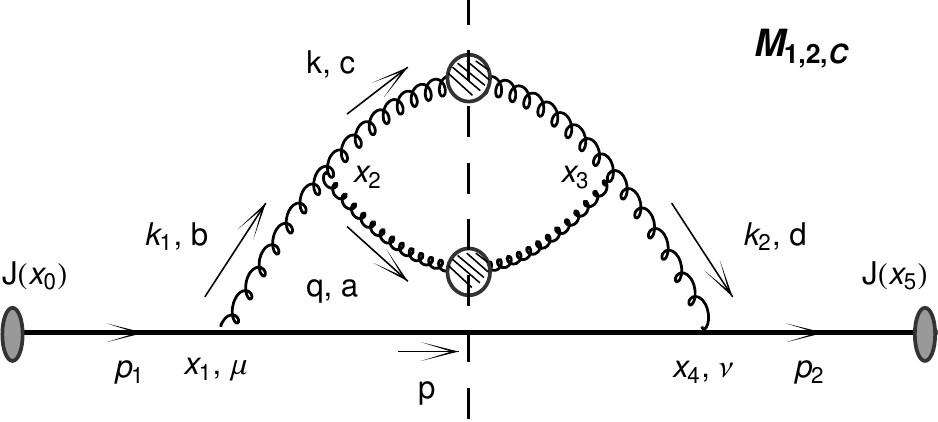}
\caption{Feynman diagram $M_{1,2,C}$, labeled in the same way as Fig.~\ref{fig_M101C}.}
\label{fig_M12C}
\end{figure}
\end{center}

In Appendices \ref{app_M12C} and \ref{app_M12RL}, we calculate the diagrams where both ends of the exchanged gluon $q$ are attached to the radiated gluon $k$, involving two 3-gluon vertices in the process. Here, we compute the diagram $M_{1,2, C}$ shown in Fig.~\ref{fig_M12C}.
\bea
M_{1,2, C}&= & \int \prod_{i=0}^{5} d x_{i} J\left(x_{0}\right) \Delta_{++}^{+}\left(x_{1}-x_{0}\right) v_{\mu}^{+}\left(x_{1}\right) D_{++}^{+\mu \rho_{1}}\left(x_{2}-x_{1}\right) v_{\rho_{1} \rho_{2} \rho_{3}}^{+}\left(x_{2}\right) D_{-+}^{\rho_{3} \sigma_{3}}\left(x_{3}-x_{2}\right) \nn \\
&& \times D_{-+}^{\rho_{2} \sigma_{2}}\left(x_{3}-x_{2}\right) \Delta_{-+}\left(x_{4}-x_{1}\right) v_{\sigma_{1} \sigma_{2} \sigma_{3}}^{-}\left(x_{3}\right) D_{--}^{-\sigma_{1} \nu}\left(x_{4}-x_{3}\right) v_{\nu}^{-}\left(x_{4}\right) \Delta_{--}^{-}\left(x_{5}-x_{4}\right) J\left(x_{5}\right) \nn \\
& &\times \theta\left(x_{1}^{+}-x_{0}^{+}\right) \theta\left(x_{2}^{+}-x_{1}^{+}\right) \theta\left(2 L-\left(x_{2}-x_{0}\right)^{+}\right) \theta\left(x_{3}^{+}-x_{4}^{+}\right) \theta\left(x_{4}^{+}-x_{5}^{+}\right)\nn\\
&&\times \theta\left(2 L-\left(x_{3}-x_{5}\right)^{+}\right) .\label{M12C}
\eea

Following Appendix~\ref{app_M1112C} and applying the following relations:
\bea
\left(p+p_{1}\right)^{\mu} P_{\mu \rho_{1}}\left(k_{1}\right) P^{\sigma_{1} \rho_{1}}(k) P_{\sigma \nu}\left(k_{2}\right)\left(p+p_{1}\right)^{\nu} & =&\left(p+p_{1}\right)^{\mu} P_{\mu \rho_{1}}\left(k_{1}\right) P^{\sigma_{1} \rho_{1}}(k) P_{\sigma_{1} \nu}\left(k_{1}\right)\left(p+p_{1}\right)^{\nu} \nn\\
&\approx&-\frac{4(\boldsymbol{k}+\boldsymbol{q})^{2}}{x^{2}} \nn\\
\left(k+k_{1}\right)^{\rho_2} D_{\rho_{2} \sigma_{2}}^{>}(q,T)\left(k+k_{2}\right)^{\sigma_{2}}  \approx   k^{+} k_{1}^{+}\theta\left(1-\frac{q_{0}^{2}}{\overrightarrow{\mathbf{q}}^{2}}\right)\frac{T}{q_0}\!\!\!\!&&\!\!\!\!\! \frac{\boldsymbol{q}^{2}}{\overrightarrow{\mathbf{q}}^{2}} 2 \operatorname{Im}\left(\frac{1}{q^{2}-\Pi_{L}(q,T)}-\frac{1}{q^{2}-\Pi_{T}(q,T)}\right), \label{M12C_reln}
\eea
Eq.~\eqref{M12C} becomes
\bea
M_{1,2, C}&=&8 \left[t_{a}, t_{c}\right]\left[t_{c}, t_{a}\right]\int_0^{L} d\tau \int \frac{d^{3} p}{(2 \pi)^{3} 2 E}|J(p)|^{2} \int \frac{d^{3} k}{(2 \pi)^{3} 2 \omega} \int \frac{d^{2} q}{(2 \pi)^{2}}  \,g^{4}\, T\, v(q,T)\nn\\
&& \times \frac{(\boldsymbol{k}+\boldsymbol{q})^{2}}{\left((\boldsymbol{k}+\boldsymbol{q})^{2}+\chi(T)\right)^{2}} \left( 1-\cos{(\zeta(T) \tau)}\right). \label{M12C_f}
\eea
\section{Calculation of Diagrams $M_{1,2, R}$ and $M_{1,2, L}$}
\label{app_M12RL}
In this Appendix, we calculate the cut diagrams $M_{1,2, R}$ and $M_{1,2, L}$ shown in Fig.~\ref{fig_M12RL}. We start with $M_{1,2, R}$:
\begin{center}
\begin{figure}[htbp]
\includegraphics[scale=0.7]{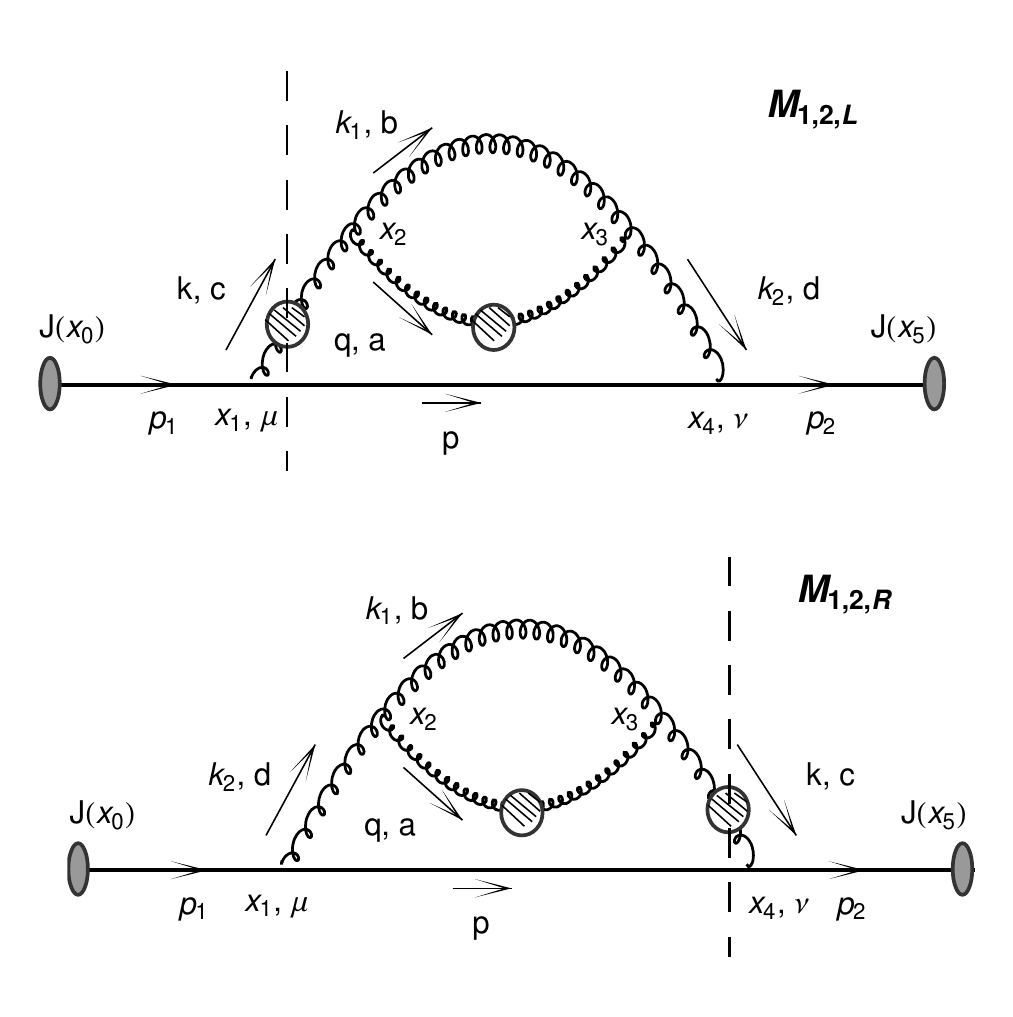}
\caption{Feynman diagrams $M_{1,2,L}$ and $M_{1,2,R}$ labeled in the same way as Fig.~\ref{fig_M101C}.}
\label{fig_M12RL}
\end{figure}
\end{center}
\bea
M_{1,2, R}&= & \int \prod_{i=0}^{5} d x_{i} J\left(x_{0}\right) \Delta_{++}^{+}\left(x_{1}-x_{0}\right) v_{\mu}^{+}\left(x_{1}\right) D_{++}^{+\mu \rho_{1}}\left(x_{2}-x_{1}\right) v_{\rho_{1} \rho_{2} \rho_{3}}^{+}\left(x_{2}\right) D_{++}^{+\rho_{3} \sigma_{3}}\left(x_{3}-x_{2}\right) \nn\\
&& \times D_{++}^{+\rho_{2} \sigma_{2}}\left(x_{3}-x_{2}\right) \Delta_{-+}\left(x_{4}-x_{1}\right) v_{\sigma_{1} \sigma_{2} \sigma_{3}}^{+}\left(x_{3}\right) D_{-+}^{\sigma_{1 \nu}}\left(x_{4}-x_{3}\right) v_{\nu}^{-}\left(x_{4}\right) \Delta_{--}^{-}\left(x_{5}-x_{4}\right) J\left(x_{5}\right) \nn\\
&& \times \theta\left(x_{1}^{+}-x_{0}^{+}\right) \theta\left(x_{2}^{+}-x_{1}^{+}\right) \theta\left(x_{3}^{+}-x_{2}^{+}\right) \theta\left(x_{4}^{+}-x_{5}^{+}\right) \theta\left(2 L-\left(x_{2}-x_{0}\right)\right) . \label{M12R}
\eea
Equation~\eqref{M12R} can be computed by following Appendix~\ref{app_M113L4R} and applying the following approximations:
\bea
&&\left(p+p_{1}\right)^{\mu} P_{\mu \rho_{1}}\left(k_{2}\right) P^{\rho_{1} \sigma_{1}}(k) P_{\sigma_{1} \nu}(k)\left(p+p_{2}\right)^{\nu}=-\frac{4 \boldsymbol{k}^{2}}{\xi(T)^{2}} \nn \\
&&\left(k+k_{1}\right)^{\rho} D_{\rho \sigma}^{>}(q, T)\left(k+k_{1}\right)^{\sigma} \approx k^{+} k_{1}^{+} \theta\left(1-\frac{q_{0}^{2}}{\overrightarrow{\mathbf{q}}^{2}}\right) \frac{T}{q_0} \frac{\boldsymbol{q}^{2}}{\overrightarrow{\mathbf{q}}^{2}} 2 \operatorname{Im}\left(\frac{1}{q^{2}-\Pi_{L}(q,T)}-\frac{1}{q^{2}-\Pi_{T}(q,T)}\right), \nn\\ \label{M12R_reln}
\eea
as
\bea
M_{1,2, R}&= & -4  g^{4}\left[t_{c}, t_{a}\right]\left[t_{a}, t_{c}\right]\int_0^{L} dl_2 \left( 1-e^{i\xi(T) l_2}\right)\int_{0}^{\infty} d l_3 e^{-i q^{-} l_3}\int \frac{d^{3} p}{(2 \pi)^{3} 2 E}|J(p)|^{2} \int \frac{d^{3} k}{(2 \pi)^{3} 2 \omega} \nn\\
&&\times \int \frac{d^{4} q}{(2 \pi)^{4}}  \frac{\boldsymbol{k}^{2}}{\left(\boldsymbol{k}^{2}+\chi(T)\right)^{2}}  \theta\left(1-\frac{q_{0}^{2}}{\overrightarrow{\mathbf{q}}^{2}}\right) \frac{ T}{q_{0}} \frac{\boldsymbol{q}^{2}}{\overrightarrow{\mathbf{q}}^{2}} 2 \operatorname{Im}\left(\frac{1}{q^{2}-\Pi_{L}(q,T)}-\frac{1}{q^{2}-\Pi_{T}(q,T)}\right) . \nn\\
\label{M12R_f}
\eea
As $M_{1,2, L}$ is a complex conjugate of $M_{1,2, R}$, one finally obtains
\bea
M_{1,2, R}+M_{1,2, L}&= &  -4 \left[t_{a}, t_{c}\right]\left[t_{c}, t_{a}\right]\int_0^{L} d\tau \int \frac{d^{3} p}{(2 \pi)^{3} 2 E}|J(p)|^{2} \int \frac{d^{3} k}{(2 \pi)^{3} 2 \omega} \int \frac{d^{2} q}{(2 \pi)^{2}}\,g^{4} \,T \,v(q,T) \nn\\
&& \times \frac{\boldsymbol{k}^{2}}{\left(\boldsymbol{k}^{2}+\chi(T)\right)^{2}}\left( 1-\cos{( \xi(T) \tau)}\right)  . \label{M12RL_f}
\eea


\end{document}